\documentclass[12pt,runningheads,book,hyperref]{llncs}

\usepackage[utf8]{inputenc}

\usepackage{amssymb}
\usepackage{graphicx}

\usepackage{appendix}

\usepackage{hyperref}
\hypersetup{
    colorlinks,
    citecolor=black,
    filecolor=black,
    linkcolor=black,
    urlcolor=black
}

\usepackage{geometry}
\usepackage{url}

\newcommand{\keywords}[1]{\par\addvspace\baselineskip
\noindent\keywordname\enspace\ignorespaces#1}

\newcommand{\beq}{\begin{equation}}
\newcommand{\eeq}{\end{equation}}
\newcommand{\la}{\langle}
\newcommand{\ra}{\rangle}
\newcommand{\e}{{\rm{e}}}

\mainmatter  

\title{Event-based complexity in atmospheric turbulence}

\titlerunning{Events in turbulence}

\author{Paolo Paradisi$^{1,*}$
\and Rita Cesari$^2$}
\authorrunning{Paradisi and Cesari}
\institute{$^1$ Istituto di Scienza e Tecnologie dell'Informazione (ISTI-CNR), \\Via G. Moruzzi 1, 56124 Pisa, Italy
\\
$^2$ Istituto di Scienze dell'Atmosfera e del Clima (ISAC-CNR), \\Strada Prov.le Lecce-Monteroni Km 1,200 73100 Lecce, Italy
\\
}

\begin{document}

\maketitle

\begin{abstract}
Since the studies of Kolmogorov and Oboukhov in $1941$, the problem of intermittent large velocity excursions was recognized to be one
of the most intriguing and elusive aspects of turbulent flows.
While many efforts were devoted since $1960$s to the magnitude intermittency, related to the statistics of the large increments in the turbulent signals, the attention towards the so-called clustering intermittency has started to increase only in the last two decades, even if some pioneering studies were carried in 
previous years. 
The low attention towards the clustering intermittency is somewhat surprising, as intermittency itself is essentially defined as an alternance of extended quiescent periods/regions and short/small high activity periods/zones. 
It is then natural to characterize intermittency by means of 
events marked along the dependent variable axes, whatever time or space.
Conversely, the concept of crucial events and of temporal complexity, related to non-trivial clustering properties of these same events, has been proposed as a general interpretative framework for the investigation of complex systems with metastable self-organizing dynamics.\\
Without any claim to being complete, in this chapter we give a review of event-based complexity approaches used in literature for the description of turbulence in the planetary boundary layer.
The main goal is to put a first bridge between turbulence studies exploiting event-based complexity approaches, such as clustering exponents and classical Oboukhov intermittency exponents,
and studies about intermittent complex systems, where concepts and ideas developed in the fields of non-equilibrium statistical physics, probability theory and dynamical system theory are jointly exploited.
\end{abstract}

%
$[*]$ Corresponding author;\ \
Email: paolo.paradisi@cnr.it

\large

\vspace{1.cm}
\noindent
{\bf 
This is a pre-print version. Please cite as:\\ 
Paradisi, P. and Cesari, R.,  Event-based complexity in turbulence, in
“Paolo Grigolini and 50 Years of Statistical Physics” (edited by Bruce J. West and
Simone Bianco), Cambridge Scholar Publishing, pp. 199-277 (2023). ISBN: 1-5275-0222-8
}

\normalsize

\tableofcontents

\section{Introduction}
\label{intro_sect}

The term {\it fluid or hydrodynamical turbulence} denotes a dynamical condition of
a fluid flow characterized by ``certain complex and unpredictable motion''
\cite{benzi-frish_2010} and, in particular, by the emergence of
vortex structures. More generally, turbulence is associated with
the emergence of self-organized structures that can have different statistical and geometrical features, depending also on the 
range of considered space/time scales and on the boundary 
conditions.
The main reference hydrodynamical model for fluid turbulence is given by the so-called {\it Newtonian fluid}, where a linear stress-strain relationship is
assumed. This led to introduce a viscosity matrix and, under the isotropy assumption, which is a natural one in many applications, a dynamic viscosity coefficient
$\mu_f$. 
By also adding an incompressibility hypothesis, the following Navier-Stokes motion equation can be derived for the 
velocity field ${\bf u}({\bf x},t)$ \cite{schlichting_2017,tampieri_2017}
of a isotropic, incompressibile Newtonian fluid\footnote{
It is worth noting that here the fluid is isotropic, in the sense that the stress-strain relationship is isotropic, but the {\it fluid flow}, i.e., the velocity field, can be anisotropic depending on the boundary and initial conditions.
}:
\beq
\frac{\partial {\bf u}}{\partial t} + {\bf u}\cdot {\bf \nabla} {\bf u} = -\frac{1}{\rho_f}{\bf \nabla} P({\bf x},t) + \nu \nabla^2 {\bf u} + {\bf F}({\bf x},t)
\label{navier_stokes}
\eeq
where $P({\bf x},t)$ is the pressure, ${\bf F}({\bf x},t)$ a
generic term encoding an external forcing such as, e.g., gravity and buoyancy, $\rho_f$ the fluid density (constant for incompressible fluids) and $\nu = \mu_f / \rho_f$ the kinematic
viscosity.
The Navier-Stokes equation is ubiquitously applied to many viscous fluids, including water and atmosphere. 
For these same Newtonian fluids, a huge number of experimental studies prove that turbulence emerges when 
the nonlinear advective term ${\bf u}\cdot {\bf \nabla} {\bf u}$ becomes dominant with respect to the 
viscous term $\nu \nabla^2 {\bf u}$, where the first term encodes momentum transport by the velocity field and the second one momentum diffusion by molecular viscosity.
This condition is parameterized by means of the 
adimensional Reynolds number:
\beq
Re = \frac{U\ L}{\nu}\ \Leftrightarrow \frac{\rm advection\ or\ inertia}{\rm viscous\ dissipation} ,
\eeq
being $U$ and $L$ a velocity and length scale coming out
from the geometrical and dynamical boundary conditions.
Being the Reynolds number a reference value for the ratio between advective and viscous terms, thus the onset of turbulence corresponds to the limit
$Re \rightarrow \infty$ and, thus, to vanishing viscosity.
This interplay or competition among inertia forces, given by the nonlinear advective term, and
viscous dissipation is the most interesting aspect of Navier-Stokes hydrodynamics, as it hides
the still elusive mechanism triggering the onset of turbulence \cite{frisch_1995}.
The most problematic point lies in the so-called {\it dissipative anomaly} \cite{benzi-frish_2010}. In fact,
the dissipation goes to zero with the viscosity $\nu$ when the fluid flow is laminar, but this does not occur in the 
turbulent case.
%
\vspace{.15cm}
{\it Atmospheric turbulence}

\vspace{.1cm}
\noindent
%
Atmospheric turbulence is a very intriguing phenomenon that
characterizes the so-called Planetary Boundary Layer (PBL) and
that has attracted the interest of many researchers since decades.
Oboukhov himself got a fundamental result from the 
analysis of atmospheric measurements \cite{oboukhov_1962},
that is, the confirmation of the Kolmogorov-Oboukhov 
$-5/3$ spectral exponent 
\cite{kolmogorov-1941b,kolmogorov-1941,oboukhov-1941} and, at 
the same time, the failure of the assumed universality of the 
pre-factor in the Power Spectral Density (PSD).\\
PBL turbulence is, in fact, the turbulent fluid flow 
reaching the higher Reynolds numbers and, thus, going
very near to the ideal case $Re \rightarrow \infty$.\\
The interest towards PBL turbulence is also related to the
problem of pollution, as turbulence near the ground
highly affect the transport properties of contaminants
and, thus, their diffusivity near the surface, thus a better understanding of PBL turbulence can give an effort in the improvement of local, regional and global forecasting, thus
impacting on pollution management policies at both local and
global levels.\\
However, the turbulent PBL is clearly not a controlled
laboratory experiment and, thus, the analysis and modeling
of turbulent motions can be quite dependent on the 
particular atmospheric condition and, especially, on the
particular site. 
For this reason, after the pioneering era
of turbulence studies in the PBL over flat terrain, the interest of the  scientific community has moved towards the so-called {\it complex terrain} problem, i.e., fluid flow over and inside vegetated and urban canopies, and/or in the region very near to the ground ($< 1\ m$). In both cases, the effect of surface geometry and of roughness is directly felt by the turbulent flow, so that it becomes much more difficult to
discriminate the direct effect of both geometry and local
sources over the ground from the internal dynamics of
the fluid flow.\\
A further complication is the modulation of ``local'' PBL turbulence dynamics,
especially by mesoscale dynamics, classically referring to scales of $10-1000$ km and $1-10$ hours\footnote{
This reference spatial and temporal scales are the classical ones and represent a quite rough classification that is nowadays almost overwhelmed by finer
classifications. We do not go into these details here, but the reader can
refer to recent literature (see, e.g., \cite{mahrt_2014,sun_2015}).
}, 
but also by synoptic-scale atmospheric dynamics, involving spatial scales of thousands km and temporal scales of days.
%

\vspace{.15cm}
{\it Richardson's energy cascade and K41 model}

\vspace{.1cm}
\noindent
The finite non-zero limit of the mean dissipation rate of Turbulent Kinetic Energy (TKE), usually denoted as $\langle \epsilon \rangle$ or $\overline{\epsilon}$, is one of the assumptions exploited by Kolmogorov  and Oboukhov to
derive the famous $2/3$ and $5/3$ laws, \cite{kolmogorov-1941b,kolmogorov-1941,oboukhov-1941}, that we will briefly introduce later and are named {\it K41 model} or {\it K41 theory} in the literature\footnote{
Considering the important contribution to this model by
Oboukhov, this should better denoted as the
Kolmogorov-Oboukhov, i.e., KO41 model.
}. 
Another concept exploited in the K41 model is the concept of turbulent energy cascade, conjectured for the first time by Richardson in 1920 \cite{richardson_1920}. This concept
is described in a famous sentence by Richardson itself that can be found at
page 66 of his book 
``Weather Prediction by Numerical Processes'' \cite{richardson_1922} and that sounds very much like a poem:

\vspace{.1cm}
\noindent
``Big whirls have little whirls\\
that feed on their velocity,\\
And little whirls have lesser whirls\\
and so on to viscosity -\\
in the molecular sense.''

\vspace{.1cm}
\noindent
Thus, energy flows from the largest scales of {\it energy-containing 
eddies}\footnote{
The term {\it energy-containing}
is related to the maximum of the turbulent energy (power) spectrum,
whose location is taken as an approximate estimate of the length scale
of energy-containing eddies themselves.
}, 
which are directly fed by some source of energy through an external 
forcing, to smallest scales, where TKE is directly 
dissipated into heat as a consequence of viscosity. 
In the Richardson picture, the passage of energy is among near scales.\\
In K$41$ model, Kolmogorov assumed that the energy flowing across scales was constant across scales themselves,
which corresponds to assume
a {\it quasi-equilibrium} picture between energy fluxes from the largest scales down to
the smallest scales, namely the scales where viscous dissipation equals production of
kinetic energy.
This quasi-equilibrium condition defines the {\it inertial subrange}, i.e., the set of scales
where the TKE coming from the large scales of motion
is in equilibrium with the TKE that is dissipated into heat at the smallest scales due to
viscosity.
Thus, according to this Richardson-Kolmogorov view of an inertial subrange at
equilibrium, the energy is transferred from larger to contiguous smaller scales of motion
without loss of TKE at intermediate scales, until the dissipation scale is reached.
By definition, the scales in the inertial subrange do not 
feel neither the large-scale forcing nor the 
viscous dissipation at small scales. This range of scales is usually
delimited by the scales $L$ and $\lambda_K$.\\
The so-called integral scale $L$ is defined as the mean correlation length
of the energy-containing eddies. The dissipation scale $\lambda_K$, also
named Kolmogorov microscale, is the smallest scale with some kind of flow structure,
i.e., the scale of the smallest eddies beyond which the viscous forces 
dissipate the TKE into heating \cite{batchelor_1953}. 

\vspace{.1cm}
{\it Intermittency and events}

\vspace{.1cm}
\noindent
The K41 model is still a
milestone in turbulence studies, both experimental and theoretical, and the
$5/3$ law a reference to which compare new models and analyses are compared.
However, as we will see in Section \ref{intermit_turb_sect}, a few years later it was already clear
that K41 model needed an
improvement due to the non-negligible presence of intermittency
in the velocity increments and, thus, in the dissipation rates
\cite{batchelor_1949}.\\
Interestingly, most studies on turbulence intermittency are
devoted to the so-called Magnitude Intermittency (MI),  while a more recent 
interest was focused on the so-called Clustering Intermittency (CI), that 
was almost neglected in the {\it first era} of turbulence studies 
\cite{bershadskii_2004,katul_2006_surf_ren,sreenivasan_2006}.\\
%
As the name says, MI is related to the amplitude variations, i.e., to the increments of
turbulent signals, and investigates the distribution of, and correlations among, such
increments.
This allowed to characterizes not only the non-Gaussianity of signal increment distribution, but also
its multiscaling features \cite{ott_2002}, being multiscaling a synonim of intermittency in
turbulence studies. In fact, as known, a non-Gaussian distribution is
the signature of large excursions in the signal that are less rare than in the Gaussian
case, a condition possibly
related to the emergence of anomalous diffusion
\cite{metzler_jpa2004_ctrw_fract}
while the multiscaling in the moment of signal increments, also denoted as
{\it structure functions}, is
a signature of so-called {\it strong anomalous
diffusion} \cite{andersen_2000,castiglione_1999}.\\
Conversely, CI looks at the temporal (or spatial) order, that is,  at the distances between
the time instants or the space points at which large excursions occur. 
As large excursions occur in a short time and are relatively rare, this clearly reminds 
the concept of {\it event}, more precisely, of large excursion event, and the system's
dynamical or geometrical properties can often be reduced to a set of point in time and/or
space.
The distribution of (time or space) distances among events and the inter-event
correlations are important examples of CI measures. When the distribution is found to
display a decaying stretched exponential or, more importantly, a slow power-law decay,
the large excursion events become {\it crucial events}
\cite{west-grigolini_2021_book_events}.
In fact, the power-law decay is a signature of self-similarity and, thus, of some kind of
(self)-organization in the turbulent flow.\\
%
Even if less studies were devoted to a event-based
approach, the interest towards the time/space 
ordering of large excursion events is increasing
in the last decade or so. Further, the event-based approaches are somewhat associated
with the
concurrent research activity on turbulent flow structures, e.g., hairpin vortices in wall
flows \cite{luchik_1987,tardu_2014,tubergen_1993}.
The availability of more accurate turbulence
data could trigger in the next future a rapid
increase in the interest of the scientific
community. 

\vspace{.1cm}
\noindent
Interestingly, similar approaches based on the concept of crucial events were recently
proposed as a general theoretical picture
of complex systems with metastable self-organizing dynamics
\cite{west-grigolini_2021_book_events}.\\
Thus, the main goal is here to put a first bridge between turbulence studies exploiting event-based complexity approaches and studies about intermittent complex systems, where the focus
is on the paradigm of cooperation and 
emergence of metastable self-organization.

\vspace{.25cm}
\noindent
Far from being complete, in this chapter we give a brief
review of the most recent research in turbulence, with particular attention to PBL turbulence, in the perspective of intermittency and focusing on approaches in the framework of so-called {\it event-based complexity}, where the detection
of single crucial events, and of associated turbulent structures, is a central component of the analysis and 
modeling of turbulence data.
We will carry out a survey of methods and results in
event-based approaches and, thus, on its different definitions and associated detection algorithms. In particular, we will
discuss some of the most used statistical indices, or event-based complexity measures.
In summary, in this review chapter we are interested in the CI more than in the MI. However, due to its importance we will also briefly review fundamental results of turbulence MI.

\vspace{.2cm}
\noindent
The chapter is organized as follows.\\
In Section \ref{intermit_complex_sect} a short introduction to intermittent complex systems and to the concept of crucial event is given, also including a brief survey of the dynamical foundations of renewal point processes.
In Section \ref{intermit_turb_sect} a brief outline of the main results about magnitude intermittency in turbulence is sketched,
including a discussion on the differences between magnitude and clustering intermittency.
To the sake of completeness, Section \ref{pbl_sect} is devoted to a short introduction of the planetary boundary layer, where turbulence emerges.
In Section \ref{structures_and_events_sect} the interplay between self-organized turbulent structures and crucial events is discussed and a convention about terminology is given.
Sections \ref{event_detect_sect} and \ref{event_based_complex_sect} are devoted to a short survey of
the most used event detection algorithms and event-based complexity measures/exponents, respectively. Section \ref{soc_eddis_sect} discusses significant relations between some event-based complexity exponents and the applications of these same relations to data analysis.
Section \ref{intermit_pbl_sect} reports an outline of some applications of event-based complexity analysis to PBL turbulence data that, in our opinion, can be considered as the most interesting ones in the recent literature.
Finally, in Section \ref{concl_sect} we sketch a brief discussion of the conclusions reached.

\section{A very short introduction to intermittent complex systems}
\label{intermit_complex_sect}

A great interest has rapidly increased in the last two decades towards  systems, also including social dynamics,
with many units that cooperate having the
goal of maximizing self-regulation capacity and flexibility under the effect of rapidly changing environmental conditions.
Essentially, there is a change in the paradigm, where the focus is on the {\it emerging self-organizing} dynamics of cooperative nonlinear systems. This paradigm is usually referred to as {\it complexity science}, a field of research that combines the experiences and expertise of many different scientific and technological fields.
This paradigm nowadays involves not only the scientific
community, but also policy makers, private
companies and other socio-economic actors,
and follows from the capability of nowadays observation systems to collect and store huge amounts of data.
In fact, the present temporal accuracy of data acquisition
is accompanied by the collection of countless aspects 
from a single system component that can be simultaneously
observed on very large numbers of these same components
without resorting to bulk measurements.
These single components are representative of working units, particles, agents, fluid regions or other depending on the specific research field.
Thus, the ``mining'' on these large datasets is
nowadays requiring increasingly refined and powerful algorithms involving so-called
feature extraction. Artificial Intelligence (AI), big data mining
and complex network analysis \cite{boccaletti_2006,roh_2021_ai,topol_2019_ai,zanin-boccaletti_2016}
are examples of hot topics ubiquitously exploited in nowadays research that involve feature extraction algorithms.

\vspace{.1cm}
\noindent
In a nutshell, we can say that complexity science investigates {\it how self-organization emerges from
cooperative dynamical systems and what functionality
of the entire complex system the above self-organizing
behavior is optimizing}.\\
Even if complexity is a somewhat elusive concept
and a universally accepted definition does not
exist, we here refer to Ref. \cite{paradisi_springer2017}, where a cooperative system is defined to be {\it complex} if some
kind of self-organized structures can be recognized by means of statistical indices
displaying self-similarity (mono- or multi-scaling) and there is not a master directly driving the great majority of the system's components, if not all, i.e., 
almost all the internal degrees of freedom 
are not slave variables of an external forcing.

\vspace{.15cm}
\noindent
The complexity paradigm has yet a long history, starting from very different scientific disciplines, such as 
sociology, psychology, cybernetics \cite{bateson_1972},
physics and chemistry \cite{nicolis_1977,prigogine_1955},
and biology \cite{maturana-varela_1980}.\\
More recently, the network science introduced new perspectives in the way  complex systems are modeled
and analyzed, e.g., introducing concept such as
connectivity measures derived from the topological
structure of the complex network \cite{boccaletti_2006,zanin-boccaletti_2016}.\\
Nowadays the complexity paradigm as emergence of 
self-organizing states 
is
exploited in many different research fields and 
applications, such as: brain
cortex organization in neurosciences \cite{bullmore-sporns_2009,buzsaki_2009,thomasyeo_2011};
graph theory with application to social networks, e.g.,
the detection of communities \cite{fortunato_2010,palla_2005,rosvall_2008}; detection of competing clusters in economy \cite{porter_1998};
resilience of social networks and activities 
\cite{duchek_2020,mahmoodi-grigolini_2018_temp_complex}, a concept that has come nowadays to the forefront in the recent period in view of a response to the pandemic effects of Covid-19 on socio-economic activities \cite{paradisi_fbn2021};
all the -omics, e.g., genomics \cite{misteli_2020}, proteomics and transcritomics 
\cite{barabasi_2004}, metabolomics \cite{jeong_2000}.
%

\vspace{.15cm}
\noindent
From the emerging self-organization paradigm follows the development, and application to large datasets (``Big Data''),
of algorithms for the recognition of emerging structures and
calculation of related emerging properties. 
This term refers to the evaluation of statistical indices related to the identification of some form of structure within the data. Depending on the scientific field and on the ultimate goal of the specific research, the identification 
of data structures takes on particular facets, 
assumes particular terminologies (sometimes indicating similar properties in different contexts) and uses different approaches derived through mathematical tools, in particular statistical tools derived from
probability theory, stochastic processes, nonlinear
dynamical system theory, network science, topological
data analysis, deep learning.\\
%

\subsection{Crucial events and complex intermittency}

In complex systems where the self-organizing behavior
dynamically evolve in time, the role of
emerging self-organized states is tightly bound
with that of {\it crucial events} marking the
transitions between two ordered, i.e., self-organized states or between a ordered and
a disordered state.
The term ``event'' comes from the ubiquitous 
observation, both experimental and theoretical,
that the transitions {\it Order} $\Rightarrow$ {\it Order} or {\it Order} $\Leftrightarrow$ {\it Disorder} occur in a time interval much shorter
than all other characteristic time scales in the
system, thus also including the so-called
{\it life-times} of the self-organized states.\footnote{
In the following, the self-organized states will also be named {\it self-organized structures}. 
In many papers on turbulence, the large scale self-organized structures, which are essentially given by the so-called energy-containing eddies, are denoted as {\it coherent structures}, e.g., eddies, vortices and so on.
In our opinion, the small scale vortex structures that, as will be explained later, are triggered by the dynamical instabilities at the sharp edges of the large scale energy-containing eddies, could be also considered as self-organized
states of the flow, but at temporal and spatial
scales that are smaller with respect to the large
scale eddies.
As it will be explained below, in order to avoid ambiguities,
we will denote both small scale and large scale
structures as {\it self-organized turbulent structures}, leaving the term {\it coherent} only for the large-scale eddies.
}
Thus, the general picture is that of an alternance
of self-organized states, with some kind of 
internal structure,
and of totally disordered states of an alternance
between two different self-organized states, in
both cases being the passage 
marked by a transition occurring very quickly and
with an abrupt change in the observed variables.\\
These large excursions are also associated with a
fast memory drop in the system, which often
reflects a drastic change in some topological
features of the system itself. In complex
networks, the transition can be given by a fast
change in the connectivity; in turbulence or
other continuum mechanical systems, it can be
given by a fast passage from short-range to long-range synchronous/correlated motion
of fluid particles.
In any case, the memory drop recalls of
a transition event occurring in a somewhat ``random'' fashion and the assumption of a
time shorter that all the time scales of interest can be well approximated by just a time point.
This opens the way to a modeling of Rapid 
Transition Events (RTEs) as point stochastic
processes \cite{cox_1980_point} and, in particular, as renewal point
processes \cite{cox_1970_renewal}, that will be briefly introduced below.
Conversely, the coherence of self-organized states is
mirrored in the emergence of a inverse power-law decay
in the Probability Density Function (PDF) of Inter-Event Times (IETs): 
\beq
\psi(\tau) \sim \frac{1}{\tau^\mu}\ ,
\label{iet_pdf}
\eeq
which thus becomes a signature of self-organization. 
In particular, the system is considered to have higher complexity for lower values of the exponent $\mu$.
%

\vspace{.1cm}
\noindent
To our knowledge, the first studies proposing a event-based modeling approach were carried out
by Montroll and co-workers in the $1960$s.
The milestone papers were a series of
four papers: ``Random Walk on Lattices I-IV''
\cite{montroll_1964_1,montroll_1969_3,montroll_1973_4,montroll_1965_2}, where the Continuous Time
Random Walk (CTRW) model was firstly introduced
(see also \cite{weiss_1983_ctrw_review} for a review).
The CTRW is a random walk where the time step is not
constant like in the Markovian random walk, but
it is randomly distributed. The underlying microscopic dynamics are those of a particle motion in the presence of potential wells, where particles are trapped,  with the superposition of a random force.
Thus, the trapping time has a strong dependence on the position  and velocity of the particle before
falling inside the well itself and the exit time becomes essentially a random variable. These exit
times corresponds to the above introduced
IETs, also denoted as Waiting Time (WTs) in the 
framework of point stochastic processes \cite{cox_1970_renewal,cox_1980_point}.
The first applications of the CTRW model were
in the context of charge mobility in disordered solids \cite{montroll_1969_3,scher_1973_conduct,scher_1973_conduct2,tunaley_prl1974_conduct} and it is nowadays still applied as a basic modeling 
approach for {\it anomalous diffusion}
\cite{klafter_1987_ctrw_anom_diff,metzler-klatfter_pr2000_ctrw_fract_review,metzler_jpa2004_ctrw_fract}, 
which is defined
by a non-Gaussian distribution of a diffusing stochastic process $X(t)$ and a nonlinear time dependence of the variance\footnote{
An exception that was recently investigated is given by
the Brownian yet non-Gaussian motion, where the variance is linear in time but the distribution is non-Gaussian
\cite{sposini_2018}.
}:
\beq 
\langle X^2 \rangle \sim t^{2H}\ ;
\label{anom_diff}
\eeq
where $H$ is the second moment scaling, which is the same
as the Hurst exponent of rescalig analysis \cite{hurst_1951}.
CTRW-based modeling attracted, and is still attracting, the
interest of many research group, both regarding applications
to experimental data analysis and theoretical/mathematical
studies.
The relations between CTRW and Fractional Diffusion Equations
(FDEs)
has attracted the attention of many research groups since more than three decades: Hilfer and Anton \cite{hilfer_1995_ctrw_fract}, Compte \cite{compte_1996_ctrw_fract}  Barkai, Klafter, Metzler and co-workers \cite{barkai-cp2002_ctrw_fract,metzler_2014_ctrw_anom_diff_bio_cell,metzler-klatfter_pr2000_ctrw_fract_review,metzler_jpa2004_ctrw_fract}, Mainardi, Pagnini, Paradisi and co-workers
\cite{paradisi_1997_laurea,paradisi_caim15_ctrw_fract,scalas_2004_ctrw_fract,sposini_2021_ctrw_fract_hetero}. 
The mathematical properties and solutions of FDEs were extensively investigated by Mainardi and co-workers \cite{mainardi_1996_fract,mainardi_2010_book_fract,mainardi-pagnini_2003_fract,pagnini_2012_fract,paradisi_pceb01,paradisi_pa01_fract} and others \cite{saichev_1997_fract}, while Grigolini, West and co-workers 
focused more on the physical aspects of FDEs
\cite{guo-west_2021_fract,niu-west_2021_fract,west_2006_fract,west_2010_fract,west_2014_fract,west-grigolini_2003_book_fract,west-grigolini_pr2008_fract_events,west-grigolini_2015_fract}.
It is worth noting that Mainardi, Pagnini, Paradisi and co-workers investigated the relationships of fractional diffusion in the framework of other modeling perspectives, e.g., subordination
\cite{gorenflo_2007_subord_fract}, generalized grey Brownian Motion (ggBM) 
\cite{molina-garcia_2016_ggbm_fract,mura_2009_ggbm_fract,pagnini_2012_ggbm_fract,pagnini_fcaa2016_ggbm_fract}, fractional random walks
\cite{gorenflo_cp2002_fract_rw,gorenflo_pa2002_fract_rw,gorenflo_nd2002_fract_rw} and heterogeneous ensemble of Brownian particles
\cite{diTullio_2019_hetero_fract,dovidio_2018_hetero_fract,pagnini_fcaa2016_ggbm_fract,pagnini_2019_hetero_fract,sliusarenko_2019_hetero_fract,sposini_2021_ctrw_fract_hetero,vitali_2018_hetero_fract}.

\vspace{.15cm}
\noindent
CTRW, event detection methods, anomalous and fractional diffusion were
and are still widely applied in many different contexts:
turbulence and turbulent transport
\cite{dentz_2016_ctrw_turb,paradisi_jpcs2015,paradisi_romp12,paradisi_npg12}, human mobility \cite{song_2010_ctrw_human_mobil}, heterogeneous media
\cite{neuman_2009_ctrw_hetero_media_nonfick,diTullio_2019_hetero_fract,sposini_2021_ctrw_fract_hetero}, network
science \cite{hoffmann_2012_ctrw_network,hoffmann_2013_ctrw_temp_net,masuda_2017_ctrw_network}, transport in biological cells \cite{bressloff_2013_ctrw_bio_cell,fox_2021_ctrw_bio_cell,hofling_2013_ctrw_bio_cell,metzler_2014_ctrw_anom_diff_bio_cell}, foraging \cite{viswanathan_2011_ctrw_foraging}, porous media \cite{sahimi_2011_ctrw_porous}, pollution studies \cite{barati_2021_ctrw_pollution,cesari_2014}.

\vspace{.1cm}
\noindent
Starting from the CTRW Model, Grigolini and co-workers 
recognized the central role of crucial events and
gave a fundamental contribution in 
understanding that crucial events underlie fundamental dynamics that lead to self-organization
\cite{allegrini_2003a_model,allegrini_2006_model,allegrini_2003b_model,allegrini_2003_dea,allegrini_2007_linear_resp,bohara-grigolini2018,grigolini_csf15_bio_temp_complex,grigolini_2001_dea,mahmoodi-grigolini_2017,paradisi_csf15_pandora,paradisi_springer2017,piccinini-grigolini_2016,scafetta_2002_dea,turalska-grigolini_pre11_critical,west-grigolini_2021_book_events}. In fact, they
found that anomalous diffusion
is only one of the important mechanisms triggered by the presence of crucial events.
%
Thus, the following definition can be given:

\vspace{.2cm}
\noindent
{\bf Definition (Crucial events):}\\
A sequence of events is defined by a birth/death point process of self-organization and they are considered ``crucial events'' if:\\
(i) the {\it renewal} condition \cite{cox_1970_renewal},
apparent or hidden, is satisfied;\\
(ii) an inverse power-law decay is seen in the IET-PDF, given in Eq. (\ref{iet_pdf}) with $1 < \mu \le 3$.
%

\vspace{.15cm}
\noindent
Following Refs. \cite{grigolini_pre15_temp_complex,grigolini_csf15_bio_temp_complex,paradisi_csf15_pandora,paradisi_springer2017,paradisi_csf15_preface,turalska-grigolini_pre11_critical}, the above conditions (i-ii) are denoted as {\it Fractal Intermittency} or {\it Complex Intermittency} and identifies the emergence of 
{\it Temporal Complexity} (TC) or {\it Intermittency-Driven Complexity} (IDC).
The range $ 1 < \mu < 3$ is related to an infinite variance of the
fluctuating IETs. In the case $1 < \mu \le 2$, also the mean IET
is infinite and the complex system is also non-ergodic due to
strong aging.\\
The presence of crucial events is then the signature
of cooperative dynamics where the emergence
of self-organization is given not by a
static but by a dynamically evolving structure and
is one-by-one associated with a kind of instability
that starts slowly and then has a rapid 
acceleration, i.e., the RTE occurrence, that destroys the self-organized
structure, but only to trigger a new one.\\
This behavior reminds the concept of {\it metastability} and
metastable states. This concept was discussed by Afraimovich and Rabinovich, who proposed and discussed a model for brain
cognitive dynamics and information flow 
\cite{afraimovich_2004,rabinovich_plr12,rabinovich_prl06,rabinovich_pcb08}. The dynamical laws have a set of equilibrium saddle points that, as known, also
involve stable and unstable separatrices and manifolds and that are combined into a single structure called {\it stable heteroclinic channel}.
This is defined by the set of trajectories moving in the
neighborhood of the separatrices, moving from one saddle
point to the other. When the system reach a region around a
given saddle point, there is a slow motion along the
stable separatrix and towards the point itself. 
During this long residence time interval, the system displays long-term memory and spatial/topological coherence, i.e., falls into a self-organized state\footnote{ 
When carrying out the statistical analysis of the signal derived from the complex system , which is based on time or ensemble averages,
long-range correlations and some kind of synchronization can be observed. 
}.
After a relatively long {\it life-time}, the
dynamical instabilities start to increase, the system moves
towards the unstable separatrix until a sudden acceleration
(the event) determines a fast motion of the system along the unstable
separatrix that then becomes the stable one of the next
saddle point, where the same temporal pattern is repeated.

\vspace{.15cm}
\noindent
Another example is given by the Manneville-Pomeau (MP) map \cite{manneville_1980,pomeau-manneville_1980}, which is a prototype of so-called Type-I intermittency and 
will be briefly discussed below.
Interestingly, the MP model was proposed as a toy model
for a fluid particle motion experiencing an alternance of long time intervals with laminar, or quiescent, motion and of short-time {\it turbulent bursting} events.
Thus, the MP map could be thought as a model for the intermediate
regime of turbulence, that is, when a Richardson's energy cascade is still not fully developed\footnote{
We recall that the mechanisms triggering the fully developed energy cascade of turbulence are
associated with instabilities at the sharp edges
of the main, energy-containing eddies. In this
sense, the meaning of transition event could 
be applied to the formation of smaller
vortex structures from the instabilities of
the larger ones. This concept is at the basis of
random cascade models (see, e.g., \cite{frisch_1995}).
}.
In the MP model, a marginally stable point determines the slow ``laminar'' motion towards the chaotic region, where
a turbulent bursting occurs. Shortly, the particle 
position comes back to the laminar region and the low
predictability of the chaotic region acts as a random 
back injection to the laminar region\footnote{
Grigolini, West and co-workers have extensively studied
a continuous-time version of the MP time-discrete model
\cite{allegrini_2003a_model,allegrini_2005_model,allegrini_2006_model,allegrini_2003b_model}.
Paolo Grigolini often used to denote this model as
``cheerful equation'' as, in italian, {\it cheerful}
translates in ``allegra'', which sounds like
Allegrini, one of the
collegues who contributed the most to the study of
this continuous-time version of the MP model.
For this reason, in the following we will denote
this model as Allegrini's model.
}.
The RTEs are here given by the passages from the 
laminar to the chaotic/turbulent region and {\it vice versa} and the IETs are given by the inter-burst times\footnote{
In the MP time-discrete model, the burst duration is not always negligible, but this
does not change the essential results. 
Alternatively, it is almost equivalent to consider the differences between the times at which two
successive bursts begin.
}. 

\vspace{.15cm}
\noindent
There is an interesting link between intermittent complex
systems and critical phenomena. In fact, it is well-known
that second-order phase transitions are associated with
long-range, i.e., power-law decaying, correlation
functions.
It was proven that these correlations follow as a 
consequence of serial MP intermittent dynamics driving the ﬂuctuations of the ``order parameters''.
This crucial result was derived both analytically 
\cite{contoyiannis_2002,contoyiannis_2000}
and numerically \cite{turalska-grigolini_pre11_critical,vanni-grigolini_prl2011_critical} and was found to have important
consequences on the critical brain dynamics
\cite{allegrini_2009,allegrini_csf13,allegrini_pre15,chialvo2010_critical_brain,fraiman_2009_critical_brain,paradisi_aipcp13}.

\vspace{.2cm}
\noindent
The concept of crucial event is recently entering
also in the so-called {\it big data} or {\it data science}, where data mining, machine and deep learning and, in general, artificial intelligence
tools are used.
Complex network analysis came to the forefront
about two decades ago and it still a widely used
tool for big data mining (see, e.g., \cite{boccaletti_2006,zanin-boccaletti_2016}).
Complex network measures, e.g., centrality, degree distribution, clustering, shortest path, are basically static and are still widely exploited in
network analysis, even when applied to
time-varying networks, where time is just used 
to build up the statistical ensemble.
However, for some time-varying networks, these static measures, or metrics, can cause some
misinterpretation of the results.
A central point is the rough approximation that
considers all links, or edges, among nodes always the same,
active or not active, thus without any change over
time. On the contrary, links can have their own dynamics possibly triggering an alternance of active periods and non-active periods.
In recent years, the concept of {\it temporal networks} was introduced and widely investigated
\cite{holme_2012,holme_2019}. The temporal network is
represented not by a single adjacency matrix, but
by a temporal sequence of adjacency matrices.
According to Ref. \cite{holme_2019}, 
``The fundamental building blocks of temporal networks are {\it events} (or contacts, links,
or dynamic links). These represent units of interaction between a pair of nodes at
specified times.''

\vspace{.15cm}
\noindent
However, it is worth noting that, with respect
to the concept of event as a local transition in the link/edge  activity, the above introduced RTEs have to be thought as a transition in the self-organizing behavior of the network, both at the local and global level. Thus, the self-organizing RTE and the single link activity transition
event are somewhat different concepts, but they are surely inter-related and this aspect should deserve further investigations.

\subsection{Renewal processes}
\label{renewal_subsect}

In the following, we sketch the main definition and
results of Cox's renewal theory and its dynamical
foundations on the basis of both Manneville-Pomeau
\cite{manneville_1980,pomeau-manneville_1980}
and Allegrini models \cite{allegrini_2003a_model,allegrini_2005_model,allegrini_2006_model,allegrini_2003b_model}.

\begin{figure}
\begin{center}
\includegraphics[width=14.cm,height=9.5cm,angle=0]{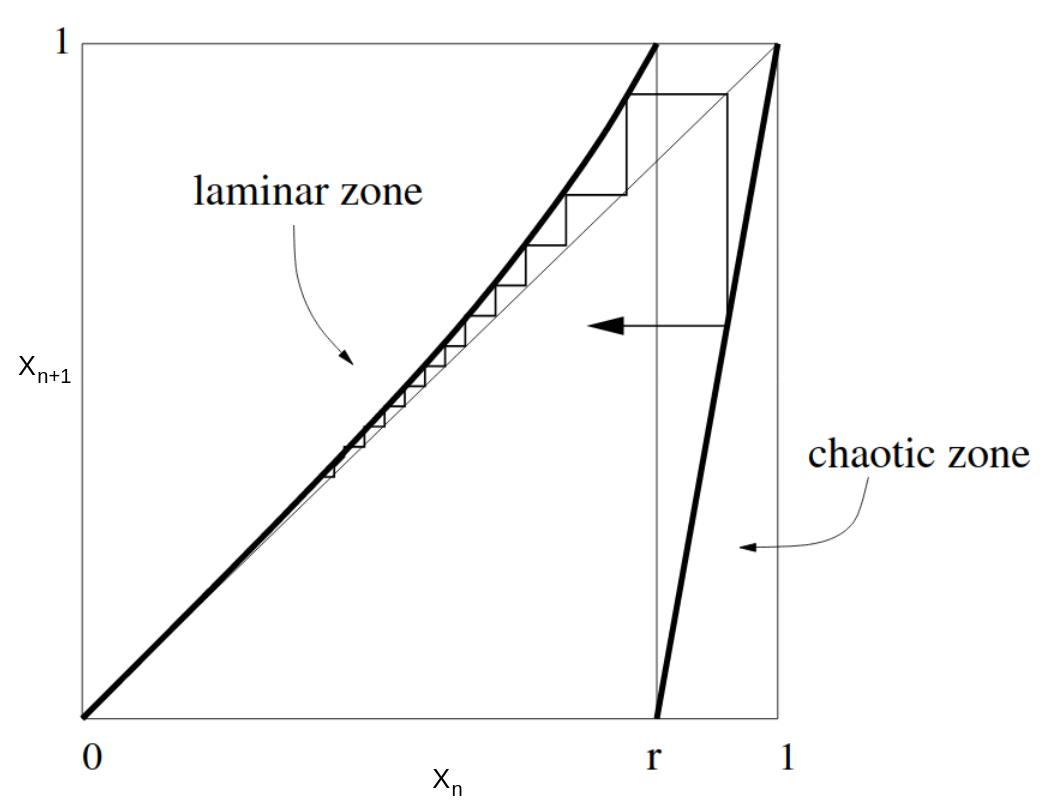}
\caption{Plot of the basic function $x_{n+1}$ vs. $x_n$ of the iterated map.}
\label{manneville_fig}
\end{center}
\end{figure}

\vspace{.1cm}
\noindent
The MP map and its continuous-time
version given by the Allegrini's stochastic 
model represent 
basic prototypical toy models of TC/IDC complex systems.
In fact, as said above, these models satisfy the renewal condition and reproduce the inverse
power-law behavior in the Probability Density Function (PDF) of IETs, in the following denoted as IET-PDF.
The MP time-discrete deterministic
model for turbulent bursting is given by the following iterated map \cite{manneville_1980,pomeau-manneville_1980}:
\beq
\left \{
\begin{array}{lll}
X_{n+1} = & X_n + k |X_n|^z\ \ & \mathrm{ if } \:\: X_n <r\\
\ \\
X_{n+1} = & (X_n - r)/(1-r)\ \ & \mathrm{ if } \:\: X_n >r\\
\end{array}
\right .
\label{manneville_map}
\eeq
being $z>1$ and $r$ is defined by the relation: 
$r + k|r|^z=1$. 
The basic function of the MP iteration map is
plotted in Fig. \ref{manneville_fig}, where the laminar and chaotic zones are reported.\\
%
\begin{figure}
\includegraphics[width=14.cm,height=8.cm,angle=0]{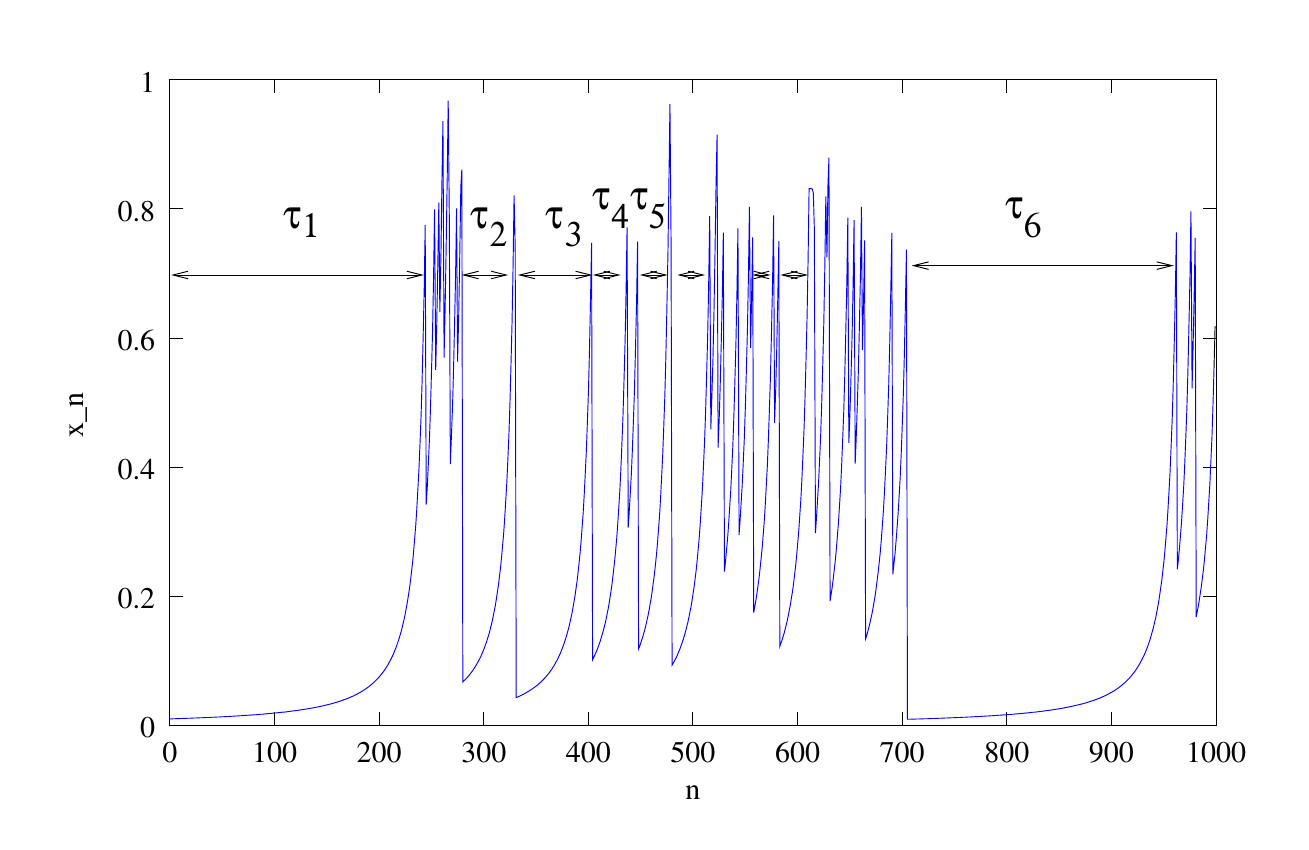}
\caption{A typical trajectory of the Manneville-Pomeau time-discrete map.}
\label{traj_manneville_fig}
\end{figure}
Fig. \ref{manneville_fig} sketches the general behavior of the MP map and Fig. \ref{traj_manneville_fig} report an example of a
typical trajectory with turbulent bursting.
Considering that $X=0$ is a marginally unstable fixed point, it can be seen that trajectories starting
near this point will spend a large amount of time near it, remaining in the laminar zone for a relatively
extended time interval. 
Comparing Figs. \ref{manneville_fig} and \ref{traj_manneville_fig}, it is easy to see that in
the laminar region the motion is smooth and highly
predictable. On the contrary, when passing through
the threshold $r$, the motion becomes chaotic and
unpredictable until a return to the laminar region
occurs that marks the end of the burst. 
This is mirrored in the inverse power-law IET-PDF given in Eq. (\ref{iet_pdf}).

\vspace{.1cm}
\noindent
Now, let us rescale the discrete time $1 \rightarrow \Delta t$ with $k=\alpha_{_0} \Delta t$ and
$y(t_n) = y(n \Delta t) = X_n$.
Thus, carrying out the continuous-time limit $\Delta t \rightarrow 0$ in Eq. (\ref{manneville_map}), we clearly get the following equation:
\beq
\left \{
\begin{array}{lll}
&\dot y = \alpha_{_0} y^z\ &;\quad \ y \in [0,1) \\
\ \\
&y(t_n^-) = 1 &\Rightarrow \  
y(t_n^+) = u(0,1]\ ,
\end{array}
\right .
\label{allegrini_model}
\eeq
where $u(0,1]$ is a uniform random variable in the
interval $(0,1]$. The initial condition is given by:
$y(t=0^+) = u(0,1]$\footnote{
In this formulation we avoid both the initial condition and the return to $y=0$, because this would
give an infinite time to reach the border $y=1$, thus
giving an infinite exit time. This does not change
the computation of probabilities as $y=0$ is a subset
with zero Lebesgue measure. In other words,
for any given $\tau_*$, there
will exist a time $t_n$ such that $t_n-t_{n-1}> \tau_*$.
}.
\begin{figure}
\begin{center}
\includegraphics[width=12.cm,height=5.cm,angle=0]{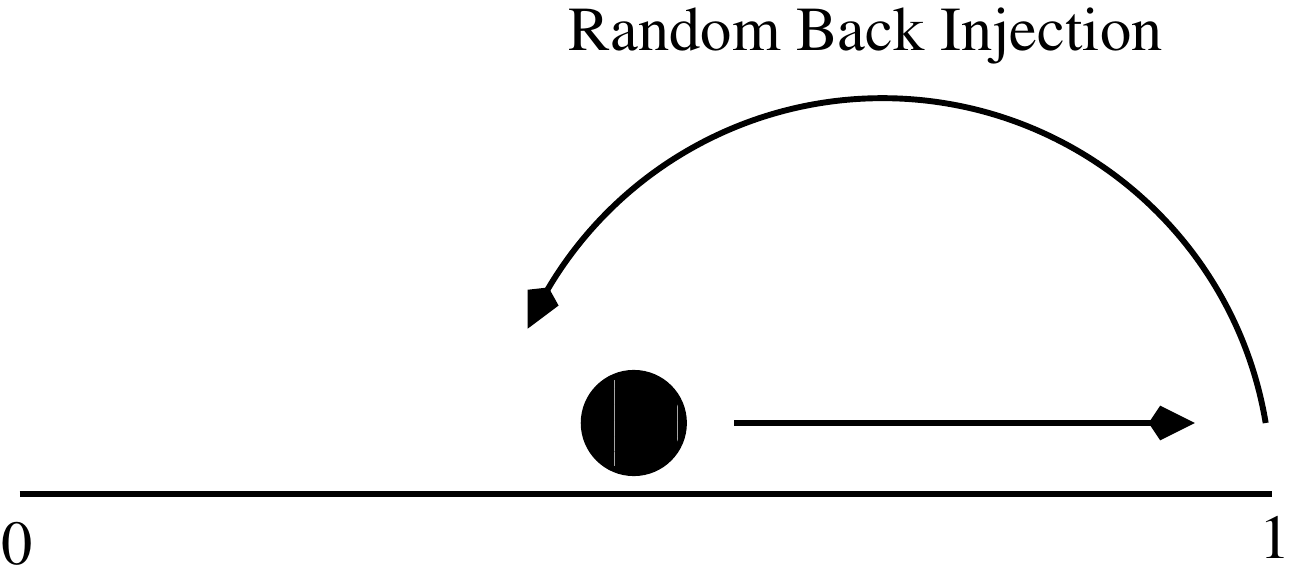}
\caption{Sketch of the Allegrini's stochastic model and of the random back injection.}
\end{center}
\label{allegrini_fig}
\end{figure}
%
The limit $\Delta t \rightarrow 0$ implies 
$r \rightarrow 1$ by definition, thus the chaotic region collapse into the boundary point $y=1$.
The random back injection, sketched in Fig. \ref{allegrini_fig}, mimics the effect of the
bursting activities, whose duration are reduced to the time occurrences $t_n$ of a stochastic point process.
Thus, the sequence of event occurrence times:
\beq
\left\{ t_n \right\}_{n \in {\cal N}}\ ;\quad   
 t_{n+1}> t_n\ ,
\label{time_sequence}
\eeq
being ${\cal N} = \left\{0,1,2, ... \right\}$ the set of positive integer numbers, is a {\it point process} that is also represented as a
stochastic counting process $N(t)$:
\beq
N(t)=\int_0^t Z(t^\prime) dt^\prime\ ;\quad Z(t)=\sum_n \delta(t-t_n)
\label{count_proc}
\eeq
where $\delta(\cdot)$ is the Dirac function. 
The IETs are given by:
\beq
\tau_n=t_n-t_{n-1}\ ;\quad t_0=0\ ;\quad n \ge 1\ .
\label{iet}
\eeq
A {\it renewal process} is defined as a point process
whose IETs are mutually independent random variables \cite{cox_1970_renewal}:\\ 
$Pr(\tau_i|\tau_j)=Pr(\tau_i)$,
or, equivalently, 
$Pr(\tau_i,\tau_j)=Pr(\tau_i|\tau_j)Pr(\tau_j)=Pr(\tau_i)Pr(\tau_j)$, for whatever couple $(i,j), i>j$. \\
%
A renewal process is also time-homogeneous, i.e., stationary, if the IETs
are not only statistically independent, but also
identically distributed. In this case, the renewal
process is uniquely defined by the IET distribution:
\beq
\left \{
\begin{array}{lll}
&\Psi(\tau) = Pr \left\{ {\rm IET} > \tau \right\}
\ ;\quad &{\rm (Survival\ Probability\ Function)}\\
\ \\
&
\psi(\tau) = -  d\Psi / d\tau \ ;
\quad &{\rm (Probability\ Density\ Function)}
\end{array}
\right .
\label{iet_surv_pdf}
\eeq
being $\psi(\tau)d\tau = Pr\left\{ \tau < {\rm IET} < \tau+d\tau \right\} = \Psi(\tau) - \Psi(\tau+d\tau)$.
Clearly, it results:
$$
\Psi(\tau) = \int_{\tau}^{\infty} \psi(s) ds =
1 - \int_0^{\tau} \psi(s) ds \ .
$$
Equivalently, the renewal process can be described by the {\it local rate of event production} $r_c(t)$, also denoted as the {\it Cox's event rate}, which is the expected number of events per
time unit in a neighborhood of the time $t$ \cite{cox_1970_renewal} (see also \cite{akin_jsmte09,paradisi_romp12,paradisi_cejp09} for more details).
More rigorously, $r_c(t)$ is defined as the conditional probability density that an event occurs in an infinitesimal time
interval $[t,t+dt]$, given that no events occurred in the time
interval $[t_n,t] (t > t_n)$:
\begin{equation}
r(t) = \lim_{dt \rightarrow 0}\ \frac{1}{dt}
\Pr \left\{ t<t_{n+1}\leq t+dt\ |\ t_{n+1} > t \right\} .
\label{rate_cox}
\end{equation}
Limiting for simplicity to the case $t_0=0$, we have:
\begin{equation}
r(t) = \frac{\psi(t)}{\Psi(t)}=-\frac{1}{\Psi(t)}
\frac{d\Psi(t)}{dt}\ .
\label{rate_def}
\end{equation}
This is exactly the definition that Cox gave starting
from the concept of a statistical ensemble of 
independent trajectories that have a unique event.
In the context of intermittent complex signals,
Eq. (\ref{rate_def}) is generalized by means of time
shift $t \rightarrow t-t_n$ and the Cox's rate becomes
trajectory dependent \footnote{
This also allows for a perturbation theory of
event-driven processes \cite{akin_pa06,allegrini_2007_linear_resp,aquino-grigolini_2007_linear_resp,barbi-grigolini_2005_linear_resp}
} \cite{akin_jsmte09,paradisi_cejp09}.
Thus, always limiting for simplicity to the first IET, so that $t_0=0$, we have the following relationship between
the event rate and the Survival Probability Function (SPF):
\begin{equation}
\label{homo_surv_prob}
\Psi(\tau) = \exp \left(-\int_0^\tau r(\tau^{\prime}) d\tau^{\prime} \right)\ .
\end{equation}
where $\Psi(0) = Pr\{IET > 0\} = 1$ was the trivial
initial condition.
The time-homogeneous Poisson process is a particular case of a renewal process that is associated with intermittent
system that do not display self-organizing behavior.
The IET-SPF of a Poisson process is an exponential function:
$\Psi(\tau)=\exp(-r_{_0} \tau)$ and the rate does not change in time: $r(t)=r_{_0}$.
At variance with Poisson processes, a renewal Non-Poisson process is characterized by a Non-Poisson distribution of the events, a non-exponential WT distribution and a changing in time event rate $r(t)$. This condition is a necessary, but not sufficient condition that a intermittent complex system should satisfy.

\vspace{.1cm}
\noindent
Coming back to the Allegrini's model, Eq. (\ref{allegrini_model}), it is possible to derive the
following exact expressions for the IET-PDF and the IET-SPF:
\beq
\psi(\tau) = \frac{\mu-1}{T_{_0}} \left(\frac{T_{_0}}{T_{_0}+\tau}
\right)^\mu\ ;\quad \Psi(\tau)=\left(\frac{T_{_0}}{T_{_0}+
\tau}\right)^{\mu-1}\ ,
\label{powerlaw_pdf_surv}
\eeq
where
\beq
\mu=\frac{z}{z-1}\ ;\quad T_{_0}=\frac{1}{\alpha_{_0}\left(z-1\right)}\ .
\eeq
A crucial aspect is that both $\psi(\tau)$ and $\Psi(\tau)$ display an inverse power-law decay in the long-time regime
$t\gg T_{_0}$, thus satisfying scale invariance and self-similarity.

\noindent
Similarly, it is possible to derive the following expression for the event rate:
\beq
r(\tau) = \frac{r_{_0}}{1+r_{_1} \tau}\ ,
\label{nonpoissonrate}
\eeq
being
\beq
T_{_0}=\frac{1}{r_{_1}}\ ;\quad \mu=1+\frac{r_{_0}}{r_{_1}}
\ ,
\label{rate_param}
\eeq 
The Non-Poisson renewal process reduces to a Poisson one
when $r_{_1}\rightarrow 0$, which is equivalent to the
the limit $T_{_0}, \mu \rightarrow \infty$ in 
Eq. (\ref{powerlaw_pdf_surv}), with the 
constrain $(\mu-1)/T_{_0}\rightarrow r_{_0}$. 
Thus, $\Psi(\tau)$ reduces to the exponential function $\exp(-r_{_0}\tau)$.

\vspace{.25cm}
{\bf Renewal Aging}

\vspace{.25cm}
\noindent
The CTRW was recognized to have aging, i.e., a not
negligible dependence on the initial condition even at
very long times, even reaching the ideal case of infinite time in the subdiffusive case, i.e., $\langle X^2 \rangle \sim t^{2H}$
with $H<0.5$, a condition determinted by a very slow 
power-law decay in the IET-PDF ($\mu<2$)
\cite{tunaley_jsp1974_ctrw_asympt,tunaley_prl1974_conduct,tunaley_1976_ctrw}.\\
Grigolini and co-workers recognized
the central role of aging in intermittent complex systems
even outside the CTRW model, that is, only referring to the
sequence of crucial events, and
developed a statistical tool, called {\it Renewal Aging} (RA) analysis \cite{allegrini_2006_model,bianco_jcp05,paradisi_aipcp05}, that can be exploited to give some clue about the validity of renewal condition\footnote{
In the application of the Renewal Aging analysis some care
is needed and some aspect have to be considered, such as
the presence of noise and modulating external 
forcings that can affect the results of the analysis
(see, for example, \cite{akin_jsmte09,paradisi_cejp09,paradisi_ijbc08})
}.

\vspace{.1cm}
\noindent
RA analysis can estimate
the level of modulation, random or deterministic, of 
a complex system that displays a non-exponential IET-PDF
and, in particular, a power-law decay $1/\tau^\mu$,
usually with $\mu<3$.
Given a sequence of IETs $\left\{ \tau_n \right\}$ and
the associated event occurrence times 
$t_n = t_{n-1} + \tau_n$,
the RA algorithm is given by the following steps \cite{paradisi_epjst09}:
\begin{itemize}
\item[(i)]
Given the aging time $t_a$ and the event sequence 
$\left\{ t_n \right\}$, let us consider the set of
sliding windows:

\hspace{2.cm}
$
{\cal T} = \left\{t_n \le t < t_n+t_a ; n = 0,1,2,...   \right\}\ .
$

Then, the starting time of sliding windows are shifted from
one occurrence time to the next one.
This is equivalent to build up a set of trajectories that
mimics a statistical ensemble of trajectories, even if the single realizations are not really independent each other.
\item[(ii)]
Given the $n$-th window, whose starting time is given by $t_n$, the first event $k$ ($k>n$) occurring after the end of the time window is considered:

\hspace{2.cm}
$
{\rm Find}\ k: t_{k-1} < t_n + t_a < t_k
$

\item[(iii)]
The $n$-th truncated or aged IET is computed as:
$\tau_n (t_a) = t_k - (t_n + t_a)$, and 
added to
the sequence of aged IETs: $\left\{ \tau_n (t_a) \right\},\ n=1,2,...$.
\item[(iv)]
A random shuffling is applied to the experimental sequence
of IETs in order to get a {\it renewal} sequence $\left\{ \tau^{^{_S}}_n \right\}$, where the prime $S$ stands for ``Shuffled''. In this way, a sequence with exactly the same
IET-PDF of the experimental sequence, but without correlations among IETs, is obtained.
\item[(v)]
The steps (i-iii) are repeated on the shuffled sequence
with the same $t_a$ to get the sequence 
$\left\{ \tau^{^{_S}}_n (t_a) \right\}$.
%
\item[(vi)]
The IET-PDFs and IET-SPFs for the three different IET sequences are computed:
%
\begin{itemize}
\item [-]
Brand-new distribution: $\psi_0(\tau)$ and $\Psi_0(\tau)$ from original sequence $\left\{ \tau_n \right\}$.
\item [-]
Aged experimental distributions: $\psi_{\rm exp}(\tau,t_a)$ and $\Psi_{\rm exp}(\tau,t_a)$ from the aged experimental sequence
$\left\{ \tau_n (t_a) \right\}$.
\item [-]
Aged renewal distributions: $\psi_{\rm R}(\tau,t_a)$ and $\Psi_{\rm R}(\tau,t_a)$ from the aged shuffled sequence
$\left\{ \tau^{^S}_n (t_a) \right\}$.
\end{itemize}
\item[(vii)]
The steps (i-vi) are carried out for several values of $t_a$.
\end{itemize}

\vspace{.1cm}
\noindent
In the following, $\Psi_{\rm exp}$ and $\Psi_R$ will be denoted as {\it Experimental Aging} and {\it Renewal Aging}, respectively\footnote{
Slightly different conventions were sometimes used, such as
Real instead of Experimental Aging, and 
$\Psi_S$ instead of $\Psi_R$.
}.
The three IET-PDFs can be graphically compared to 
get a clue about the amount of memory in the IET experimental sequence. The Renewal Aging acts as a reference
curve for the aging time $t_a$. The comparison between
$\Psi_0$ and $\Psi_R$ gives an estimation of how
far the IET distribution is from an exponential one,
while the comparison between $\Psi_{\rm exp}(\tau,t_a)$ and $\Psi_R(\tau,t_a)$ gives a measure of the departure from the renewal condition and, thus, on the memory of the
event sequence.
A quantitative evaluation of the {\it renewal content} can
be carried out by introducing the Aging Intensity Function
(AIF) \cite{bianco_jcp05,paradisi_aipcp05}:
\begin{equation}
I_a (\tau,t_a)  = \frac{\Psi_{\rm exp}(\tau,t_a) - \Psi_0(\tau)}
{\Psi_R(\tau,t_a) - \Psi_0(\tau)}\ .
\label{aif_def}
\end{equation}
In the purely renewal Non-Poisson case, $I_a (\tau,t_a)  = 1\ \forall \tau,t_a$. 
This property was verified comparing
renewal Non-Poisson and slowly modulated Poisson models.
The Non-Poisson model was a renewal process with the
same IET-PDF of the Allegrini's model, given in Eq. (\ref{powerlaw_pdf_surv}). The slowly modulated model
was derived in such a way to get the same IET-PDF. 
Clearly, the modulation introduces memory in the
time ordering of the events. 
Thus, Renewal Aging was proven to detect this kind of memory in event sequences.
In fact, the AIF showed a well-defined
behavior with an asymptotic constant that was used as
a measure of the renewal content of the artificial event sequences.
The theoretical analysis gave the following general results \cite{allegrini_2006_model,allegrini_csf07,bianco_jcp05,paradisi_aipcp05}:
%
%
\begin{itemize}
%
\item [-]
For time-homogeneous Poisson processes there's no aging: 
$\psi_0(\tau)=\psi_{\rm exp}(\tau,t_a)=\psi_R(\tau,t_a) \forall t_a$.
%
%
\item [-]
For time-homogeneous renewal Non-Poisson processes:
$\Psi_0(\tau,t_a) < \Psi_{\rm exp}(\tau,t_a)=\Psi_R(\tau,t_a)$.
%
This is the condition identifying a renewal process, but only if it
is satisfied for any aging time $t_a > 0$. 
\item [-]
For Non-Homogeneous Poisson Processes with a random rate modulation 
generating Non-Poisson events, the Experimental Aging $\Psi_{\rm exp}(\tau,t_a)$
decreases 
from the Renewal Aging $\Psi_R(\tau,t_a)$ to $\Psi_0(\tau,t_a)$ as the
modulation becomes slower and slower. In other words, the Renewal Aging
decreases as the ratio between Non-Poisson and Poisson events decreases.
%
%
%
%
%
\item [-]
Very slow modulation has zero Renewal Aging,
which means $\Psi_0(\tau,t_a) = \Psi_{{\rm exp}}(\tau,t_a) \ne \Psi_R(\tau,t_a)$, even if the brand new WT distribution is not an
exponential function but an inverse power-law.
%
%
%
\end{itemize}
%
%
RA analysis was successfully applied to the fluorescence intermittency of Blinking Quantum Dots \cite{bianco_jcp05,paradisi_aipcp05}.
In this application to real data, the AIF was not really constant, but is was
found to reach an asymptotic behavior, even if with some
fluctuations\footnote{
It is worth noting that AIF has to be treated with some care due to the possible small differences in the denominator of 
Eq. (\ref{aif_def}) that can cause large variations in the
AIF.
}
\cite{bianco_cpl07}.

\section{A brief history of intermittency in turbulence}
\label{intermit_turb_sect}

\subsection{Small-scale intermittency in ideal homogeneous and isotropic turbulence}
\label{homo_iso_turb}

The theoretical picture given by the K41 model, which still
represents a guideline for all the turbulence theoretical research, is limited to the ideal case of {\it homogeneous and isotropic turbulence} \cite{monin_1975,zhou_2021}, also named {\it fully
developed turbulence}.
As said above, the basic assumptions are: \\
(i) the dissipation rate remain finite and different from zero in the limit $Re\rightarrow \infty$\\
and \\
(ii) a quasi-equilibrium condition across the scales of the inertial subrange, which naturally follows from the global energy conservation:

\vspace{.2cm}
mean dissipation rate at small scales 
($\langle \epsilon\rangle$)\ $=$\ 
energy flux at large scales

\vspace{.2cm}
\noindent
The K41 model and almost all turbulence studies are developed in the random field formalism, thus characterizing the
statistics of the velocity increments:
\beq
\Delta{\bf u}(d{\bf x},dt) = {\bf u}({\bf x}+\Delta{\bf x},t+\Delta t) - {\bf u}({\bf x},t)
\label{vel_incr}
\eeq
Thus, multi-dimensional probability distributions are 
evaluated. The {\it p-th order structure functions} and, in particular, the {\it longitudinal p-th order structure functions}, are of particular interest:
\beq
S_p(\Delta x, \Delta t) = \left\langle \left| \left( 
{\bf u}({\bf x}+\Delta{\bf x},t+\Delta t) - {\bf u}({\bf x},t)
)
\right) \cdot \frac{\Delta {\bf x}}{\Delta x} \right|^p\  \right\rangle
\label{struct_funct_p}
\eeq
being $\Delta x = |\Delta {\bf x}|$ the distance between the
two points at which velocity is measured\footnote{
Notice that homogeneous and steady conditions are implicit in the notation $S_p(\Delta x, \Delta t)$.
}.
In addition to the above assumptions, Kolmogorov made two
basic similarity hypotheses that were assumed to be valid in 
the limit $Re \rightarrow \infty$ and $\Delta x \ll L$, where 
$L$ is the integral scale of the energy-containing eddies introduced above:
\begin{itemize}
\item[]
{\it First similarity hypothesis}:\\
In isotropic conditions, the multi-dimensional distribution of the velocity increments
are uniquely defined by mean dissipation rate $\langle \epsilon \rangle$ and the kinematic viscosity $\nu$.
\end{itemize}
From this hypothesis follows that the length, velocity and
time scales (Kolmogorov microscales) are defined as:
\beq
\lambda_K = \left( \frac{\nu^3}{\langle \epsilon \rangle} \right)^{1/4}\ ;\quad 
v_K = \left( \nu \langle \epsilon \rangle \right)^{1/4}\ ;\quad 
\tau_K = \left( \frac{\nu}{\langle \epsilon \rangle} \right)^{1/2}\ .
\label{kolmo_1}
\eeq
Then, we have the
\begin{itemize}
\item[]
{\it Second similarity hypothesis}:\\
The multi-dimensional distribution 
$P(\Delta{\bf u}_1, ..., \Delta{\bf u}_N)$
at small scales, i.e., in the inertial subrange\footnote{
$U$ is the mean large-scale velocity.
}:
\beq
\lambda_K \ll \Delta x_n  \ll L\ ;\quad 
\tau_K \ll \Delta t_n  \ll L/U
\ ;\quad \forall n\ ,
\nonumber
\eeq
and satisfying the conditions:
\begin{eqnarray}
&&\Delta x_n \gg \lambda_K\ ;\quad |\Delta{\bf x}_n - \Delta{\bf x}_m| \gg \lambda_K\ ;\quad n \ne m\ ,  
\label{kolmo_hp2_1}\\
&&\Delta t_n \gg \tau_K\ ;\quad |\Delta t_n - \Delta t_m| \gg \tau_K\ ;\quad n \ne m\ , 
\label{kolmo_hp2_2}
\end{eqnarray}
are uniquely determined by mean dissipation rate $\langle \epsilon \rangle$ and are independent of $\nu$.
\end{itemize}
From the above assumptions and the two Kolmogorov similarity
hypotheses follows the well-known scaling relationships
of homogeneous and isotropic turbulence:
\beq
S_p(\Delta x, 0) \sim \left( \langle \epsilon \rangle
\Delta x \right)^{p/3}\ ;
\label{k41_scaling_0} 
\eeq
\beq
S_2(\Delta x, 0) = C_1\ \left( \langle \epsilon \rangle
\Delta x \right)^{2/3}\ ;
\label{k41_scaling_1} 
\eeq
\beq
E(k) = C_2\ \langle \epsilon \rangle^{2/3} k^{-5/3}\ ;
\label{k41_scaling_2} 
\eeq
\beq
S_2(0,\Delta t) = C_0\ \langle \epsilon \rangle
\Delta t\ ,
\label{k41_scaling_3}
\eeq
 being $E(k)$ the spatial PSD in the wave number
$k = 1/\Delta x$, whose $5/3$ scaling is valid in the
range $1/L \ll k \ll 1/\lambda_K$.
The K41 model was experimentally verified in his essential aspects in many experimental studies, 
but, even if the basic approach is still valid, some of the assumptions were questioned, both experimentally and theoretically.\\
In fact, Landau reported his objection on the universality assumption in the 1944 Russian edition of the book ``Fluid mechanics''\cite{landau_1959}:

\vspace{.15cm}
\noindent
``
... averaging these expressions is dependent on the variation of $\epsilon$ over times of large-scale motions (scale $L$), and
this variation is different for different specific flows. Therefore, the result of the averaging cannot be universal.
''

\vspace{.15cm}
\noindent
which means that the ``universal'' constants $C_0$, $C_1$ and $C_2$ in Eqs. (\ref{k41_scaling_1}-\ref{k41_scaling_3}) are
not really constant, but functions of space and time.
%
This objection was confirmed by Oboukhov's findings in atmospheric turbulence \cite{oboukhov_1962}.
Analysis of different experimental samples confirmed the
K41 $5/3$ and $2/3$ scaling exponents given in Eqs. 
(\ref{k41_scaling_1}-\ref{k41_scaling_2}), but found that the pre-factor has significant changes depending on the large scale dynamics and, thus, on the external forcing.
This confirmed Landau's objection and proves that the fluctuations in the energy dissipation rate are not negligible. 
%
%
Another similar objection came from the finding of intermittent behavior of velocity increments and dissipation rates at small scales, which results in the
failure of the quasi-equilibrium  assumption\footnote{
The neglect of small scale intermittency was later remarked also by Landau himself, and it seems that Kolmogorov, trying to overcome this K41 limitation, interpreted Landau's objection as mainly referred to the lack of intermittency in the K41 model.
}. 
In fact, in 1949 Batchelor and Townsend \cite{batchelor_1949} carried out experiments that revealed the intermittent nature
of the instantaneous dissipation\footnote{
According to Ref. \cite{benzi-frish_2010}, ``the fact that small-scale activity in high-Reynold number turbulence becomes increasingly clumpy and that self-similarity is broken is generally referred to as {\it intermittency}''.
However, it is worth noting that different 
definitions of turbulence intermittency can be found in the literature.
}.
By analyzing the velocity derivatives, they found that
energy dissipation at small scales (i.e., large wave-numbers)
is very heterogeneous in space, with active regions displaying large excursions in the turbulent signals
alternating with quiescent regions. Further, the 
fluctuations tend to an on-off intermittent discontinuous
process, a behavior that becomes more relevant as
the Reynolds number increases. This spatial intermittency 
determines the emergence of non-Gaussian shapes in the distributions of dissipation rates and velocity increments.

The finding of intermittency led to the {\it refined similarity hypothesis} proposed by Kolmogorov and Oboukhov \cite{kolmogorov-1962,oboukhov_1962}, where the fluctuations of the dissipation rate are taken into account and, further, 
a direct local relationship between velocity increments
and dissipation rates is assumed.
As a consequence, multidimensional probability 
distributions of velocity increments and associated structure functions are directly related to the 
probability distribution of the local dissipation rate 
$\epsilon$.
In more detail, the refined similarity model assumes that the dissipation rate $\epsilon$ 
remains constant inside  regions of the  inertial subrange with size $r \gg \lambda_K$, 
and change passing from one region to another. 
Following this approach, Kolmogorov and
Oboukhov introduced a local spatial average of the dissipation rate:
\beq
\epsilon_r ({\bf x},t) = 
\langle \epsilon \rangle_r ({\bf x},t) = \frac{1}{V_r}
\int_{|{\bf z}| < r} \epsilon({\bf x+z},t)d{\bf x} \ ,
\label{k62_local_diss}
\eeq 
being $V_r = 4\pi r^3/3$ the volume of the sphere of radius $r$.
Given the distance $\Delta x$, it follows that Eq. (\ref{k41_scaling_1}) is still valid on the local scale 
$r = \Delta x$, so that it can be mediated to provide the following expression:
\beq
S_p(\Delta x,0) = C^\prime_p \langle \epsilon_r^{p/3}
| r = \Delta x \rangle \Delta x^{p/3}
\ 
\label{k62_scaling}
\eeq
In this formulation, the averaged dissipation rate
$\epsilon_r ({\bf x},t)$ depends on the
large scales, i.e., on the structure of large-containing
eddies, thus partially addressing the Landau's objection\footnote{
The derivation of Eq. (\ref{k62_scaling}) follows from
two modified similarity hypotheses involving conditional
probabilities instead of total probabilities.
}.
In particular, Kolmogorov exploited this dependence to
address the intermittency problem and, to this goal, he considered the Oboukhov's proposal of a log-normal distribution for $\epsilon_r$, which is often referred to as the {\it third similarity hypothesis}\footnote{
A log-normal distribution of a random variable $X$
is defined as a Gaussian distribution of the random
variable $Y=log(X)$. Reading the original 1962 paper,
Kolmogorov write explicitly that \\
``...Oboukhov has now discovered how to refine our 
previous results in a way which takes Landau's comments into consideration''\\
before introducing the basic modifications of the new model, that is, the local spatial average defining the random variable $\langle \epsilon \rangle_r ({\bf x})$ and
the assumption of the associated log-normal distribution.
Thus, Kolmogorov seems to have basically followed the
ideas proposed by Oboukhov.
}.
In fact, in this approach, a proper model for the distribution of $\epsilon_r$ is sufficient to 
describe the velocity increment statistics.
The third similarity hypothesis includes also an {\it ad hoc} assumption for the dependence of variance on distance
$r=\Delta x$:
\beq
\sigma_r = \langle log( \epsilon_r({\bf x},t ) \rangle =
A({\bf x},t ) + \mu^* log\left( \frac{L}{r} \right)\ ,
\label{sigma_lognorm}
\eeq
where $\mu^*$ is assumed to be a universal constant.
Applying the third hypothesis, it follows:
\beq
S_p(\Delta x,t) = D_p({\bf x},t) \langle \epsilon \rangle^{p/3} \Delta x^{\zeta_p}\ , 
\label{k62_multiscaling_1}
\eeq
where
\beq
D_p({\bf x},t) = C_p({\bf x},t) L^{^{\mu^* p (p-3)/18}} 
\label{k62_multiscaling_2}
\eeq
depends on the large scale structure of the energy-containing eddies and
\beq
\zeta_p = \left[ \frac{1}{3}+\frac{\mu^*}{6} \right] p
-\frac{\mu^*}{18} p^2\ .
\label{k62_multiscaling_3}
\eeq
The nonlinear dependence of $\zeta_p$ on p is a signature
of multiscaling\footnote{
The case of a linear $\zeta_p$ with a constant (affine
function) also falls into the multiscaling case.
}, 
i.e., the system's self-similarity is not described only by one exponent, but by a set of exponents
that often depend on one or two adimensional
parameters, 
such as $\mu^*$ in Eq. (\ref{k62_multiscaling_3}).
The parameter $\mu^*$ is related to the 
autocorrelation function of the dissipation rate:
\beq
\langle \epsilon(x+r) \epsilon(x) \rangle
\sim \left( \frac{L}{r} \right)^{\mu^*}
\nonumber
\eeq
and it can be also evaluated by the 6th-order velocity
structure function \cite{frisch_1978}.

%
The K62 model is probably the first model  derived to reproduce intermittency of turbulent fluctuations by 
means of multiscaling exponents in the structure functions $S_p$.
Since then, a very large number of studies faced the 
problem of turbulent intermittency and multiscaling 
carrying out Direct Numerical Simulations (DNS) of the
Navier-Stokes equations or proposing different modeling approaches, mainly focused on local statistical features
of the dissipation rate or on the energy cascade.
This is the focus of so-called random cascade modeling approaches.
In fact, an extended research line was, and is still, focused on the phenomenological
modeling of intermittency in the framework of random cascade models mimicking the Richardson's turbulent energy cascade
(see, e.g., \cite{benzi_1984} and, for a brief review,  \cite{falkovich_2006,seuront_2005,sreenivasan_1997}).
An example is given by the $\beta$-model 
\cite{frisch_1995,frisch_1978}, where the multiscaling
exponent is given by:
    \beq
\zeta_p = \frac{\mu^*+1}{3} p -\mu^*
\ 
\label{frisch}
\eeq
Alselmet et al. (1984) \cite{anselmet_1984} carried out
a famous experiment and an accurate evaluation of
high-order velocity structure functions and they found
that:

\vspace{.2cm}
\noindent
(i) the linear and affine (linear+constant) behavior 
of K41 and $\beta$ models, respectively, do not reproduce
the experimental behavior of structure functions;\\
(ii)
k41 departs from the data starting from $p=4$, while 
the $\beta$-model performs slightly better (up to $p\approx 6-7$);\\
(iii) the K62 log-normal model reproduces the structure
functions up to the $12$-th order\footnote{
The orders greater than $12$ have less accuracy, but
the scaling relationship is preserved, so that the departure of K62 from the data for $p>12$ is considered 
genuine and definitive.
};\\
(iv)
the parameter $\mu^*$ was evaluated and given by:
$\mu^* = 0.2 \pm 0.05$.

  \vspace{.2cm}
\noindent
She and Leveque (1994) \cite{she-leveque_1994} proposed a stochastic model where the log-normal assumption is
 substituted with a log-Poisson one, thus getting:
 \beq
 \zeta_p = \frac{p}{3}-\frac{2p}{9}+2\left(
 1-\left( 2/3 \right)^{p/3}  \right)
 \ 
 \label{she_leveque}
 \eeq
These intermittency corrections much improved the agreement
with both Navier-Stokes simulations and experiments
\cite{she_2009,sreenivasan_1997}, e.g., the multiscaling 
exponents given in Ref. \cite{anselmet_1984} were much 
better reproduced by the She-Leveque log-Poisson model than
by the K62 log-normal model.

In summary, many authors went on addressing the problem of intermittent fluctuations in turbulence, as it is at the heart of the main theoretical and mathematical issue of turbulence, namely, the existence of a solution of the Navier-Stokes equation in the limit $Re \rightarrow \infty$, i.e., for vanishing viscosity.\\
%
Regarding the investigation of intermittency by DNS of Navier-Stokes equations, interesting
advancements were achieved about two decades ago by means of theoretical and numerical studies in the transport of passive scalars \cite{celani_2000,frisch_1998} .
These results are considered to be a milestone in the study of intermittency in homogeneous and locally isotropic turbulence as they found some probably
conclusive results on the universal character of the multiscaling spectrum in scalar intermittency
\cite{benzi-frish_2010}.
%
%
Remarkably, these numerical investigations on passive
scalar transport in Navier-Stokes turbulence started from
the conjecture that Kraichnan developed in 1994 with a random field model mimicking the spatial 
Kolmogorov spectrum \cite{kraichnan-prl-1994}, but with short-time correlation, thus suggesting that 
intermittency is mainly related to the spatial correlations of the turbulent velocity field.\\
In summary, the discovery of intermittency in 
small-scale turbulence triggered a bunch of studies,
both experimental and theoretical. 
It is interesting to notice that intermittency refers
to {\it large excursion events} that occur randomly in the
fluid flow. Thus, it seems natural to jointly investigate
the statistics of large excursion statistics (Magnitude
Intermittency) with that of the event spatial and/or temporal ordering (Clustering Intermittency).
%

\subsection{Magnitude Intermittency vs. Clustering Intermittency}

As anticipated in the Introduction, since Kolmogorov and Oboukhov papers in 1941
the great majority of the studies about intermittency in
small-scale turbulence refer to the
$p$-th order structure functions that characterize the
Magnitude Intermittency (MI) without any link to the
Clustering Intermittency (CI).\\
This gap is probably related to the intrinsic difficulties
in defining events unambiguously and detecting them in
experimental time series. In fact, the MI approach is a purely statistical approach that only needs the values of measured turbulent signals after a proper pre-processing, e.g., noise removal/reduction, of raw data.
Conversely, the detection of single structures and
related single transition events in the data needs some
evaluation of the local geometrical shape of random
signals and fields \cite{alfonsi_2006,ouellette_2012,primavera_2008,tardu_2014,zhong_1998}. This is often a challenging task as it 
depends on recipes of signal processing that, even if 
starting from quite general concepts, involve {\it ad hoc}
choices \cite{paradisi_epjst09,paradisi_npg12}.
The reasoning leading to many of these choices
follows from the specific problem, as they depend on the specific geometry of the structure to be detected within the signal or field.
To give an intuitive idea, given the two-dimensional graphic of a signal with time/space on the horizontal axis and the  signal intensity on the vertical one, MI (CI) is referred to the vertical (horizontal) axis. 
Being the horizontal (time or space) axis uniformly sampled, both
the statistical distributions of  this variable and of its
increments are trivial.
It is then clear that, at variance with MI, that only needs
the signal values to derive statistical distributions,
the CI approach need to refer to some definition of 
event and to the related algorithm of event detection, 
so that some assumptions must be made on the shapes that
 we want to detect in the light of their meaning linked
 to the underlying dynamical counterpart.
%
In this sense, the CI approach is more difficult that
the MI one.
\vspace{.1cm}
\noindent
To our knowledge, the focus on the distinction between MI and CI was firstly introduced
and investigated in Bershadskii et al. \cite{bershadskii_2004}.
This involved the direct detection of episodes/events in the turbulent flow in order to evaluate the CI structure.
In fact, the CI is directly linked to a viewpoint about turbulence that gives a primary 
relevance to the concept of  {\it crucial event}.
%
As far as we know, the application of {\it event-based} approaches to turbulence was first introduced or, at least, brought into the attention of the scientific community,  by Narasimha and co-workers \cite{kailas_1994,narasimha_1995,narasihma_1987,narasimha_1990,narasimha_2007},
who discussed this aspect and questioned about the need of an alternative ``episodic'' view
of turbulence with respect to classical approaches based on concepts such as the eddy diffusivity\footnote{
Eddy diffusivity modeling is an approach based on so-called Reynolds averaging of
Navier-Stokes equations where, due to nonlinearities, the averaged dynamical equations
must be closed with some assumptions about mean transport properties of eddies, i.e., of
turbulent fuctuations. Estimation of local eddy diffusivity through a modeled relationship
between fluxes (momentum, mass, heat) and gradients of mean fields (velocity,
concentration, temperature) is the first-order development of this approach.
The evaluation of eddy diffusivity comes from two approaches: (i) scaling arguments, following the prescription of the
Buckingham $\pi$ theorem of dimensional analysis;
(ii) spectral approaches, where eddy diffusivity is related to the correlation features of the turbulent fluctuations (Taylor's theorem \cite{taylor_1921}).
}.
In particular, Narasimha recalls and extends a statement by
Jeffrey (1926) \cite{jeffreys_1926} for which events ``are not to be thought of as perturbations on the background general circulation but as essential to maintaining it'' \cite{narasimha_1990}, where Jeffrey has in mind
a background modeled as a superposition of harmonic random waves, according to the spectral approach to turbulence.
Narasimha grounded his reasoning on experimental studies that highlight the episodic structure of  turbulence (see, e.g., \cite{corino_1969,kline_1967}).

\vspace{.15cm}
\noindent
Since 2004, many studies applied event-based approaches to turbulence studies in order to investigate the
role of CI with respect to the more studied MI \cite{bershadskii_2004,sreenivasan_2006}. 
Nowadays, this is especially a hot topic in the context of experimental and modeling studies of atmospheric turbulence near the ground, i.e., in the so-called Planetary Boundary Layer (PBL), 
where turbulence is known to emerge. In particular, the case of turbulent flows over complex terrain \cite{kaimal_and_finnigan_1994,finnigan_2000}
is taking momentum in the scientific community
\cite{cava_2009,cava_2012,huang_2013,huang_2021,katul_2006_surf_ren,poggi_2007,poggi_2009}.

\vspace{.25cm}
\noindent
Being both linked to the occurrence of large excursions
in the signal, CI and MI are two strictly interconnected aspects of the same intermittency phenomenon.
However, they give somewhat different information about the complex system, in this case the turbulent flow, and it is not yet clear if the parameters, i.e., the scaling exponents, derived from both approaches are strictly related to each other by a unique relationship only related to the dynamical flow equations or if, on  the contrary, their relationship changes according to the particular geometrical and dynamical conditions.

\section{A few words about Planetary Boundary Layer meteorology}
\label{pbl_sect}

The  Planetary (or Atmospheric) Boundary Layer (PBL or ABL) is an important part of natural environment where humans live.
The role of turbulence is central in PBL dynamics.
Thus, before passing to discuss the application of complexity concept to PBL turbulence, in this section we present a brief survey of the main PBL features and related terminology.

%
\vspace{.3cm}
\noindent
{\bf Basic concepts:}
The PBL concept is related to the more general one of boundary layer in fluid flows. This term was introduced in the literature by Prandtl in 1904 \cite{prandtl_1904} for a flow of a fluid of low viscosity close to a solid boundary. 
As regard atmospheric flows, a precise definition of boundary layer is not so easy. It is useful to define the boundary layer as ``that part of the troposphere that is directly influenced by the presence of the earth's surface, and responds to surface forcing with a time scale of about an hour or less'' \cite {stull_1997}. 
The simplest and more {\it idealized} boundary layer is over an infinite flat surface. It is equivalent to a flow horizontally homogeneous \cite {kaimal_and_finnigan_1994}. In addition to horizontal homogeneity, for most applications the flow is assumed to be stationary. 
The PBL can be roughly divided in two region, in analogy to the two-dimensional boundary layer generated in a wind tunnel:
\begin {itemize}
\item [1.] 
The {\it inner region}, called {\it atmospheric surface layer} (ASL) where the interaction between the atmospheric flow and the ground surface is very strong. The wind structure is mainly affected by surface friction and by the vertical exchanges of heat and moisture, and the flow is little affected by the earth's rotation. This region extend above the roughness obstacle, on average, up to $10\div 100$ m. In this region the effect of viscosity and of the structure of the roughness elements are negligible, the flow is fully turbulent and the vertical turbulent fluxes are mainly constant (their magnitude vary less than $10\%$) .
\item [2.] 
The {\it outer region}, sometimes referred as {\it Ekman layer} where the effect of the Coriolis force due to the Earth's rotation and of the pressure gradient are dominant. In this layer the surface fluxes decrease with height, that can reach $500-2000$ m. 
\end {itemize}
An overlap region is present between the two layers. Above the PBL is the free atmosphere, the flow is nearly geostrophic and no longer influenced by the surface. An {\it interfacial sublayer} (or {\it microlayer}) between the earth surface and the bottom of the ASL extends up to about three or four times the height of the roughness obstacles. This layer may be {\it viscous} in the case of smooth ground surface, or a {\it canopy sublayer} over rough surface. In the interfacial sublayer the flow is very inhomogeneous and three-dimensional.\\
PBL turbulence depends on the atmospheric large-scale dynamics and on the momentum and heat exchanges at the surface, strictly related to the radiation balance and to the heat flux at the ground. In addition, moisture effects can be taken into account \cite{tampieri_2017}. Turbulent motion can be described by a superposition of vortexes of different size, ranging from few millimeters to the depth of the boundary layers. Time series of meteorological variables can be used to represent spatial evolution of the variable under investigation, if the Taylor hypothesis of frozen turbulence is verified. This is true when the time life of a vortex is longer than the time the vortex needs to overpass the measurement instrument, that is the mean wind speed is large compared to turbulent intensity.

\vspace{.3cm}
\noindent
{\bf Stability of the Planetary Boundary Layer:}
The features of the PBL and its turbulent characteristics depend on the relative balance between the different process that generate turbulence, mainly due to buoyancy and mechanical forces. 
Mechanical turbulence production is generate from wind shear, while the buoyancy force are a consequence of the variation of air density with the altitude. This variation is produced by the thermal stratification due to heating of the ground by solar radiation during clear days, and its radiative cooling during the night.\\
In an adiabatic atmosphere the potential temperature is constant. %
The potential temperature is the temperature that a fluid parcel, with absolute temperature $T$  (in $K$ degrees) and air pressure $P$ at a height $z$, takes if it moves adiabatically to a reference height $z_0$ with pressure $P_0(z_0)$ and is given by:
\beq
\Theta = T \Big(\frac{P}{P_0}\Big)^{-R/c_p}
\label{potential_T}\ 
\eeq
$P_0$ is a reference pressure, usually set to 100 kPa or to the surface pressure, $R$ is the gas constant of air and $c_{p}$ is the specific heat capacity at a constant pressure (for air $R/c_{p}=0.286$).
For thermal stratification a {\it static stability} criterion can be adopted:
\begin{eqnarray}
\textrm{unstable:}\quad && {{\partial\Theta\over\partial z} < 0}\  \\
\textrm{neutral:}\quad && {{\partial\Theta\over\partial z} = 0}\  \\
\textrm{stable:}\quad  && {\partial\Theta\over\partial z} > 0\ 
\end{eqnarray}
\par\noindent
In a nutshell, air parcels tend (i) to oscillate around a given height in stable atmosphere, (ii) to go far fron the initial height in an unstable atmosphere and (iii) to remain in the same point in neutral atmosphere.\\
The static stability does not depend on wind. Since 
statically stable wind shear can generate turbulence, a comparison between the relative magnitudes of the shear production and buoyant consumption terms of the Turbulent Kinetic Equation (TKE) equation can be used to estimate when the flow might became {\it dynamically unstable}. An indicator is given by the dimensionless flux Richardson number,  defined as the ratio between the buoyant and the mechanical production terms of the TKE budget equation\footnote{
By convention: $U_1=U$ and $U_2=V$ are horizontal component of mean velocity, while $U_3 = W$ is the vertical one. Similarly for the velocity fluctuations: $u'_1=u'$ and $u'_2=v'$ and 
$u'_3 = w'$.
}:
\beq
R_f=\frac{(\frac{g}{\overline{\Theta}_V}) \overline{(w'\Theta'_v)}}
{ \overline{(u'_i u'_j)}          \frac{\partial\overline{U}_i}{\partial x_j}}\ 
\label{richardson}
\eeq
where $\Theta_v$ is the virtual potential temperature, that is the potential temperature that dry air must have to equal the density of moist air at the same pressure.
The Richardson number can be expressed as a function of the gradients of mean velocity and temperature. The gradient Richardson number is then derived by the Richardson flux number by assuming horizontal homogeneity and neglecting the vertical component.
This is defined by:
\beq
Rg=\frac{\frac{g}{\overline{\Theta}_V} \frac{\partial\overline{\Theta}_V}{\partial z}}{ \Big[\big(\frac{\partial\overline{U}}{\partial z}\big)^2 +       \big(\frac{\partial\overline{V}}{\partial z}\big)^2\Big]}\ 
\label{Richardson_gradient}
\eeq   
Based on the Richardson number, the following classification is usually given:
\begin{eqnarray}
&&R_f < 0 \quad \textrm {for unstable stratification }\ \\ 
&&R_f > 0 \quad \textrm{for stable stratification} \ \\
&&R_f = 0 \quad \textrm{for neutral stratification}
\ 
\end{eqnarray}
The value $R_f = 1$  is a theoretical limit beyond which atmospheric turbulence is suppressed. $R_f = 1$ is the value at which the turbulent energy production by shear stress is consumed by buoyancy forces, and 
$R_f << 1$ indicates that the heat flux is negligible in the TKE balance. In stable conditions, a critical value of $R_f$ can be defined and empirically estimated (0.20 - 0.25), below which stationary turbulence is possible. Above this value, intermittent or decaying turbulence can occur. This is the case of stably-stratified turbulent flows \cite{tampieri_2017}.
\par\noindent
Stability in the PBL can be measured by a scaling parameter, defined as follows. \\
Following the Monin-Obukhov Similarity Theory (MOST), it is assumed that the PBL dynamics is driven by the surface turbulent fluxes \cite{monin_1971}, assumed to be constant with the height. 
\par\noindent
A velocity scale $u_*$, called friction velocity, is defined as:
\beq
u^2_* =\Big[\overline {u'w'_s}^2+\overline{v'w'_s}^2\Big]^{1/2}
\eeq
and a length scale $L$, called Oboukhov length, is defined as:
\beq
L=-\frac{ \overline{\Theta_v }u^3_* }{ \kappa g ( \overline{w'\Theta'_v } )_s} \ 
\eeq
where $\kappa$ is a universal constant (the von Karman constant).\\
The Oboukhov lenght can be interpreted as a measure of stability. The sign of $L$ is determined by the sign of the heat flux, and it is the same of $R_f$:
\begin{eqnarray}
L<0 \quad \textrm {for unstable flow }\  \\ 
L>0 \quad \textrm{for stable flow}\  \\
L\to\infty \quad \textrm{for neutral flow}\ 
\end{eqnarray}
$L$ is proportional to the height above the surface at which buoyant factors first dominate over mechanical production of turbulence, and it is used to define the surface layer scaling parameter: 
\begin{equation}
\zeta=\frac{z}{L}\ 
\label{z_over_L}
\end{equation}
where $z$ is the height above the surface. As $\zeta$ decrease from ideal value $0$ (neutral condition, $L\rightarrow \infty$) to $-1$ (slightly unstable) the buoyancy effects become more important.

%
\vspace{.3cm}
\noindent
{\bf Evolution of the PBL:}
The evolution of the PBL over land, at mid-latitude, 
has a well-defined structure, related to the diurnal cycle. The major components are the mixed layer (ML), the residual layer (RL), and the stable boundary layer (SBL). 
\par\noindent
In cloud free situation, the growth of the mixed layer starts about a half hour after sunrise, due to solar heating of the ground. The depth of the convective boundary layer (CBL) increases during the day by entraining the less turbulent air from above and reaches its maximum value in late afternoon. Although a nearly ML can form also in region of strong winds, the turbulence is usually convective, i.e., buoyancy-driven. Thermals of warm air (updraft) rise from the ground, their growth is limited by a stable layer at the top of ML, called inversion layer. With the passing of thermals ramp-cliff structures are observed in the temperature time series.
Thermals of cool air (downdrafts) are related to the entrainment process at the top of the ML.\\
Turbulence is dominated by these large-scale coherent structures that extend vertically over the whole CBL, and the turbulent mixing is very strong. During cloudy days the growth of ML is slow, the intensity of thermals is reduced, and for high overcast, the ML may became non turbulent or neutrally-stratified. In the ML, mechanical turbulence is generated by wind shear across the top of the layer, with consequent formation and breakdown of so-called Kelvin-Helmholtz waves.
In late afternoon, about half an hour before sunset, the surface buoyancy flux decreases and changes sign. As a consequence, turbulence decays in near-adiabatic remnant of the daytime boundary layer. This layer is sometimes called residual layer (RL). After sunset, turbulence in the upper part of the layer continues to decay, whilst at low levels both a surface inversion and shallower and stable nocturnal boundary layer (SBL) develops \cite {garratt_1994}. \\
During the night the bottom portion of the RL is transformed into a stable boundary layer and the thickness of the nocturnal stable boundary layer gradually increases. The RL is entrained into the new ML in the next morning. In the SBL the turbulence has low intensity and sporadic behaviour. A low-level wind maximum, called nocturnal jet or Low Level Jet (LLJ), generally forms at the top of surface layer and enhance wind shears that tend to generate turbulence. The buoyancy force, instead, acts to suppress turbulence. As a result, turbulence sometimes occurs in sporadic bursts.

%
\vspace{.3cm}
\noindent
{\bf Stratified Atmospheric Boundary Layer:}
The PBL with weak stratification is well described by the similarity theory \cite{monin_1971} and by numerical models that assume stationary and homogeneous conditions. As stability increases, the structure and the description of the PBL  becomes  more  complex as the balance between mechanical and buoyancy force varies from case to case. The process at the top of the PBL can become more important than the surface forces in driving the PBL turbulence, and the SL is very shallow, if any. The interaction between small-scale turbulence structures, non-turbulent motion and mean flow leads to several different scenarios where turbulence is often intermittent \cite {mahrt_2014}. 
Following Mahrt \cite{mahrt_2014} it would be useful to adopt a simple scheme to distinguish the stable boundary layer in two regimes:
\begin{itemize}
\item [1.] 
Weakly stable regime, that refers to a boundary layer where SL, outer layer and entrainment zone can be defined. In this regime, turbulence decreases with height and is nearly continuous in time and space. This regime occurs with either cloud cover or strong wind and generally follows the similarity theory (``excluding nonstationarity and heterogeneity''). 
\item [2.] 
Very stable regime, that occurs with strong stratification and weak wind. In this regime the boundary layer is thinner than the weakly stable boundary layer (typically of an order of magnitude), similarity theory is not satisfied and several vertical structures (usually non stationary) are possible. For example, turbulence can increases with height, and the maximum value can be reached in a layer intermittently coupled to the surface.
\end{itemize}
As already highlighted, a critical flux Richardson number can be used to separate between the two regimes: the turbulence decreases with increasing Richardson number, after which it varies slowly or remains stationary. 

\vspace{.1cm}
\noindent
Some peculiar features of turbulence in the very stable regime \cite{mahrt_1999,mahrt_2010,mahrt_2014} are given in the following:
\begin {itemize}
\item [-] The range of turbulent scales decreases with increasing stability. In strong stability case, the turbulent eddies can be confined to small scale not directly interacting with the ground. 
\item [-] 
The main source of turbulence can be at the top of the surface inversion.
\item [-] 
Submeso motions are always present, with scales ranging from the main turbulent eddies scale ($\sim \rm{O(100 m)}$) to the smallest meso-gamma scale ($\sim \rm{2 km}$). 
\item [-] 
A separation between turbulence and waves is not possible. There are intermediate ranges of scales having characteristics between those of turbulent and non turbulent motion (hybrid motion). Furthermore, a temporal overlap of turbulent and non turbulent is possible. 
\item [-] 
The turbulence is typically not in equilibrium with the non-turbulent motion. The non-stationarity contributes to intermittent behavior of the  turbulence. 
\item [-] 
Small scales characterize vertical mixing, which is weak, and the impact of individual roughness elements may be important. 
\item [-] 
Small correlations are present between vertical velocity fluctuations and scalars.
\end {itemize}

%
\vspace{.3cm}
\noindent
{\bf Intermittency:}
According to Mahrt (1999) \cite {mahrt_1999} all turbulence can be considered 
intermittent to the degree that the fine scale structure occurs intermittently within larger eddies, and a definition of intermittency in the PBL depends on the considered scale \cite {mahrt_2014}. In any case, intermittency can be seen as a strong variability of the turbulence in space and time. 
Tsinober (2009) \cite{tsinober_2009}, refers definitions of intermittency in the PBL:
\begin{itemize} 
\item [-] 
{\it ``At any instant the production of small scales is ... occurring  vigorously in some places and only weakly in the others''} (Tritton (1988) \cite{tritton_1988})
\item [-] 
{\it ``Typical distribution of scalar and vector fields is 
one in which there appear characteristic structures 
accompanied by high
peaks or spikes with large intensity and small duration of
spatial extent. The intervals between the spikes are 
characterized by small intensity and large extent''} (Zeldovich et al. (1988) \cite{zeldovich_1988})
\item [-] 
{\it ``Intermittency is a phenomenon where Nature spends little time, but acts vigorously''} 
\end{itemize}
The intermittency in the PBL is traditionally divided into
three different kinds depending on the causes from which it originates:
\begin{itemize}
\item[-]
{\it Internal} or {\it small-scale intermittency} 
\item[-]
{\it External} or {\it large-scale intermittency} 
\item[-]
{\it Global intermittency} 
\end{itemize}

\vspace{.25cm}
\noindent
{\it Internal Intermittency}

\vspace{.25cm}
\noindent
As already said above in Section \ref{homo_iso_turb},
small-scale, i.e., internal intermittency has been extensively studied 
since first experiments performed by Batchelor and
Townsend (1949) \cite{batchelor_1949}. Small-scale 
intermittency, also called microscale or fine-scale 
intermittency, affects small subregions of main/large-scale eddies. 
It arises 
from overall
modulation of turbulence by the main eddies or in
connection with sharp edges of the main eddies themselves, 
where dynamical flow instabilities occur \cite{mahrt_1989}. 
For thermals in the CBL, small-scale turbulence is more intense inside than outside the eddies and is concentrated in small sub-regions of the eddies. 
The regions of concentrated shear at the edge of the main eddies are called microfronts, gust fronts, pulses. In a very stable boundary layer, where the turbulence is non-stationary for the external forcing by submeso motions, it is difficult to isolate internal intermittency. However, internal intermittency has been identified in such stability conditions both experimentally \cite {kit_2017}, than with numerical simulations \cite {pardyjak_2002,rorai_2014}. 
%
%

\vspace{.25cm}
\noindent
{\it External Intermittency}

\vspace{.25cm}
\noindent
External intermittency 
is associated with the coexistence and the alternation of laminar and turbulent flow regions,
in particular with the random movements of the boundary between turbulent and non-turbulent regimes,  and to the continuous transition of laminar flow into turbulent via the boundary \cite {ansorge_2016b,tsinober_2009}.
Significant regions of the outer layer of a boundary layer characterized by external intermittency may be non-turbulent \cite{ansorge_2014}.

\vspace{.25cm}
\noindent
{\it Global Intermittency}

\vspace{.25cm}
\noindent
When a laminar patch of fluid extends down to the surface, the external intermittency may become global. 
The global intermittency occurs on scale larger than the scale of the main turbulent eddies. Global intermittency is often observed in the very stable regime, as extensively discussed in Mahrt (1999) \cite{mahrt_1999}, who also used the term global intermittency to indicate the turbulence intermittency at high levels of a weakly stable boundary layer. 
Direct numerical simulations of the turbulent Ekman layer performed by Ansorge and Mellado (2014) \cite{ansorge_2014} reproduce global intermittency. Using as initial condition a neutrally stratified flow, they found that, beyond a certain stability, global intermittency is intrinsic to the stable stratified boundary layer.

\vspace{.2cm}
\noindent
Interestingly, the Manneville-Pomeau map introduced in Section \ref{intermit_complex_sect} could resemble some of the features
of both external and global intermittency, in particular the
alternance of turbulent and ``laminar'' periods.

\vspace{.3cm}
\noindent
{\bf Wind gusts:}
%
During clear night, wind gusts are the main responsible for turbulent exchange between the surface and the upper boundary layer, thus contributing to mixing processes of scalar quantities. 
Accordng to Acevedo et al. (2003) \cite{acevedo_2003}, 
``weak wind gusts below a threshold (approximately 1.5 m/s) mix the air down to the colder ground, cooling the surface layer'' and
``wind gusts above this threshold promote mixing with upper levels, warming the surface''.\\
In order to identify wind gusts, Cheng et al. \cite {cheng_2011} divide the turbulent fluctuations $s^\prime$ into two parts: the high frequency part, defined as turbulence in a narrow sense, and another that is traditionally defined as low-frequency turbulence. 
The frequency of turbulence fluctuations is larger than $0.017\ Hz$ (time scales less than one minute), while the period of low-frequency turbulence is in the range [$1$ to $10$ minutes]. The large-scale flow is calculated as a time average over $10$ minutes.
Then, the low-frequency turbulence identifies the so-called gusty wind disturbance. 
This definition is in agreement with that given by {\it World Meteorological Organization (WMO)} (1983), that defines gusty wind as a fluid flow fluctuation at time scales between $1$ and $10$ minutes. 
During strong wind, turbulence and wind gusts have different features. As highlighted by Cheng et al. \cite {cheng_2012}, in such situations the turbulence is nearly isotropic, with weak and random coherent structures, while gusts are always anisotropic, and strongly correlated coherent structures are present. \\
%

\section{Self-organized structures and crucial events in turbulence}
\label{structures_and_events_sect}

An interesting general observation about coherent structures in turbulence is given in Robinson (1991) \cite{robinson_1991}:
\begin{itemize}
\item[-] 
{\it ``it can be said that a {\bf coherent motion} is
defined as a three-dimensional region of the flow over which at least one
fundamental flow variable (velocity components, density, temperature, humidity, vorticity, other scalars or vector fields)
exhibits significant correlation with itself or with another
variable over a range of space and/or time that is significantly
larger than the smallest local scales of the flow.''}
\end{itemize}
Further, according to Tsinober \cite{tsinober_2009} :
\begin{itemize}
\item[-]
{\it ``The nature and characterization of the structure(s) of turbulent flows are among the most controversial issues in turbulence research''};
\item[-]
{\it ``intermittency [...] is intimately related to some aspect of the structure(s) of turbulence''};
\item[-]
{\it ``the difficulties of defining what the structure(s) of turbulence are (mean) are of the same nature as the question about what is turbulence itself''}.
\end{itemize}
The following sentence is of particular interest:
\begin{itemize}
\item[-]
{\it ``Fluid dynamical turbulence (even homogeneous and isotropic) has structure(s), i.e. contains a variety of strongly localized events, which are believed to influence significantly the properties of the turbulent flows''}.
\end{itemize}
The above sentences clearly show that the central role of coherent structures in turbulence and the intimate link between coherent structures and ``strongly localized events'' is widely recognized in the scientific community.
However, the last sentence shows that the term {\it event} in the turbulence community is sometimes used to denote the coherent structure itself and, in general, they seem to refer to large-scale coherent structures.
Then, considering also the discussion given in Section \ref{intermit_complex_sect}, it seems that
there could be some slightly different terminology used in
the turbulence and statistical physics communities.
To avoid ambiguities, herein we will assume the conventions described in the following.

\vspace{.25cm}
\noindent
{\bf Self-organized structures}

\vspace{.25cm}
\noindent
The term ``coherent'' is associated with ``self-organized'' or ``self-organizing''. 
In other fields, 
such as neuroscience, a similar term is ``synchronization''.
However, the term {\it self-organization} will be here used as an abstract concept that could be seen as a
general condition characterizing a class of cooperative
multidimensional systems (infinite-dimensional in the fluid dynamics case).
Coherence or synchronization belong to this general class and they involve
some more specific mathematical definition of self-organization, e.g., phase 
coherence among different degrees of freedom.\\
%
Thus, in agreement with the jargon often used in the turbulence community, the term ``coherent structure'' will be used to denote large-scale eddies or large-scale vortex motions, which are
examples of {\it self-organized turbulent structures}, a more generic term used hereafter.\\
We can then exploit the sentence by Robinson (1991) \cite{robinson_1991} to give the following

\vspace{.2cm}
\noindent
{\bf Definition (self-organized turbulent structures)}\\
A self-organized turbulent structure is an emerging metastable state of the turbulent flow. It is defined as a three-dimensional region of the flow characterized by strong memory (time coherence) and a spatially coherent motion between different fluid particles
over a range of space and/or time that is significantly
larger than the smallest local scales of the flow, i.e.,
the Kolmogorov microscale of viscous dissipation.

\vspace{.2cm}
\noindent
The emerging self-organization of these structures is triggered by 
the strong interactions among different fluid particles and, in particular,
by competition between inertia and viscosity and by the nonlinear mechanism of momentum transport. The non-linearities trigger the emergence of self-organized metastable states with coherence lengths that are much larger than the Kolmogorov microscale, that is, 
even if the interactions are local in space, i.e., between nearby fluid particles, self-organization emerges, a condition that was often found in
complex systems (see, e.g., \cite{turalska-grigolini_pre11_critical}).\\
%
%
Depending on the considered range of space/time scales, this terminology can include not only large-scale
structures, such as energy-containing eddies, but also
smaller structures,
such as the small-scale eddy motions generated by the 
instabilities at the sharp edges of the large-scale energy-containing eddies. 
Conversely, this definition includes also other structures that are not interpreted inside the concept of Kolmogorov-Oboukhov-Richardson energy cascade and that could also be non-turbulent in the strict sense and acting on time/space scales larger than the energy-containing turbulent eddies.\\
%
%
The term ``episode'' could be sometimes used to denote the overall occurrence of a self-organized structure; 
this terminology also includes the time duration (life-time) of the self-organized structure.

\vspace{.25cm}
\noindent
{\bf Turbulent events}

\vspace{.25cm}
\noindent
According to the paradigm of TC/IDC systems introduced in
Section \ref{intermit_complex_sect}, 
the terms ``event'', ``crucial/critical event'' or ``transition event''  will denote a dynamical condition corresponding to a 
strong variation in some flow variables occurring on a time interval shorter than both external/driving and internal dynamical time scales of the turbulent flow, i.e., the Kolmogorov microscale.\\
Similarly to TC/IDC systems, the event occurrence is associated with a
fast memory drop corresponding to a fast decay of the spatial coherence of the flow metastable structure. 
This is a somewhat different concept with respect to events that are defined on the basis of the crossing through a threshold.
This kind of events will be named Threshold Crossing Events (TCEs), of which 
a particularly interesting case is given by Zero Crossing Events (ZCEs).\\
%
%
Going back to the RTE concept, this includes transitions given by a fast sequence of very short-time large excursion events, resulting in a {\it bursting} behavior. 
This is likely to be recognized as a short-time transition 
event, but it could be also considered as a self-organized
structure when the total duration time of the bursting 
behavior is comparable or longer of the shortest time scale
of the fluid flow.
%
%
In the turbulent boundary layer, bursts are separated in two categories \cite{robinson_1991}: single-event (or isolated) burst and multiple event bursts. The first is associated with a local instability, while in the latter a more persistent ejection of near-wall fluid is due to the passage of a large-scale vortex structure. 
The main difference is in the degree of temporal intermittency and in the time period between two subsequent events.
It is important to underline that the main difference between self-organized structures and 
events lies  in the time duration of these two different dynamical conditions: relatively
long duration time, also named ``life-time'', for the self-organized structures and time duration shorter than all time scales of the system for crucial events.\\
%
%
and structure detection algorithms (see next Sections).

\vspace{.15cm}
\noindent
After the detection of events in the signals, there are many features that can be used to characterize both each single self-organized structure and each single event.
Limiting to consider a single fluctuating variable (e.g., velocity, temperature or other), the main interesting features are:
%
\begin{itemize}
\item[-]
maximum and minimum values and related maximum excursion of the variable inside a turbulent structure;
\item[-]
time instants in which the self-organized structure is born (arising/emergence, birth RTE occurrence time) and dies  (decay/fallout, death RTE occurrence time). In some complex systems, the
death time of a self-organized structure could practically coincide with the birth time of the successive one. In this case, there is simply a sequence of crucial events marking the passage between two successive structures\footnote{
Another equivalent view is that the disordered state is included or coincide with the RTE and, in any case, their overall time duration (transition event$+$disordered state) is so short that it can be represented
as a point process \cite{cox_1980_point}.
}
\end{itemize}
Associated to the birth and death times we can derive different kind of Inter-Event Times (IETs):
\begin{itemize}
\item[-]
time duration of the structure (life-time), i.e., difference between the death and birth times of the same self-organized structure;
\item[-]
time interval between birth times of two successive structures;
\item[-]
time interval between death times of two successive structures;
\item[-]
time interval between successive events, without distinguishing if birth or death events;
\item[-]
others: time interval between peak values of two successive structures, time interval between minimum
and maximum value inside the same turbulent structure, steepness of the signal at the event occurrence time.
\end{itemize}
Since they depend on each other, only a subset of these IETs is usually analyzed, depending on the feature we are interested in and/or that characterizes better the sequence of events and of self-organized structures.

\vspace{.25cm}
\noindent
{\bf Coherent turbulent structures}

\vspace{.25cm}
\noindent
In atmospheric boundary layer, large-scales self-organized, i.e., coherent turbulent structures are usually described involving the following elements \cite{robinson_1991}:
\begin {itemize}
\item [-] 
{\it vortex}, that is any rotating fluid region, with horizontal orientation (both stream-wise direction or curved and tilted) or vertical $z$ direction;
\item [-] 
{\it sweep} (wallward motion of high momentum fluid) and {\it ejection} (upward motion of low momentum fluid);
\item [-] 
{\it low-speed} and {\it high-speed} perturbations from the mean value at any {\it z-location}.
\end {itemize}
According to the quadrant-splitting classification scheme \cite{wallace_1972}, the instantaneous product signal $u^\prime w^\prime$ is used to define four different contributions to the Reynolds stress $\langle u^\prime w^\prime \rangle$, each one corresponding to one of the above elements:
\begin {itemize}
\item [1.]
Q1: $u'>0$ and $w'>0 \Rightarrow$ high-speed fluid reflected outwards from the wall;
\item [2.] 
Q2: $u'<0$ and $w'>0 \Rightarrow$  Ejection;
\item [3.] 
Q3: $u'<0$, $w'<0 \Rightarrow$ Low-speed fluid deflected towards the wall;
\item [4.]
Q4: $u'>0$, $w'<0 \Rightarrow$ sweep.
\end {itemize}
In the boundary layer, local shear-layer instabilities near the boundaries can be responsible for the birth of vortexes.
Low speed streaks between patterns of streamwise vortexes are often formed. This is an important mechanism of turbulent kinetic energy production \cite{robinson_1991,schoppa_2002}. \\
Outer-layer structures with large streamwise length scales can modulate near-wall structures. These amplitude modulation effects become progressively stronger as the Reynolds number increases, as demonstrated in laboratory experiments and in atmospheric surface layer measurements \cite{mathis_2009}. 
In a mixing-layer flow spanwise vortexes develop
at a plane of high shear \cite{lasheras_1988,rogers_1992}.\\
Coherent structures have been extensively studied under different stability conditions, both experimentally and by numerical simulations.
Khanna and Brasser (1997) \cite{khanna_1998} explored the relative roles of buoyancy and shear in the formation of coherent structures in the PBL. A wide range of atmospheric stability conditions was analyzed using a LES model with
different values of the scaling parameter $\zeta=z/L$ (Eq. \ref{z_over_L}). In particular, Watanabe et al. (2019) \cite{watanabe_2019} used direct numerical simulation (DNS) to investigate the flow structures in a stably stratified shear layer. They found structures similar to those of wall-bounded shear flows, as hairpin-shaped vortexes. Laima et al. (2020) \cite{laima_2020} found a change in the form of the dominant
turbulent eddy at the change of stability: small scale structures are present in strong stable conditions, large scale structure in neutral conditions and then thermal plumes in strongly unstable conditions. \\
According to Robinson (1991) \cite{robinson_1991}, the main classes of coherent motions in wall-bounded shear flow are: ejection of low-speed fluid outward from the wall, sweeps of high-speed fluid inward toward the wall, low-speed streaks, hairpin-shaped vortex structures, sloping near-wall shear layers, near-wall pockets.  
Large vortex turbulent structures are observed below three-dimensional ``bulges'' that form in the outer region at the interface between turbulent and non-turbulent portions of the fluid flow.
%

\section{Event detection techniques in turbulent signals} 
\label{event_detect_sect}

As previously said, the identification, analysis and modeling of self-organized structures and of associated transition events are the fundamental elements in the physical description of the flow in terms of self-organized motions, i.e., in the event-based approach. \\
For example, in convective
turbulence, two self-organized states are naturally  given by updrafts and downdrafts that alternate each other; similarly, in wall turbulence at high Reynolds number, ejections and sweeps are characterized by RTEs 
marking the rapid momentum exchanges between 
inner and outer layer.
Regardless of the scales of motion involved and the causes leading to the alternance ``self-organized $\rightarrow$ self-organized'' or ``self-organized $\leftrightarrow$ disorder/laminar'', intermittency is related to both amplitude variability (MI) and
space/time ordering of events, RTEs or TCEs (CI).

\vspace{.1cm}
\noindent
The presence of self-organized states can be revealed and globally characterized by means of power-law decaying long-range time/space correlations.
In fact, let us assume to have a set of Eulerian measures of the fluid flow (e.g., velocity, temperature and other scalars) and that crucial events emerges in the flow field. 
Given the availability of a single realization of the fluid flow random field, memory (time coherence) and spatial coherence can be globally evaluated by means of time/space two-time two-point correlations, computed as time or space averages. 
Let us assume a stationary condition, so that we can limit to consider time averages. Then, time averaged space/time correlations depend only on time lag and can be computed
by a time moving average over the Eulerian data.
For sufficiently long time lags, this surely includes
time scales much larger than those of the rapid transition events.
Given a generic zero mean signal $s^\prime(t)$, the calculation of the time average is carried out as follows: 
\beq
C_{s^\prime}(\Delta t) = \langle s^\prime(t)s^\prime(t+\Delta t) \rangle = \lim_{T\rightarrow\infty}
\frac1T \int_0^{T} s^\prime(t) s^\prime(t+\Delta t) dt 
\ 
\label{time_corr}
\eeq
Let us limit to the case where self-organization death and birth events coincide.
Then, considering that memory drops out passing through an event, the
correlation in Eq. (\ref{time_corr}) receives contributions from all those 
windows remaining inside the same turbulent structure, i.e., between two successive
events. On the contrary, the contribution from time windows including at least
one transition event is zero when carrying out the time average.
For this reason, the time averaged correlation functions can give some
statistical global information about the self-organizing ability of the
system, thus characterizing the presence of self-organized structures. \\
Conversely, it is important to underline that the long-range correlation functions
cannot be exploited to detect events. 
In fact, the correlation functions are global statistical features, defined by means of time averaging, that depend on
a time lag and not on the actual time in which events occur.
Conversely, memory is a dynamical feature,
affecting the local dynamics of the system and, thus, also the evolution and occurrence of RTEs.
%
%
We could try to see if the availability of a statistical ensemble of flow realizations allows to derive a correlation function that is sensitive 
to single event occurrences. In other words, could events be identified by
looking for abrupt falls and climbs of the ensemble averaged correlation
function?
Actually, the answer is again negative. As in the case of time average, ensemble, or space, averaged correlation functions are statistical features over the ensemble of
flow realizations, each one having its own sequence of RTEs that are averaged out in the correlation functions themselves.

\vspace{.1cm}
\noindent
The above reasoning proved that correlation functions can give a global picture of the self-organized structures in the turbulent flow, but lose the detailed flow structure, which is averaged out, and cannot be used to detect single events and recognize self-organized structures.
The episodic description of turbulence is therefore based on characterizing self-organized structures in terms of local geometrical properties of the measured
signals (e.g., velocity steepness). This approach includes the techniques to detect the short-time transition events and their occurrence times.
%
%

\vspace{.1cm}
\noindent
The detection of events is a delicate task that require proper algorithms. Different algorithms refer to slighlty
different operational definitions of transition event. Event detection algorithms could be divided into two
main classes: the algorithms based on the crossing of one or two thresholds by the analyzed variable and the
algorithms based on evaluating the local steepness of the signal. In the second case, some care has to be taken in the
evaluation of the signal derivative, which is usually affected by noise much more that the original signal.\\
Limiting to the case of temporal signals, after the event detection algorithm has been applied we get a sequence of
event, each one labelled by an integer number and associated
with a {\it event occurrence time}, so that the following
sequence of times is derived:
\beq
{\cal E} = \left\{ t_k \right\}_{k\in A}\ ,
\label{event_seq}
\eeq
where $A$ is the set of the first integer numbers:
$A = \left\{0,1,2, ...,L \right\}$ and $t_0=0$ by convention. For an infinite duration, it results $L=\infty$.
This event sequence can then be exploited to derive
different complexity exponents.

\vspace{.1cm}
\noindent
Among the most used algorithms there are
the threshold-crossing schemes such as $U$-level \cite{luchik_1987}, modified $U$-level \cite{lu_1973}, Variable Interval Time Averaging (VITA) technique \cite{blackwelder_1976}, the
quadrant analysis technique \cite{wallace_1972}. The effectiveness of several detection algorithms was discussed in Bogard (1986) \cite{bogard_1986}. The dependence on the operational parameters, i.e. threshold levels and averaging or window times, is minimized  comparing the number of detected events with those identified by flow visualization.

\vspace{.25cm}
\noindent
{\bf U-level method and modified U-level method}

\vspace{.25cm}
\noindent
The $U$-{\it level} and modified $U$-{\it level} methods use the measurements of
the fluctuating streamwise velocity component ($u^\prime$). Tubergen and Tiederman (1993) \cite{tubergen_1993} affirm that among single-component burst detection scheme, the modified $U$-{\it level} method is to be preferred for  its threshold independence. 
Metzger et al. (2010) \cite {metzger_2010} apply the $U-level$ algorithm to wind-tunnel measurements. They identify a burst event by the occurrence of an ejection, here identified as a strong negative value of $u'$. The crossing through a lower threshold identify the leading edge of the burst event, while the trailing edge of the burst is detected by the crossing of an upper threshold. %
Bogard (1986) \cite{bogard_1986} recommends to set the lower threshold at one standard deviation of the signal
($ - \sigma_u$), while the upper threshold at 
$ - 0.25 \sigma_u$.\\
With the modified U-level threshold algorithm
two time scales can be examined: 
\begin{itemize}
    \item [1.] the time period between events, $T_e$, defined as the time between the leading edges of successive events. Histograms of $T_e$
at each fixed wall-normal location in the boundary layer for different
Reynolds number data sets, 
follows a Poisson-like distribution; 
    \item[2.] the event duration, 
$\Delta T$, calculated as the time between the trailing and leading edges of a single event. The general shape of the
histograms of $\Delta T_m$ 
are also well-modelled by 
a Poisson distribution.
\end{itemize}

\vspace{.25cm}
\noindent
{\bf The Telegraph Approximation (TA)}

\vspace{.25cm}
\noindent
The TA belongs to a class of thresholding techniques, such as the first passage time, often used in dynamical systems and stochastic processes theories.
TA is defined through the Zero-Crossing Events (ZCEs) of the signal fluctuation \cite{bershadskii_2004}:
\beq
TA_s(t) = {1 \over 2} \Bigg( {s^\prime(t)\over \left |s^\prime(t)\right | }+1 \Bigg)
\label{ta}
\eeq
where $s^\prime(t) = s(t) - \langle s \rangle(t)$.
The method finds the passages of the 
entire signal through the mean value, and it assumes
value $TA_S = 1$ when $s^\prime$ is positive and 
$TA_S = 0$ when $s^\prime$ is negative. Thus, $TA_s(t)$
is made on sequences of $0$ and $1$ and the transitions
$0 \rightarrow 1$ and $1 \rightarrow 0$ corresponds
to the ZCEs, which are the same as the original signal.
The mean value $\langle s \rangle(t)$ used to evaluate the ZCEs is evaluated in different
ways: if the signal is stationary, the average can be computed over the
entire time series, while in the general case it is computed as
a local average over a sliding time windows of proper length $T$:
\beq
\langle s \rangle(t) = \int_{t-T/2}^{t+T/2} s(t^\prime) dt^\prime\ .
\nonumber
\eeq
Another often used alternative definition of mean value is given by
carrying out a local trend over time intervals $[t,t+T]$, which is
derived by means of the minimum square method, usually with a
linear fitting function.\\
This method was firstly applied to turbulent thermal convection
by Sreenivasan and co-workers \cite{bershadskii_2004,sreenivasan_2006}. 
This event definition is simply related to a threshold passage that could also include transitions with low 
signal steepness. 
The independence from the threshold in a given range was
sometimes used as a quite reasonable signature that ZCEs are genuine transition events.
Authors exploiting TA claim that it allows to isolate the clustering from amplitude variability. However, this is only true for that part of clustering related to zero-crossing events (ZCEs). Thus, the usefulness and significance of this event detection algorithm is 
related to the presence and the leading role  of updraft/downdraft or ejection/sweep alternance.

\vspace{.25cm}
\noindent
{\bf VITA method}

\vspace{.25cm}
\noindent
The bandpass filter VITA method can be applied to either $u^\prime$ \cite{metzger_2010} or $u^\prime w^\prime$ \cite{narasimha_1990}.
This method was developed to detect large-scale
self-organized, i.e., coherent structures, which
were identified as turbulent bursts. 
As usual, the beginning and ending time of the structure
defines the crucial transition events\footnote{
As already observed, here the authors use the term ``event''
to denote the motion that is here conventionally named ``self-organized structure''.
Regarding VITA method, we will use the term ``burst'' to denote the structure.
}.\\
Thus, for any zero mean fluctuating turbulent signal 
$s'(t)$, the short-term variance $D$ is defined as:
\beq
D(s';t,t_{av)}\equiv {1 \over t_{av}} \int_{t-t_{av}/2}^{t+t_{av}/2}|s'(t')|^{^2}   dt' - {1 \over t_{av}} \left( \int_{t-t_{av}/2}^{t+t_{av}/2} s'(t)dt \right)^2
\label{vita}
\eeq
where $t_{av}$ is the averaging time interval. 
The burst is identified when a peak in $D$ exceeds a discriminatory level $k \sigma_s^{^2}$, where $k$ is the prescribed threshold and $\sigma_s^{^2}$ is the mean square value of $s'(t)$, defined as:
\beq
\sigma_s^{^2} = \langle (s')^2 \rangle = \lim_{t_{av\to\infty}}  {D(s';t,t_{av})}\ 
\label{vita_thr}
\eeq
According to the terminology of these authors, in addition to {\it ordinary bursts}, it is possible to detect a {\it super-burst} when a number of bursts coalesce into one single burst at longer $t_{av}$. At shorter $t_{av}$ we have a cluster of bursts, each of which is called {\it sub-burst}. 
\par\noindent
A flux bursting structure is detected if $s'$ is the 
turbulent flux $u'w'$. The method can be applied to detect an atmospheric burst, that occurs in the times when the calculated instantaneous flux have larger value than at others, as defined by Kline et al. (1967) \cite{kline_1967}. These authors also studied the time interval between two bursts as the time from the centre of two successive bursts. This clearly is a different feature with respect to the
inter-event time, but these two definitions could be equivalent.
Narasimha and Kailas (1990) \cite {narasimha_1990} found that, in a nearly-neutral atmospheric boundary layer, the probability distribution of the time intervals between bursts
reveals two distinct types of bursts: lone (isolated) bursts and bunched bursts.
This aspect was highlighted by Bogard (1986) \cite{bogard_1986} in a flow-visualization study. Successive ejections are spatially close if they come from the same streak, while they are separated when coming from different streaks. A burst is then a self-organized structure formed by a organized cycle of ejections. 
A burst is evidenced by VITA at longer averaging time.
\par\noindent
The duration of a burst of intensity k is defined as the time during which $D(u'w')$ is greater than the threshold $k$.

\vspace{.25cm}
\noindent
{\bf Geometric shape method} 

\vspace{.25cm}
\noindent
Starting from the evidence that visual inspection of atmospheric boundary layer time series reveals common shapes, despite the different timescale and physics phenomena involved, an attempt to recognize the geometric shape from time series has be done by
 Belu\v{s}ic and L. Mahrt, 2012 \cite{belusic_2012}. The authors try ask themselves the question: ``{\it is geometry more universal than physics in atmospheric boundary layer?}''  The method is then based on the recognition of the geometry of the structure more than on its amplitude, that varies across different scales. The main steps are the following:
\begin{itemize}
\item [(i)] 
A predefined shape function is moved through the time series. At each time point the linear correlation coefficient $r$ is calculated. The procedure is repeated over several shape functions.
\item [(ii)] 
The properties (lenght, location, etc) of the structure are saved and the corresponding part of the series is removed.
\item[(iii)]
After selecting the shape function with the next largest $r$, the previous step is repeated until all shape functions with $r>0.9$ are selected.
\item [(iv)] 
The procedure is repeated for different scales.
\end{itemize}
The threshold value  $r=0.9$ allows to distinguish between the different shapes. The shapes chosen are: a simple sine function, a step function, a ramp-cliff function, and a cliff-ramp function (or a reversed ramp-cliff). These shapes correspond to physical features of the boundary layer:  a wave (sine), a (micro)front (step) \cite{mahrt_2010a}, and differently oriented turbulent ramp-cliff patterns \cite{antonia_1982}. \\
Since some definitions of self-organized structure are based on the existence of spectral phase correlation \cite{chian_2008,provenzale_1992}, the method was tested on a portion of the original series  where self-organized structures were removed by phase randomization. Then, the method detects only events not based on the existence of phase correlation. The number of detected events in the surrogate time series was about $90\%$ less compared to the original data. Furthermore, a visual inspection of the captured structures was performed. 
The analysis was applied to two boundary layer stability regimes: unstable and stable. Results 
highlighted that the geometry of the structures is nearly independent from the time scale, although the physical process generating the shape changes with the considered scale.

\vspace{.25cm}
\noindent
{\bf Noise test method} 

\vspace{.25cm}
\noindent
Kang et al. (2014) \cite{kang_2014a} proposed a method to extract and classify self-organized structures\footnote{
The authors use the term ``coherent structure'' to indicate fluid elements with given features, and the term ``event'' to indicate the trace of this same coherent structure in the time series. In order to avoid confusion, here we will use our convention, which was introduced above, thus denoting the Kang ``event'' as ``self-organized turbulent structure''.
}
from time series of turbulence data, without an {\it a priori} knowledge of the dynamics and of the generating mechanisms. The method was applied both to artificial and to atmospheric data and is based on determining
the characteristics (i.e., colors) of background noise, which were
assumed to be known {\it a priori}. \\
Two different noise models were considered: white noise and red noise, 
used to represents the background noise of atmospheric turbulence.
In a white noise process, the elements of the time series at different times are not correlated. The Ljung-Box test \cite{box_1970} is applied to check if the data points are independently distributed. The null hypothesis $H_0$ (i.e. the data are independently distributed) is rejected when the $p$ value is less than  the significance level $0.05$. In this case the data points do not resemble a white noise.\\
The red noise is modeled as a first-order autoregressive process AR(1), where the error term is given by a white noise. A time series $s^\prime(t)$ is then represented as:
\beq
s^\prime(t)=\Phi\, s^\prime(t-1)+\epsilon (t) 
\ 
\eeq
being $\Phi$ ($0<\Phi<1$) the first-order autocorrelation coefficient and $\epsilon(t)$ a with noise process. Only stationary processes can be modeled as an AR(1) process, then a stationary test is first applied to $s^\prime(t)$\footnote{
The authors use the so-called Philips-Perron (PP) unit root test 
\cite{banerjee_1993,perron_1988}.
}.  
If $s^\prime(t)$ is found non-stationary, then the time series is different from a red noise. If $s^\prime(t)$ is stationary, the time series is fitted with the AR(1) model 
$\tilde s(t) = \Phi \tilde s(t-1)$. 
The white noise test is performed on the residuals $\epsilon(t)=s^\prime(t)-\tilde s(t)$. Finally, the given time series is a red noise if the residuals behave according to a white noise. 
Being AR(1) a stationary linear process that does not support oscillations, the events detected as non-AR(1) processes are non-stationary and/or oscillatory and/or nonlinear processes.
The method  consists of two steps: 
\begin{itemize}
\item [(i)] 
{\it  First step: self-organized structure detection}\\
Given a sampled signal $s^\prime(t)$ of total length $m$, sub-sequences of given len $l<m$ are considered.
The $q$th sub-sequence $s_q(t)$ is defined as:
\beq
s^\prime_q(t)=[s^\prime(t_q),...,s^\prime(t_{q+l-1})] 
\ 
\label{noise}
\eeq
with $1\leq q\leq m-l+1$. 
Sub-sequences that are significantly different from noise identifies a potential ``local structure''. A structure is defined only if a sequence of consecutive potential local structures is long enough. A real self-organized structure starting at time $t_0$, with time scale $\Delta_t$, is defined when the noise test identifies consecutive potential local structures in the time interval $(t_0 - \Delta_{t_1},t_0 + \Delta_{t_2})$, with $\Delta_{t_{1,2}}\leq \Delta_t$. \\
The basic assumption of this method is that individual structures or trains of structures are separated by noise regions. It follows that the initial and final times of the period where the structure is identified corresponds to the crucial transition events introduced above.
\item [(ii)] 
{\it Second step: structure clusterization}\\
Self-organized structures with similar features are grouped together: each structure is described using a feature vector, then the Euclidean distance among vectors is used to cluster them. 
This allows shapes with similar features but different lengths, or time lags, to be cluster together.  \\
Being $n_e$ the number of structures extracted in the first step, and $d$ the number of features of every group of structures, the second step groups together the $n_e$ structures in $k < n_e$ clusters, in a $d$-dimensional space.
\end{itemize}
The method was tested on a subsequence of the original series  where self-organized structures were removed by phase randomization. The number of detected structures in the surrogate time series was smaller than that in the original one.
The authors use four basic shapes to generate artificial data: box, ramp-cliff, cliff-ramp and a sine function. In atmospheric real data, since some of the chosen features used for clusterization are correlated, a principal
component analysis (PCA) is applied to the feature vector to reduce both correlation and dimension. An application to the method in stable atmospheric boundary is described in \cite{kang_2015}.

\vspace{.25cm}
\noindent
{\bf Velocity increments method}

\vspace{.25cm}
\noindent
The method is based on the following steps \cite{paradisi_jpcs2015,paradisi_epjst09,paradisi_romp12,paradisi_npg12}:
\begin {itemize}
\item [(i)]
Given the non-zero mean signal $s(t)$, the local average ${\overline s}(t) = \langle s \rangle_{_T}(t)$ is computed by applying a moving average over sliding windows 
$[t-T/2,t+T/2]$ or a linear detrending over disjoint time intervals
$[nT,(n+1T]$. In atmospheric turbulence, the duration $T$ is usually between $15$ and $30$ minutes in order to minimize the
non-stationarities modulating the local turbulence and due to
large- and meso-scale meteorological dynamics.
\item [(ii)] 
The corresponding variance $\sigma^2 = \langle (s(t) - {\overline s}(t) \rangle_{_T}$ is computed. This is constant inside the 
time window and changes only passing from one window to the next
one.
\item [(iii)] 
The turbulent fluctuations $s(t)$ are then transformed into a sequence of normalized fluctuations:
\beq
s^\prime (t) = \frac{s(t)-\overline{s}(t)}{\sigma}
\label{norm_fluct}
\
\eeq
\item [(iv)] 
The crucial event is defined as passage of the signal increment through a threshold $D_0$:
\beq
|\Delta s^\prime(t)|=\left| 
s^\prime(t + \Delta t) - s^\prime(t)
\right| > D_0\ .
\label{parad_event}
\eeq
In this formulation, positive and negative increments are not distinguished, but clearly the method can be generalized 
by removing the absolute value, thus introducing two kind of events: $+ \rightarrow -$ and $- \rightarrow +$.
The threshold $D_0$ is chosen according to the $n$-th percentile of the $|\Delta s^\prime|$ distribution. Alternatively, a comparison between different thresholds is carried out.
\end {itemize}
\noindent

\vspace{.25cm}
\noindent
{\bf PDF analysis method}

\vspace{.25cm}
\noindent
In applying this method, events are defined as signal extremes caused by a physical mechanism different from background turbulence and noise. Thus, this method belongs to the class of TCEs.\\
To extract the extreme fluctuations of the signal, a threshold is defined based on the behavior of the PDF tails. Chosen a mathematical distribution as a reference PDF, the position where the PDF of the signal deviates from the reference PDF identifies the threshold value. 

\noindent
The method was proposed by Katul. et al. (1994) \cite{katul_1994} and applied to sensible heat flux measurements $w'T'$. 
The procedure adopted in \cite{katul_1994} is described by the following steps:   
\begin{itemize}
\item [(i)] 
To obtain a time series with zero mean and unit variance,
the measurements were normalized by removing the mean turbulent flux $\la w'T' \ra$ and dividing by the standard deviation $\sigma_{wT}$ of the flux measurements.
\item [(ii)] 
The PDF of the normalized time series was compared to a gaussian PDF with zero mean and unit variance.
\item [(iii)] 
The threshold value $H_c$ was defined as $H_c=\sigma_{wT}(H_n)+ \la w'T' \ra$, where $H_n$ is the most distant point at which the normalized $w'T'$ PDF intersects the reference PDF. 
\end{itemize}
Liu et al., 2014 \cite {liu_2014} used a L\'evy stable distribution \cite{levy_1954} as reference PDF of the vertical velocity in the unstable PBL. This choice is justified by the study of Liu et al., 2011 \cite {liu_2011}, that found that the PDF of vertical velocity in the unstable boundary layer can be fitted by a truncated L\'evy stable distribution. The difference between the truncated stable and the stable PDF is only in the tails of the PDF. The truncated stable and the stable distributions have three common parameters: a characteristic exponent, a skewness parameter and a scale parameter. The two distributions differ for a cut-off parameter and the
truncated stable distribution becomes the stable distribution when the cut-off parameter equals zero.\\
The main steps to find the threshold value are the following: 
\begin{itemize}
\item [(i)] 
A range $G = \left[s_{max} - D/2,s_{max}+D/2\right] $ is defined, where $s_{max}$ is the maximum point defined by:
\beq
f(s_{max}) = {\rm max} f(s)
\label{pdf_range}
\eeq
$f(x)$ is the PDF (original stable truncated PDF or the stable one). 
\item [(ii)] 
In this range, the normalized PDF $f_G$ is defined as 
\beq
f_G(s)={1\over c}f(s+s_{max})\ ;\quad
c=\int_G f(s)ds
\label{pdf_norm}
\eeq
where $c$ is the normalization coefficient.
\item [(iii)] 
The absolute value of relative deviation $RD$ between maxima of the normalized truncated stable and of stable PDF is computed by:
\beq
RD=\frac{ |{\rm max} f_G - {\rm max} \hat f_G| }{{ \rm max } f_G} \cdot 100\%
\label{pdf_dev}
\eeq
where $f_G$ and $\hat f_G$ are normalized truncated stable and stable PDFs.
\item [(iv)] 
The dependence on $D$ of $RD$ has a linear regime and
the threshold is defined as the end point of the linear regime of RD:
\beq
T_\pm=s_{max}\pm D/2\ .
\label{pdf_thr}
\eeq
Then, $s > T_+$ and $s < T_-$ represents extreme fluctuations and the passages through this threshold
are used to select the transition events.
\end{itemize}

\section{Event-based complexity measures: intermittency, clustering and friends}
\label{event_based_complex_sect}

In the following we give a sample of some event-based 
measures that are extensively used, and sometimes also
introduced {\it ex novo}, in the analysis of 
turbulent flow in the PBL \cite{cava_2009,cava_2012,huang_2021,katul_2006_surf_ren,paradisi_jpcs2015,paradisi_epjst09,paradisi_romp12,paradisi_npg12,poggi_2009}.

\vspace{.25cm}
\noindent
{\bf Non-Gaussianity measures and imbalance} 

\vspace{.25cm}
\noindent
In general, the non-Gaussianity of a distribution 
$p(s^\prime)$ of a zero mean signal $s^\prime(t)$ with variance $\sigma^2 = \langle (s')^{^2} \rangle$ is evaluated by means
of normalized moments of order higher than $2$:
\beq
{\rm Skewness:}\ \ Sk = \frac{\langle (s')^{^3}
\rangle}{\sigma^3}\ ;\quad {\rm Kurtosis:}\ \ 
Ku =  \frac{\langle (s')^{^4} \rangle}{\sigma^4}\ .
\label{skew_kurt}
\eeq
The Gaussian value of the kurtosis, $Ku=3$ is a reference
value to define non-Gaussianity. Clearly,  the Gaussian
skewness is zero.\\
The imbalance of $TA_s(t)$ is the same as the original
zero mean signal $s^\prime(t)$ and it can be evaluated by
considering the relative time duration of the two states
$0$ and $1$ \cite{huang_2021}:
\beq
\Delta \Gamma = \Gamma_+ - \Gamma_-\ ;\quad
\Gamma_+ = Pr\left\{ TA_s(t)=1 \right\}      \ ; \quad
\Gamma_- = Pr\left\{ TA_s(t)=0 \right\}
\
\label{imbalance_1}
\eeq
The imbalance of a distribution is related to the skewness
of the distribution by the following formula \cite{huang_2021}:
\beq
\Delta \Gamma = -\frac16 \sqrt{\frac{2}{\pi}}\ Sk\ 
\label{imbalance_2}
\eeq

\vspace{.25cm}
\noindent
{\bf Spectral Exponents $\beta_{TA}$ and $\beta_s$} 

\vspace{.25cm}
\noindent
The PSD of a zero mean signal $s^\prime(t)$ is given by:
$$
E(f) = 2 \lim_{T\rightarrow \infty} \frac1T |\hat s_T (f)|^2\ 
$$
being $s_T$ the signal up to the time $T$, $\hat s(f)$
the Fourier transform of the signal $s^\prime(t)$ and $f$ the
frequency\footnote{
Clearly, experimental time series have a finite duration
time $T$, so that the limit is not
carried out. Further, the data are sampled in time and
the Fourier transform becomes the time-discrete Fourier
transform. The factor $2$ follows from considering
that the Fourier spectrum is symmetric for real-valued
signals.
}.
As known, the interest in the PSD lies in the Parseval's
theorem stating:
$$
\int_0^\infty E(f) df = \int_0^\infty |s^\prime(t) |^2 dt\ ,
$$
where the integral on the right-hand side represents the
total energy of the signal. Thus, the PSD is interpreted as the distribution of the signal energy over the frequencies,
i.e., on the time scale $\tau_f = 1/f$.
The same reasoning can be done for the spatial Fourier
transform, wave numbers and spatial scales.

\vspace{.1cm}
\noindent
In turbulence, PSDs display some typical power-law, such as
the Kolmogorov-Oboukhov $-5/3$ law: $E(k) \sim k^{-5/3}$.
In event-based approaches, the power-law behavior of the
TA-PSD is of particular interest:
\beq
E_{TA}(f)\sim f^{-\beta_{TA}}\ ,
\label{beta_ta}
\eeq
where $\beta_{TA}$ is the spectral exponent of the TA 
dichotomous signal.
This was firstly investigated by Sreenivasan and co-workers
\cite{bershadskii_2004}, who also found
a relationship with the scaling of the original signal:
\beq
E_{s}(f)\sim f^{-\beta_{s}}\ ,
\label{beta_s}
\eeq
in both experimental data and some simple stochastic
models \cite{sreenivasan_2006}. This relationship 
is given by:
\beq
\beta_{TA} = \frac{\beta_s+1}{2}
\label{beta_ta_vs_s}\ 
\eeq
Thus, according to this formula\footnote{
Here we have the $-5/3$ law because the temporal signals represent Eulerian velocities or other turbulent quantities. Then, under proper conditions, the Taylor's frozen hypothesis holds and the frequency spectrum is actually a wave number spectrum.
}, when 
$\beta_s=5/3$,
it should result
$\beta_{TA}=4/3$.

\vspace{.25cm}
\noindent
{\bf Clustering Exponent $\alpha_c$} 

\vspace{.25cm}
\noindent
The clustering exponent is based on the computation of the
{\it event density} $\rho_\tau(t)$. As we deal with experimental
sampled signals with sampling time $\Delta t$, we can
rewrite for simplicity: $\rho_m(n)$ with 
$\tau=\tau_m = m\Delta t$ and $t=t_n = n \tau_m = n m \Delta t$.\\
Let us now consider the sequence of events given in Eq. 
(\ref{event_seq}): ${\cal E} = \left\{ t_k \right\}_{k\in A}$, $A=\left\{ 0, ...,L \right\}$ that
is derived by the application of some event detection
algorithm to the turbulent signal.
Let us define $N = t_L / \Delta t$ as the total number of sampling times in the original time signal.
Then, the algorithm works as follows\footnote{
The clustering exponent seems to have been introduced ot the community of atmospheric turbulence in Ref. \cite{sreenivasan_2006}.
The algorithm here presented is a guess taken from
Refs. \cite{sreenivasan_1997,cava_2009,huang_2021}.
%
}: 
\begin{itemize}
    \item[(i)]
A statistical ensemble is derived by dividing the event sequence $\left\{ t_k \right\}$ into
$M$ subsequences of total duration $\tau_m = m \Delta t$, so that: 
\beq
{\cal E}_n = \left\{k: n\, \tau_m \le t_k < (n+1) \tau_m   \right\}\ ;\quad 0 \le n < M\ ;\quad M = [N/m] \ ,
\label{subseq_ensemble}
\eeq
being $[a]$ the integer part of $a$.
Each subsequence is a sample in the statistical ensemble
and $m \ll N$ is chosen is such a way to get enough statistical samples in the ensemble ($M \gg 1$).
Notice that: ${\cal E}_n \subset {\cal E}$ and, when 
$[N/m] = N/m$, ${\cal E} = \cup_n {\cal E}_n$.
\item[(ii)]
Compute the event density by simply counting the size of 
${\cal E}_n$, i.e., the number of events in the $n$-th time interval, and dividing by the number of sampling time
inside the subsequence, which is given by $m$:
$$
\rho_m(n) = \frac{\#{\cal E}_n}{m}
\ 
$$
\item[(iii)]
Compute the mean and variance of $\rho_m(n)$ by averaging 
over the statistical ensemble:
\begin{eqnarray}
\langle \rho_m \rangle &=& \frac1M \sum_{n=0}^M \rho_m(n) \\
&&\ 
\\
\sigma_\rho^2 (m) &=& \langle \left(\rho_m(n)-\langle \rho_m(n) \rangle \right)^2 \rangle = \frac1M \sum_{n=0}^M \left(\rho_m(n) - \langle \rho_m(n) \rangle \right)^2
\label{event_density_var}
\end{eqnarray}
\end{itemize}
Then, the {\it clustering exponent} $\alpha_c$ is defined by the following power-law scaling:
\beq
\sigma_\rho^2 (m) \sim m^{-2\alpha_c} \quad {\rm or, equivalently,} \quad
\sigma_\rho^2 (\tau_m) \sim \tau_m^{-2\alpha_c}
\ 
\label{cluster_exp}
\eeq
In general, higher clustering corresponds to lower $\alpha_c$ and {\it vice versa}.
Regarding TA and, thus, a sequence of ZCEs, a white noise
has clearly no clustering, because two successive jumps
are completely independent from each other.
The clustering exponent of the white noise is given by
$\alpha_c = 0.5$, which is a reference value to which 
compare the experimental $\alpha_c$ value.

\vspace{.25cm}
\noindent
{\bf Oboukhov intermittency Exponents $\gamma_q$} 

\vspace{.25cm}
\noindent

This is probably the first measure of intermittency in
turbulence and it was firstly proposed by Oboukhov 
in 1962 \cite{oboukhov_1962}. Turbulent intermittency found by Batchelor and Townsend \cite{batchelor_1949} referred to the dissipation rate $\epsilon \sim |du/dx|^2$, thus
it is possible to introduce the analogous temporal
dissipation rate of a generic zero mean turbulent signal $s^\prime(t)$:
\beq
\epsilon(t) = \left| \frac{ds^\prime(t)}{dt}  \right|^2\ .
\label{temp_diss_rate}
\eeq
In the case of Eulerian measures, this can be related to the real dissipation rate when Taylor's frozen 
turbulence hypothesis can be assumed.
In analogy with the Oboukhov's spatial local average,
Eq. (\ref{k62_local_diss}), a time average can be
defined as:
\beq
\epsilon_\tau = \frac{1}{\tau} \int_t^{t+\tau} \epsilon(t^\prime)\ dt^\prime
\ 
\label{time_average_diss}
\eeq
The Oboukhov intermittency exponents are then defined
by the following relationships:
\beq
\frac{\langle \epsilon_\tau^{^q}  \rangle}{\langle \epsilon_\tau 
\rangle^{^q}   }  \sim  \tau^{-\gamma_q}
\ 
\label{oboukhov_intermit_exp}
\eeq

\vspace{.1cm}
\noindent
In general, if $\gamma_q = 0\ \forall q$, there is no
Oboukhov intermittency, i.e., intermittency related to the signal dissipation rate.
The Oboukhov intermittency was applied to both the original signal
and the associated sequence of crucial events, in
particular to the TA signal.
Given a sequence of events $\left\{ t_k \right\}$, TCEs, RTEs o TA, a sequence of pulses can be build exploiting the Dirac $\delta$ functions:
$$
e(t) = \sum_{k \in A} \delta(t-t_k)\ .
$$
Then, when Oboukhov intermittency is applied to a sequence of pulses, it results:
\begin{eqnarray}
&&\gamma_q \ne 0 \Rightarrow {\rm clusterization\ of\ pulses}\\
&&\gamma_q = 0 \Rightarrow {\rm no\ clusterization} 
\end{eqnarray}
Then, also the Oboukhov intermittency exponent applied to
a sequence of pulses is an estimate of 
clusterization related to crucial events.
Of particular interest are the $2$nd-order exponents of
both original and TA signals: $\gamma_{2,s}$ and 
$\gamma_{2,TA}$. To simplify the notation, these $2$nd-order exponents will be denoted as $\gamma_{s}$ and $\gamma_{TA}$.
The inter-comparison of these two exponents gives 
an estimate of the relative complexity of the
clustering features encoded in the event sequence,
often given by the TA signal, with respect
to the amplitude variability.

\vspace{.1cm}
\noindent
It is worth noting that higher intermittency is given by a higher Oboukhov intermittency exponent $\gamma_s$ and the same applies when considering the  Oboukhov exponent $\gamma_{TA}$. Conversely, higher clustering is associated with a lower clustering exponent $\alpha_c$.

\vspace{.25cm}
\noindent
{\bf TC/IDC exponent $\mu$}

\vspace{.25cm}
\noindent
We here refer to the already introduced temporal or intermittency-driven complexity (TC or IDC), following
the convention introduced by Grigolini and co-workers \cite{grigolini_pre15_temp_complex,grigolini_csf15_bio_temp_complex,paradisi_springer2017,turalska-grigolini_pre11_critical}\footnote{
This is named as persistence exponent and indicated with the $\gamma$ greek letter in some of the cited papers devoted to intermittency in turbulent flows
\cite{cava_2009,cava_2012,cava_2019,huang_2021}.
}.

\vspace{.25cm}
\noindent
Given the usual event sequence ${\cal E} =\left\{ t_k
\right\}$ and the associated IETs 
$\left\{ \tau_k \right\}$, the TC or IDC exponent is defined by the
power-law decay in the IET-PDF:
\beq
\psi(\tau) \sim \tau^{-\mu}
\ 
\label{tc-idc_exp}
\eeq
%
Notice that $t_0$ has to be the occurrence time of the first detected event and not the start time of the signal\footnote{
By convention, a time shift is usually applied to get
$t_0=0$.
}.
We recall that a power-law decay in the IET-PDF with $1 < \mu \le 3$ is the signature of metastable self-organized structures
and define an intermittent complex system in the sense explained above in Section \ref{intermit_complex_sect}.\\
Interestingly, in the turbulent trasport of scalars, the power-law scaling in the IET-PDF, Eq. (\ref{tc-idc_exp}), also denoted as inter-pulse period PDF, is claimed to be a signature of ``active turbulence''  (e.g., temperature in unstable convective turbulence), while a log-normal PDF is associated with ``passive'' scalars \cite{cava_2009,sreenivasan_2006}:
$$
\psi(\tau) \sim \e^{-|a|\theta^2+b\theta+c}\ ; \quad \theta = log(\tau)\ ,
$$
being $a$, $b$ and $c$ proper constants depending
on mean and variance of $\theta$.

\vspace{.25cm}
\noindent
{\bf Diffusion exponents $H$ and $\delta_*$}
\label{diff_exp}

\vspace{.25cm}
\noindent
A widely used approach to evaluate self-similarity exploits the properties of diffusion laws.
As diffusion is originated by the sum of many 
random contributions, this method has a robust theoretical foundation in the limit theorems of probability theory, i.e., the Gaussian central limit
theorem and the L\'evy generalized limit theorem
\cite{gnedenko-kolmogorov_1968,levy_1954}.

\vspace{.1cm}
\noindent
In general, the diffusion method works as follows:
\begin{itemize}
\item[(i)]
Let us consider an experimental time series $\xi(t)$,
which can be a turbulent velocity fluctuation or
other zero mean turbulent signal.
\item[(ii)]
The diffusion variable is computed as:
\beq
X(t) = \int_0^t \xi(t^\prime) dt^\prime
\Rightarrow X_N = \sum_{n=0}^N \xi_n\ ,
\label{diff_variable}
\eeq
being $\xi_n = \xi(n \Delta t)$, $X_N = X(N \Delta t)$ and $\Delta t$ the 
experimental sampling time.
\item[(iii)]
Different statistical indices can be computed on the
diffusion variabile $X(t)$. The averages are carried
out as time averages and, thus, similarly to clustering
and Oboukhov exponents, sliding windows are used to
define a set of sub-trajectories.\\
In particular, two
complexity exponents related to diffusion
are of wide interest:

\vspace{.25cm}
\noindent
\begin{itemize}
\item[1.]
{\bf Second moment exponent H
(Detrended Fluctuation Analysis) }\\
In analogy with Eq. (\ref{subseq_ensemble}), let us
consider signals sampled at $t_k = k \Delta t$ and a
window lenght $\tau_m = m \Delta t$.
Thus, we have the following sub-trajectories:
\begin{eqnarray}
&&\widetilde{X}^n_{k} = \widetilde{X}^n_{k^\prime} = X(k^\prime \Delta t)\ ; \quad
k^\prime = k-mn\ , \quad k \in {\cal Q}_n (\tau_m)\ ;\quad 
\label{subtraj_ensemble}  \\
&&\ \nonumber \\
&&{\cal Q}_n (\tau_m) = \left\{ k:  n \tau_m \le t_k < (n+1) \tau_m   \right\}\ ,\quad 0 \le n \le M\ ,\quad M = [N/m]\ ,
\nonumber
\end{eqnarray}

being $N$ the total number of sampling times in the data. When not ambiguous, in the following we will use the simplified notation ${\cal Q}_n$ in place of ${\cal Q}_n(\tau_m)$.\\
Notice that it always results: $0 \le k^\prime < m$
and that the set
$$
\left\{ \widetilde{X}^n_{k};\ k^\prime = 0,...,m \right\} =  \left\{ \widetilde{X}^n_{k^\prime};\ k \in {\cal Q}_n \right\}
$$
is a statistical ensemble
of stochastic trajectories that are assumed to be
approximately independent and over which statistical
features, and the associated scaling
exponents, can be computed. 

\vspace{.15cm}
\noindent
The second moment scaling $H$ is computed by applying the Detrended Fluctuation Analysis (DFA) \cite{peng_n92,peng_pre94}
(see also \cite{paradisi_npg12} for details).
Given the time lag $\tau_m$, the DFA firstly compute the variance over each single trajectory in the above statistical ensemble (average over $k^\prime$ or, equivalently, over $k$). At this stage, an important step is the estimation of the trend inside each time window $[n \tau_m, (n+1) \tau_m)$, usually by means of linear
or quadratic regression methods.
After that, an average over 
the ensemble is carried out 
(average over $n$). In formulas:
\beq
\sigma^2_X(\tau_m) = \langle \langle  
X^n_k - {\overline X}^{n}_k 
\rangle_{_{{\cal Q}_n}} \rangle_{_{\cal Q}} 
\label{dfa}
\eeq
The averages $\langle \rangle_{_{{\cal Q}_n}}$
and $\langle \rangle_{_{\cal Q}}$ are carried out over the
single trajectory and over the entire ensemble, respectively. The mean trend
${\overline X}^{n}_k = \langle \widetilde{X} \rangle^{\tau_m,n}_k$ is
evaluated over the $n$th trajectory (${\cal Q}_n$) for the window length $\tau_m$.\\
This calculation is carried out for several $\tau_m$
and, for self-similar processes, we expect to get:
\beq
F(\tau_m) \sim \tau_m^H\ ;\quad F(\tau_m) = \sqrt{\sigma^2_X(\tau_m)}\ .
\label{dfa_hurst_exp}
\eeq
The exponent $H$ is essentially the Hurst exponent of
rescaling analysis \cite{hurst_1951}. The case of normal (Gaussian) diffusion is given by $H=1/2$ and is a consequence of the central limit theorem. \\
Then, the range $H>1/2$ is denoted as {\it fast diffusion} or {\it super-diffusion}, while $H<1/2$ is named {\it slow diffusion} or {\it sub-diffusion}.
\vspace{.25cm}
\item[2.]
{\bf PDF self-similarity exponent (Diffusion Entropy analysis) }\\
For a given time lag $\tau_m = m \Delta t$, let us 

consider a statistical sample with the trajectory
increments:
\begin{eqnarray}
&&\Delta X^m (k) = X_{k+m} - X_k\ = X(t_k+\tau_m) - X(t_k) ; \quad 
k \in {\cal P}_m
\nonumber
\\
&&\  \label{increments_ensemble} \\
&&{\cal P}_m = \left\{ k: 0\le k \le N-m \right\}\ ,
\nonumber
\end{eqnarray}
being, as said above, $N$ the total length of the sampled diffusion variable $X_n = X(n \Delta t)$.
Then, the PDF of this statistical sample of increments
can be evaluated for several $\tau_m$: $p(x,\tau_m)$.
In the self-similar case, we have:
\beq
p(x,\tau_m) = \frac{1}{\tau_m^{\delta_*}} f\left( \frac{x}{\tau_m^{\delta_*}} \right)
\ .
\label{pdf_selfsimilar}
\eeq
The variable $z = x / \tau_m^{\delta_*}\delta$ is named 
``self-similarity variable'' and it comes from
a self-similarity transformation (see, e.g., \cite{paradisi_caim15_ctrw_fract}) that leaves the 
probability of interval unchanged.
When Eq. (\ref{pdf_selfsimilar}) comes into action,
the diffusion is said to be self-similar and, in 
particular, monoscaling.
An efficient way to detect the scaling
exponent $\delta$ was found to be the Diffusion 
Entropy (DE) analysis developed and widely applied by Grigolini and co-workers \cite{allegrini_2003_dea,grigolini_2001_dea,grigolini_2002_dea,scafetta_2002_dea}.
The main idea of DE is to evaluate the Shannon entropy
of the diffusion process $X(t)$:
\beq
S(t) = -\int p(x,\tau_m) \log (p(x,\tau_m)) dx\ ,
\label{diff_entr_1}
\eeq
which, under the self-similarity condition given in Eq. (\ref{pdf_selfsimilar}), has the following behavior:
\beq
S(t) = A + \delta_* \log t\ ;\quad A = -\int f(y) \log f(y) dy
\ .
\label{diff_entr_2}
\eeq
\end{itemize}
\end{itemize}

\section{Some relations between complexity exponents: SOC and EDDiS method}
\label{soc_eddis_sect}

The relationship between the spectral exponents of
the original signal and of the TA signal, which is given by the sequence of zero-crossing events, is
already given in Eq. (\ref{beta_ta_vs_s}). 
As we will see in the next Section \ref{intermit_pbl_sect}, this relation is far from being verified in atmospheric turbulence. Actually, Sreenivasan and Bershadskii (2006) \cite{sreenivasan_2006} derived this formula in a
heuristic way and verified it on some turbulence datasets, but they did not claim a wide validity of
the formula itself. In this sense, the studies of Katul
and co-workers was and is still relevant in understanding the limits of Eq. (\ref{beta_ta_vs_s})
and of its variants.
%

\vspace{.25cm}
\noindent
{\bf Self-Organized Criticality} 

\vspace{.25cm}
\noindent
An interesting aspect that has recently begun to attract the interest of some authors is the 
relationship between Self-Organized Criticality 
(SOC) \cite{bak_1987,bak_1988,jensen_1998} and 
turbulence intermittency in the sense of Oboukhov.
A classical SOC formula relates the spectral exponents
of the TA signal $\beta_{TA}$ with the TC exponent $\mu$, related to IET-PDF.
For ``nonintermittent'' cases (in the Oboukhov sense), the relation between the exponents 
$\beta_{TA}$ and $\mu$ is given
by \cite{jensen_1998}:
\beq
\beta_{TA} = 3 - \mu
\ 
\label{beta_ta_vs_mu_soc}
\eeq
Discrepancy from this relation, since there are no amplitudes involved, should be entirely related to clustering, 
that is considered to be a part of intermittency by many authors and that we would rather say that it is an aspect of intermittency directly linked to crucial events rather than to large variability in the 
signal increments\footnote{
This seems to be related to the previouly introduced
magnitude intermittency (MI) \cite{liu_2020}, that some authors also denote
as amplitude variability \cite{cava_2009,huang_2021}, while the aspects related to TA and crucial events in
general are associated with clustering intermittency (CI).
}.
Bershadskii et al. (2004) \cite{bershadskii_2004}
found that the above relationship between the TA spectral and TC exponents must be modified to
take into account the Oboukhov intermittency related
to the signal dissipation rate:
\beq
\beta_{TA} = \left( 3 - \frac{\gamma_{TA}}{2} \right) - \mu
\ 
\label{beta_ta_vs_mu_oboukhov}
\eeq
where, as said in Section \ref{event_based_complex_sect}, $\gamma_{TA}$ is 
the TA-based Oboukhov intermittency $2$nd-order
exponent and it is related to clustering features.
We will discuss the application of this relationship to atmospheric turbulence in Section \ref{intermit_pbl_sect}.

\vspace{.25cm}
\noindent
{\bf The EDDiS method} 

\vspace{.25cm}
\noindent
The Event-Driven Diffusion Scaling (EDDiS) approach is based
on two main ingredients:
\begin{itemize}
\item[(i)]
the sequence of crucial events $\left\{ t_k \right\}$;
\item[(ii)]
the CTRW model.
\end{itemize}
According to the view of temporal complexity \cite{grigolini_pre15_temp_complex,grigolini_csf15_bio_temp_complex,mahmoodi-grigolini_2018_temp_complex,paradisi_csf15_preface,paradisi_springer2017,west-grigolini_pr2008_fract_events}, a working hypothesis of EDDiS is given by the renewal condition for the crucial events.
The renewal aging analysis gives some effort in evaluating the renewal
condition and, conversely, an estimate of the memory content of a event sequence referred to the presence of random or deterministic modulation. However, being based on the brand-new and aged IET-PDFs, this analysis can be affected by noise, so that,
when renewal condition is not verified, this could be due to both
modulation or noise \cite{allegrini_pre10}. Conversely, some
theoretical studies on artificial data showed that, even without
noise, the amount of memory, i.e., the departure from the renewal
condition, can be linked to different underlying dynamics. In other
words, renewal aging does not always unambiguously identify the correct dynamical model.

\begin{figure}
\begin{tabular}{c}
\includegraphics[height=.35\textwidth,width=.5\textwidth,angle=0.]{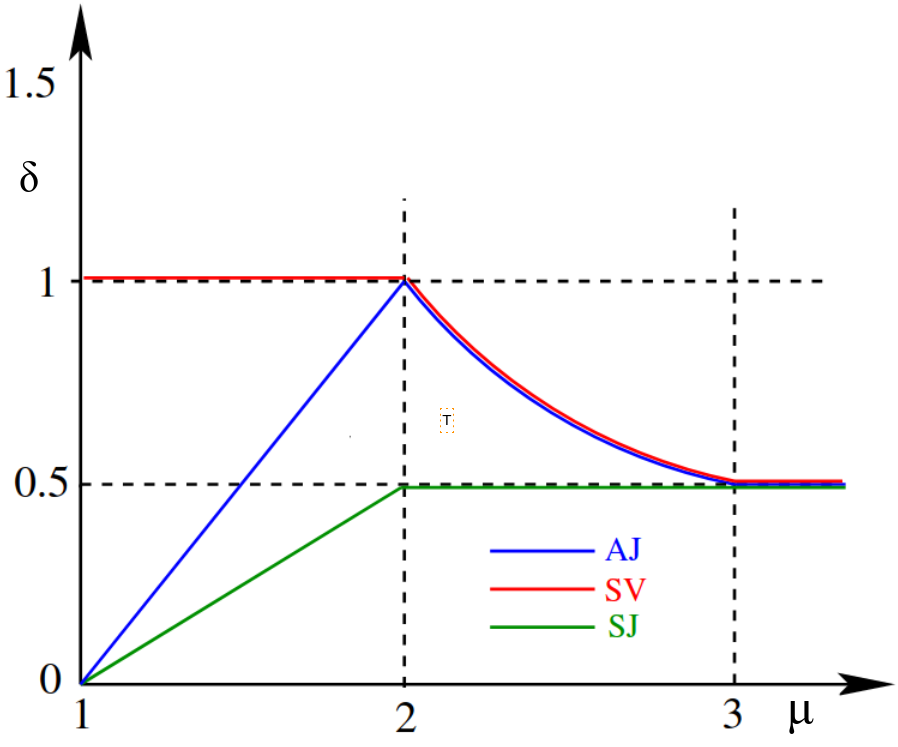} 
\\
(a)
\end{tabular}
\begin{tabular}{c}
\includegraphics[height=.35\textwidth,width=.5\textwidth,angle=0.]{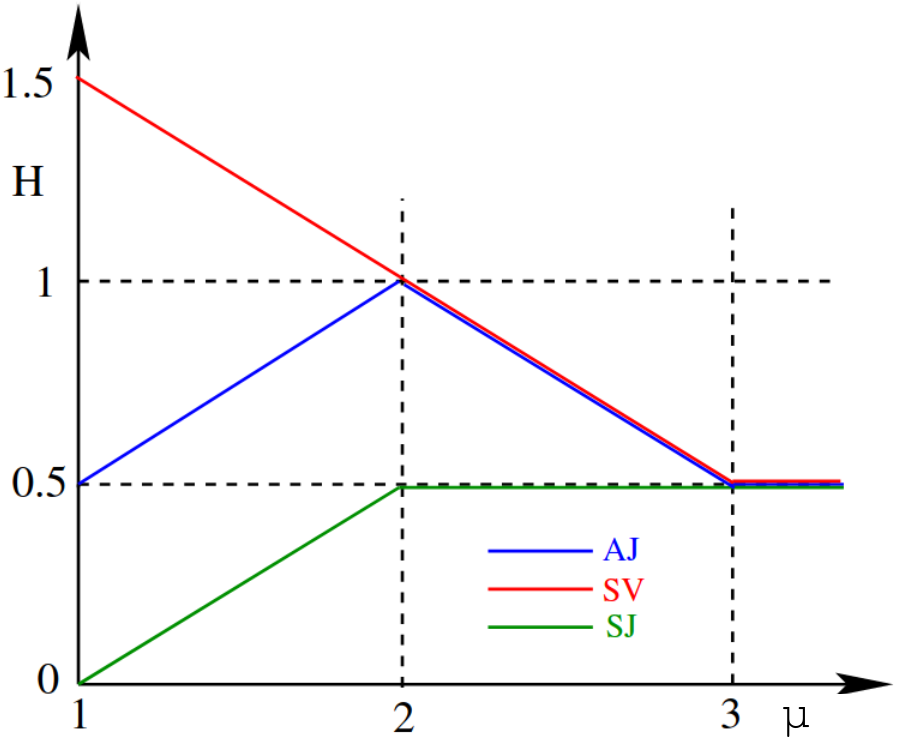}
\\
(b) 
\end{tabular}
\caption{Scaling $\delta_*$ (panel (a)) and $H$ (panel (b)) vs. complexity index $\mu$ for the three
walking rules: AJ (intermediate blue line), SJ (bottom green line) and SV (top blue line).}
\label{renewal_diff_scaling}
\end{figure}

\vspace{.1cm}
\noindent
In order to overcome this limitation and to get a robust estimation
of the TC/IDC exponent $\mu$, Allegrini et al. (2009) \cite{allegrini_2009} applied a tool, later named EDDiS by 
Paradisi and Allegrini (2015,2017) \cite{paradisi_csf15_pandora,paradisi_springer2017},
that puts together a set of
analyses based on three different CTRWs, on DFA and on DE, combining
and comparing different results already known in literature about
the relatioship between $\mu$, Hurst exponent $H$ and PDF self-similarity $\delta_*$
(for a review, see \cite{allegrini_2009,paradisi_springer2017,weiss_1983_ctrw_review}
and references therein).
The main aspect is that the above results are derived under the renewal condition. Thus, the working hypothesis of the EDDiS method
is that the event sequence is a renewal process.\\
Thus, the most general version of the EDDiS algorithm works as follows (for a review, 
see \cite{allegrini_2009,paradisi_csf15_pandora,paradisi_romp12,paradisi_npg12}):
\begin{itemize}
\item[(i)] 
Let us consider the event sequence $\left\{ t_k \right\}_{k \in A}$
of Eq. (\ref{event_seq}), with $A = \left\{0,1,2, ...,L \right\}$, and the following three walking rules:
\begin{itemize}
\item[(a)] 
{\it Asymmetric Jump (AJ) rule}
The walker makes a jump ahead in correspondence of each event occurrence time $t_k$ (sequence of pulses):
\beq
\left\{
\begin{array}{l}
\xi(t)=0 \quad {\rm if}\ \  t \ne t_k\ \ \forall k 
\\
\ \\
\xi(t_k) = 1 
\end{array}
\right.
\label{aj_rule}
\eeq
\item[(b)]
{\it Symmetric Jump (SJ) rule} \\
Given the random dichotomous variable $u\in\left\{-1 , +1 \right\}$
(coin tossing prescription) and a sample of $u$, $\left\{u_k \right\}_{k \in A}$, that is, a sequence of $+1$ and $-1$, the SJ
rule is:
\beq
\left\{
\begin{array}{l}
\xi(t)=0 \quad {\rm if}\ \  t = \ne t_k\ \ \forall k 
\\
\ \\
\xi(t_k)=u_k
\end{array}
\right.
\label{sj_rule}
\eeq
The walker makes a unitary jump in a random direction in correspondence of an event: $\xi(t_k)=\pm 1$.
\item[(c)]
{\it Symmetric Velocity (SV) rule} \\
The walker moves with constant unitary velocity $V = 1$
and has the possibility of changing directions in correspondence
of a crucial event. Taking the same sequence $\left\{u_k\right\}$
as in the SJ rule: 
\beq
\xi(t_k)=u_k\ ;\quad  t_k \le t < t_{k+1}\ ;\ \  k \in A\setminus \left\{L\right\}
\label{sv_rule}
\eeq
This is actually the walking rule of the so-called {\it L\'evy walk}
\cite{shlesinger_1987_ctrw_turb,zaburdaev-klafter_2015_levywalk}.
\end{itemize}
\item[(ii)] 
From the three walking rules three different CTRWs are built by
simply applying Eq. (\ref{diff_variable}), thus getting
$X_{_{AJ}}(t)$, $X_{_{SJ}}(t)$ and $X_{_{SV}}(t)$.
\item[(iii)]
The second moment and PDF self-similarity exponents $H$ and $\delta_*$ are evaluated by applying DFA and DE to the three
random walks. Under the working hypothesis of TC/IDC,
 i.e., a renewal process without any noise and with power-law
 decay in the IET-PDF: $\psi(\tau)\sim\ 1/\tau^\mu$, the following relationships were found in the literature
 \cite{grigolini_2001_dea,klafter_1987_ctrw_anom_diff,montroll_1964_1,montroll_1965_2,scafetta_2002_dea,shlesinger_1974_ctrw_asympt,tunaley_jsp1974_ctrw_asympt,tunaley_1975_ctrw_asympt} (for a review see also \cite{metzler-klatfter_pr2000_ctrw_fract_review,paradisi_springer2017,weiss_1983_ctrw_review,zaburdaev-klafter_2015_levywalk}):
\begin{itemize}
\item[(AJ)]
\beq
\delta_{_{AJ}}(\mu)=\left\{
\begin{array}{ll}
\mu-1\ ; &\quad 1 < \mu < 2 
\\
1/(\mu-1)\ ; &\quad 2 \le \mu < 3 \\
1/2\ ; &\quad \mu \ge 3
\end{array}
\right.
 \quad
H_{_{AJ}}(\mu)=\left\{
\begin{array}{ll}
\mu/2\ ; &\quad 1 < \mu < 2 
\\
2-\mu/2\ ; &\quad 2 \le \mu < 3 \\
1/2\ ; &\quad \mu \ge 3
\end{array}
\right.
\label{delta_h_aj}
\eeq
\item[(SJ)]
\beq
\delta_{_{SJ}}(\mu)=H_{_{SJ}}(\mu)=\left\{
\begin{array}{ll}
\vspace{.1cm}
\left(\mu-1\right)/2\ ; &\quad 1 < \mu < 2 \\
1/2\ ; &\quad \mu \ge 2
\end{array}
\right.
\label{delta_h_sj}
\eeq
\item[(SV)]
\beq
\delta_{_{SV}}(\mu)=\left\{
\begin{array}{ll}
1 \ ; &\quad 1 < \mu < 2 
\\
1/(\mu - 1)\ ; &\quad 2 \le \mu < 3 \\
1/2\ ; &\quad \mu \ge 3
\end{array}
\right.
 \quad
H_{_{SV}}(\mu)=\left\{
\begin{array}{ll}
2-\mu/2\ ; &\quad 1 < \mu < 3 \\
1/2\ ; &\quad \mu \ge 3
\end{array}
\right.
\label{delta_h_sv}
\eeq

\vspace{.3cm}
These renewal-based relationships 
between diffusion exponents and the TC exponent are summarized
in Fig. \ref{renewal_diff_scaling}.
\end{itemize}
\end{itemize}
%
%
The three rules give a normal scaling $\delta_*=H=1/2$ for $\mu\ge3$, thus giving normal (Gaussian) diffusion, in agreement
with the L\'evy generalized limit theorem \cite{levy_1954}.
%
According to Eq. (\ref{dfa_hurst_exp}), AJ and SV rules are super-diffusive ($H>1/2$) for $\mu<3$, while SJ rule is subdiffusive for $\mu<2$. All walking rules give a normal
scaling $H=1/2$ for $\mu>3$\footnote{
For a Poisson process, i.e., an exponential IET-PDF $\psi(\tau) \sim \exp(-\tau/{\overline \tau})$, 
the asymptotic diffusion for $t \gg {\overline \tau}$ is normal: $\delta_* = H = 1/2$. This is not surprising as the Poisson process corresponds to the limit $\mu\rightarrow\infty$ of
the IET-PDF given in Eq. (\ref{powerlaw_pdf_surv}).
}.
Interestingly, for $1<\mu<2$ the AJ rule, even if super-diffusive,
displays a scaling $\delta_*$ less than  $1/2$ for $\mu<3/2$.

\vspace{.1cm}
\noindent
Regarding both theoretical findings and experimental applications of the EDDiS method, some observations are in order:
\begin{itemize}
\item[(1)]
Check of renewal condition:\\
By inverting the above relationships, it is possible to have
more indirect estimations of $\mu$. Being the AJ the most
robust rule, we can invert $H_{_{AJ}}$ and $\delta_{_{AJ}}$ to get \cite{allegrini_2009,paradisi_romp12}:
\beq
\delta_{_{AJ}} =
\left\{
\begin{array}{ll}
     &  2 H_{_{AJ}} -1\ ;\quad      1 < \mu < 2  \\
     &  1/(3-2 H_{_{AJ}})\ ;\quad   2 \le \mu < 3  \\
     &  H_{_{AJ}} = 1/2   \ ;\quad  \mu \ge 3  \\
\end{array}
\right.
\label{eddis_check_renewal}
\eeq
When this condition is fulfilled within a given statistical significance,
the renewal hypothesis can be accepted.
\vspace{.2cm}
\item[(2)]
AJ rule can split the contribution of noisy events, acting in the short-time regime, from that of genuine complex events, i.e., 
displaying fractal intermittency\footnote{
We recall that, in the TC/IDC framework, fractal intermittency
is associated with both inverse power-law IET-PDF and renewal condition.
} 
and triggering fast diffusion in the long-time limit.
This was seen in many experimental datasets \cite{allegrini_2009,allegrini_pre10,allegrini_fp2010,allegrini_csf13,allegrini_pre15} and theoretical analyses \cite{allegrini_pre10,paradisi_csf15_pandora}.
\vspace{.1cm}
\item[(3)]
Robust estimation of $\mu$:\\
A direct estimation of $\mu$ can be carried out by
a best fit on the IET-PDF. However, even in the presence of
a small, but not negliglble, percentage of noisy events, the
slope of the IET-PDF can be drastically blurred and it can even 
display an apparent power-law that is different by the underlying true one \cite{allegrini_pre10}.
This problem is cured by the diffusion scaling.
In fact, when renewal condition is verified, the value of $\mu$ can be
estimated by the three walking rules. For $\mu > 2$, SJ rule gives
normal diffusion $H=\delta_*=1/2$ and only two rules are available
and four diffusion exponent, two for each rule.
\vspace{.2cm}
\item[(4)]
The SV rule gives a counter-intuitive normal diffusion at long
times when complex and noisy events are superposed. 
Allegrini et al. (2010) \cite{allegrini_pre10} and Paradisi and Allegrini (2015) \cite{paradisi_csf15_pandora} proposed
a modeling approach and a data analysis method, later named Poisson Added Noise DiffusiOn Rescaling Analysis (PANDoRA) \cite{paradisi_springer2017}, that are based on the superposition of genuine complex events (fractal intermittency) and Poisson events and that gives an explanation of the above behavior.
\vspace{.2cm}
\item[(5)]
The EDDiS method does not necessarily need the estimation of
$\mu$. In fact, in most datasets the SJ rule gives a normal
diffusion, a signature of $\mu>2$\footnote{
Paradisi et al. (2012) \cite{paradisi_npg12} confirmed this 
finding in a turbulence dataset by also applying a fractal dimension analysis by box counting method.
}.
Further, the best fit on the SV rule has often a low statistical
significance due to the presence of noise.
Conversely, the AJ rule is usually reliable and, as said above, robust under the presence of noisy events.
Thus, the $H$ or $\delta_*$ exponents can be used also
beyond the validation of the renewal hypothesis and, thus,
in a somewhat extended framework with respect to the more
restricted assumptions of TC/IDC framework\footnote{
In our opinion, in complex self-organized systems the presence of
renewal crucial events is probably ubiquitous, even if there are often some difficulties in validating the renewal assumption and
in getting a robust estimation of $\mu$. 
However, the investigation of system's complexity by the 
diffusion exponent is still possible, especially by means of
the AJ rule and DFA exponent $H$, due to the above cited ability of separating the contributions of noisy and complex events.
}. 
For example, 
in Allegrini et al. (2010) \cite{allegrini_fp2010} only the
DE analysis was used, while other works of Allegrini, Paradisi and
co-workers \cite{allegrini_csf13,allegrini_pre15,paradisi_npg12,paradisi_aipcp13,paradisi_jpcs2015} reported only on the DFA results\footnote{
By definition, the statistical ensemble of DFA is often much
larger than that of DE analysis, even considering superposed
realizations in the DE ensemble, that can also introduce some
spurious effects in the averaging procedure. Conversely, the
DFA, even in the multiscaling DFA version, is referred to the moments of the distribution and not to some information measure
such as entropy,which could include some effect not retained
in the moment analysis.
}.
\end{itemize}

\section{Intermittency, clustering and complexity in the PBL}
\label{intermit_pbl_sect}

Intermittency models of 
ideal homogeneous and isotropic turbulence, while 
remaining a benchmark, appear to be not applicable in a straighforward way to atmospheric turbulence. This is
the case not only for the classical models, K41 monoscaling or K62 log-normal multiscaling models
\cite{kolmogorov-1962,oboukhov_1962}, but also for more recent findings about intermittency and multiscaling in
turbulence modeling \cite{she-leveque_1994,she_2009}.
For this reason, in the last two decades or so, a wide 
literature was devoted to the complexity 
analysis of high time-resolution wind data, 
%
the goal being
the investigation of a general formulation for turbulent flows over complex terrain in terms of a few complexity measures, related to both MI and CI \cite{huang_2021,liu_2020}.

\noindent
In fact, complexity features become particularly
relevant, as they are related to turbulent
structures and events triggered by a particular mixture
of internal, external and global intermittency that
simultaneously affect clustering and rapid large excursions \cite{ansorge_2014,ansorge_2016b,mahrt_1999,mahrt_2014,sun_2012}.
The studies in PBL turbulence over complex terrain from the viewpoint of event-based complexity investigated many different exponents and the general picture is still not clear.
In the following we will discuss some of the open issues. 

\vspace{.25cm}
{\bf At the beginning of event-based complexity in PBL turbulence}

\vspace{.25cm}
\noindent
\vspace{.1cm}
\noindent
Milestone papers in this research field are the
studies of Bershadskii et al. (2004) and Sreenivasan and Bershadskii (2006) \cite{bershadskii_2004,bershadskii_2004b,sreenivasan_2006}, where 
fundamental results were found in experimental data
both from laboratory experiments and
field experimental campaign in the atmospheric surface layer\footnote{
In \cite{sreenivasan_2006} the authors use the 
atmospheric surface layer data given in Ref. 
\cite{sreenivasan_1998} and the wind tunnel data from
\cite{pearson_2002}.
In Ref. \cite{bershadskii_2004} laboratory data of thermal convection taken from \cite{niemela_2003} were analyzed. Notice that collecting the
data inside the surface layer (about $35$ m over the ground) means that the observed 
velocity fluctuations around the mean logarithmic profile should display a behavior quite near to the ideal homogeneous and isotropic energy cascade  \cite{kolmogorov-1941,oboukhov-1941,richardson_1920}.
}.
To our knowledge, these authors firstly applied the TA
to turbulence data. In particular, they evaluated
the TA spectral exponent jointly with the classical
one and proposed the relation (\ref{beta_ta_vs_s}) \cite{sreenivasan_2006} between
spectral exponents, whose  validity was found to be
true for the analyzed datasets. 
Further, Bershadskii et al. (2004) \cite{bershadskii_2004} investigated both clustering
and TC-IDC and Oboukhov intermittency exponents $\mu$ 
and $\gamma_{TA}$ derived from the TA signal. They
firstly proposed the formula given in Eq. (\ref{beta_ta_vs_mu_oboukhov}), extending the SOC
relationship, Eq. (\ref{beta_ta_vs_mu_soc}), to 
include the effect of Oboukhov intermittency associated
with the signal dissipation rate.
Sreenivasan et al. (2006) \cite{sreenivasan_2006} 
proposed and investigated by means of both numerical
models and experimental data, the relationship given
in Eq. (\ref{beta_ta_vs_s}) between the spectral exponents $\beta_ {TA}$ and $\beta_s$. They also
applied the clustering exponent $\alpha_c$ to 
different experimental datasets.\\
The main experimental findings of the above studies were the following:
\begin{itemize}
\item[-] 
For temperature in laboratory turbulent convection, an experimental value $\beta_{TA} \simeq 1.38$ was found to be not compatible with the SOC prediction:\\ 
$\beta_{TA}(SOC) = 3 - \mu_{\rm exp} \simeq 1.63$, \\
being
$ \mu_{\rm exp} \simeq 1.37$ \cite{bershadskii_2004}. The reduction given by the difference $\beta_{TA}(exp) - \beta_{TA}(SOC) = -0.25$
was claimed to be associated with (Oboukhov) intermittency. 
\item[-] 
The solution to the previous result was given by the same authors \cite{bershadskii_2004}, who 
derived, and validated on experimental data, the formula
given in Eq. (\ref{beta_ta_vs_mu_oboukhov}), which
generalizes the SOC relationship (\ref{beta_ta_vs_mu_soc}).
In particular, they found the experimental value
$\gamma_{TA} = 0.47 \pm 0.03$, which is compatible
with the difference between the experimental exponent $\beta_{TA}$ and that predicted by SOC.
\item[-] 
Sreenivasan et al. (2006) \cite{sreenivasan_2006} derived an asymptotic expansion 
of $\alpha_c$ for $Re_T \rightarrow \infty$, being 
$Re_T = \lambda_T U /\nu$ the Taylor microscale Reynolds
number\footnote{
The Taylor microscale $\lambda_T$ is the smallest size of the large eddies, i.e., the size at which viscous dissipation is no more negligible. We recall that the Kolmogorov microscale $\lambda_K$ is the scale at which the TKE is entirely dissipated into heat due to viscosity. Clearly, $\lambda_K \ll \lambda_T \ll L$.
}, 
and found an asymptotic minimum value:
$$
\alpha_c \rightarrow 0.1 \ \ {\rm for}\ \ Re_T \rightarrow \infty
$$
\item[-]
Conversely, for $200 < Re_T < 20000$, $\alpha_c$ range is from $0.25$ to $0.4$ for scales between the dissipative and inertial ranges, while $\alpha_c \simeq 0.5$ (white-noise) for scales larger than the integral scale $L$ of energy-containing eddies. Then, there 
is a pronounced clustering at short times, but no clustering at long times.
%
\end{itemize}

\vspace{.25cm}
{\bf Is there universal complexity in the PBL ?}

\vspace{.25cm}
\noindent
\vspace{.1cm}
\noindent
Following the approach firstly introduced in Refs. \cite{bershadskii_2004b,bershadskii_2004,sreenivasan_2006}, several authors carried out similar TA-based
analyses on turbulence fluctuations, e.g, streamwise and spanwise velocity, temperature, over complex terrain such as vegetated canopies \cite{cava_2009,cava_2012,cava_2008b,huang_2021,liu_2020}.
It is worth noting that the common jargon is that of
``separating'',  ``distinguish'' or ``isolating'' the
role of CI from the MI and that MI is found to 
``mitigate'' or ``amplify'' intermittency\footnote{
MI is sometimes named ``amplitude variability'' or ``amplitude intermittency''. This concept seems to be
associated with the Oboukhov intermittency of the original signal and, thus, related to the ``dissipation rate'' of the signal itself, actually to the derivative of the energy signal \cite{cava_2009,huang_2021}.
}, 
referring to the 
comparison between $\gamma_{TA}$ and 
$\gamma_s$ (see above, Section \ref{event_based_complex_sect}).
Other similar sentences are of the kind ``amplitude intermittency [here MI] might mitigate clustering'' \cite{bershadskii_2004}, which seems to refer to the clustering exponent $\alpha_c$, but it
could also refer generically to intermittency.
In particular, some authors claim that, when 
$\gamma_{TA} \simeq \gamma_s$, the ``observed 
(Oboukhov) intermittency may be due to clusterization
not amplitude variations'' \cite{cava_2009}\footnote{
In our opinion, this kind of terminology could sound
slightly ambiguous. In fact, the Oboukhov intermittency
exponent, and also the other exponents, give some
indications of how (how much) the analyzed signal is self-organizing (self-organized) over different temporal or spatial scales and if there is some kind
of self-similarity, whatever monoscaling or multiscaling. All these exponents refer to the single
analyzed signal, thus comparing event-based complexity measures with those of the original signal is surely interesting, but it should be treated carefully
and conclusions derived from these comparisons will surely deserve further attention. 
}.
\\
The overall picture is that of the emergence of a
event-driven dynamical process where crucial events play a crucial role. However, the underlying mechanisms of event triggering and associated birth/decay of self-organized turbulent structures is still unclear.

\vspace{.1cm}
\noindent
The above cited studies about PBL turbulence over complex terrain apply the TA technique firstly 
proposed in Bershadskii et al. (2004) \cite{bershadskii_2004}
in order to compare some complexity exponents of the original signals with those of the TA.  
In particular,
they compare spectral exponents ($\beta_{s}$ and $\beta_{TA}$) and Oboukhov intermittency exponents ($\gamma_{s}$ and $\gamma_{TA}$) for both signal and its Telegraph Approximation (TA), 
and, by definition only for TA, the clustering exponent $\alpha_c$ and the temporal complexity, or
persistence, exponent $\mu$. 

\vspace{.1cm}
\noindent
Over flat terrain quite far from the ground and above vegetated canopies, the evaluation of the spectral exponents $\beta_{TA}$ and $\beta_s$ was found to be in quite good agreement with Eq. (\ref{beta_ta_vs_s}). In fact, the 
analyses estimate the small-scale usual values $\beta_s = 5/3$ and $\beta_{TA} = 4/3$, which are also typical of the already discussed internal intermittency. 
However, some authors \cite{cava_2009} found a net failure of the relationship 
(\ref{beta_ta_vs_s}) inside the canopy. Statistically 
significant linear regressions with coefficient 
different from $1/2$ were found, but different datasets gave different linear regression coefficients.
Further, the spectral relationship (\ref{beta_ta_vs_s}) was found to be no more
valid in the region very near to the ground ($< 1$m)
\cite{huang_2021} and in the quiescent layer developed in a nocturnal high stable boundary layer \cite{cava_2019}.

%

\vspace{.1cm}
\noindent
Regarding clustering behaviour isolated from amplitude variations, this has rarely been considered in PBL turbulence except in a few studies
\cite{cava_2009,cava_2012,cava_2019,chowdhuri_pf2020,liu_2020,poggi_2009}.
Interestingly, different values of the clustering exponent $\alpha_c$ were found at small and large
scales, thus giving a double-regime for clustering. In particular, we have:
\begin{itemize}
\item[-]
For horizontal velocity, there is essentially no significant clustering at large scales, i.e., $\alpha_c \simeq 0.5$ (compatible with white noise).
\item[-] However, at large scale $\alpha_c \approx 1$ at different height, after the development of a low-level jet (LLJ) in a nocturnal stable boundary layer.
\item[-]
Clustering is higher (lower $\alpha_c$) at small scales than large scales: $\alpha_c \sim 0.24-0.30$. 
This is in agreement with laboratory experiments
at $Re_T\sim 200-20000$\footnote{ 
We recall that $Re_T$ is the Reynolds number based
on the Taylor microscale, i.e., the shortest size of 
the large scale eddies.
}, where it was found:
$\alpha_c \sim 0.25-0.40$ \cite{sreenivasan_2006}.
\item[-]
$\alpha_c$ is less scattered at small than large scales
(signature of a quasi-universal behavior at small scales)
\item[-]
Inside the canopy, the temperature clustering $\alpha_c$ was the most vertically heterogeneous.
\item[-]
At the canopy bottom, for temperature it results
$\alpha_c \simeq 0.1$ in the stable case, which
corresponds to the minimum limit value found in Ref.
\cite{sreenivasan_2006}, but for velocity\footnote{
In fact, for temperature, the limit value was found to
be $\alpha_c \simeq 0.07$.
}.
\end{itemize}

\vspace{.1cm}
\noindent
Regarding the Oboukhov intermittency exponent, a 
comparison between $\gamma_s$ of the original signal and $\gamma_{TA}$ of its TA counterpart is carried out
in recent papers \cite{cava_2009,cava_2012,cava_2019,huang_2021,li_2013}.
Streamwise/spanwise velocity fluctuations and
temperature fluctuations were the main investigated
signals.
In summary, a universal relationship
between the clustering exponent $\alpha_c$ and the Oboukhov intermittency exponents $\gamma_s$ and $\gamma_{TA}$ was not found within vegetated canopies \cite{cava_2009,cava_2012} and in a high stable boundary layer \cite{cava_2019}. \\

Over flat terrain, but very near to the ground ($< 1$m),
Huang et al. (2021) \cite{huang_2021} investigated
the difference between small and large scale, finding that:
\begin{itemize}
\item[-]
At small scale, $\gamma_{TA}/\gamma_s$ is homogeneous across heights and stability conditions both for T and u.
\item[-]
Both velocity and temperature Oboukhov intermittency
exponents are more sensitive at large than small scales when considering the height of about $1$ m. 
\end{itemize}

\vspace{.1cm}
\noindent
Regarding the TC exponent $\mu$, Cava and Katul (2009)
\cite{cava_2009} found the following results:
\begin{itemize}
\item[-]
Above the canopy, the IET-PDFs are well approximated
by a log-normal distribution for all turbulent signals
and all stability conditions. This is in agreement with
Ref. \cite{sreenivasan_2006}.
%
\item[-]
Very near to the ground there are no significant best 
fit neither power-law nor log-normal.
\item [-] 
In the stable case, the IET-PDF is well described by a power-law distribution mainly in the short-time regime. A quite good linear regression is found between measured $\beta_{TA}$ and calculated $\beta_{TA}$ by Eqs. \ref{beta_ta_vs_mu_soc} and \ref{beta_ta_vs_mu_oboukhov}. The best fits show similar linear coefficient and correlation coefficient.
\end{itemize}

\vspace{.1cm}
\noindent
In summary, a universal behavior of TA-based
complexity exponents was not found and a general parameterization, e.g., dependence on
site geometrical parameters, has not yet been derived.
Some kind of universal behavior was found in the region
just above the canopy, where exponents, and relations
between exponents, similar to those of Refs.
\cite{bershadskii_2004,sreenivasan_1997}, were found.
On the contrary, no clear and universal scaling and multiscaling seems to emerge in the PBL both inside vegetated canopies and over flat terrain at a distance less than one meter from the ground, thus pointing out a
possibile, not negligible, dependence on the particular site.
For example, inside vegetated canopies, the presence of energy short-circuiting and wake production in the crown and in the trunk space affects the PSD and a clear power-law decaying slope is not always evident
\cite{cava_2009,cava_2012}.
When a power-law is seen, the PSD has a lower decay with respect to the theoretical prediction of Sreenivasan and Bershadskii (2006) \cite{sreenivasan_2006}.

\vspace{.25cm}
{\bf Neutral, unstable, stable}

\vspace{.25cm}
\noindent
Different stability conditions are expected to display
different complexity exponents, as they are associated
with different kinds of self-organizing mechanisms of
the turbulent flow and, thus, to the different
topologies of the associated turbulent structures.
As an example, both laboratory experiments and field campaigns found that hairpin vortices dominate the vertical fluxes in the neutral PBL, while thermal plumes play the major role in unstable conditions \cite{li_2011}.

\noindent
The investigation of the stable and very stable boundary layers is becoming a particular interesting challenge. 
The stable and strongly stable conditions is the one that, more than the others, presents a intricate superposition or alternance of different types of intermittency, 
which can also act at similar time and space scales \cite{mahrt_1999,sun_2012}.
As mentioned previously, a feature of the stable boundary layer is the emergence of turbulence patches at different temporal and spatial scales (external or global intermittency), which then play a crucial role in the dynamics of self-organized turbulent structures \cite{ansorge_2014,mahrt_2014}. In addition, turbulence variability can modify the shape of various sized eddies, or their small subregions (internal, or small-scale, intermittency)\footnote{
We recall that internal intermittency is associated with
the inertial subrange and, by definition, expected to 
consistent with the Kolmogorov-Oboukhov similarity modeling 
approach, modified to include intermittency effects, such as in the She-Leveque model \cite{she-leveque_1994}.
}.
Although the nature of the observed intermittent turbulence is still not well understood, it seems triggered by the various non-turbulent submeso motions (whose characteristic scale range between the largest turbulence eddy scale
($\sim \rm{O(100 m)}$) and the smallest meso-gamma scale ($\sim \rm{2 km}$)) 
\cite{mahrt_2009,mahrt_2012}. Submeso motions include a multitude of processes, such as meandering and gravity waves, and can have different shapes, such as  steps, ramps, pulses, waves.
Conversely, in neutral and unstable conditions, the self-organized structures come from the sharp edge instabilities of the main energy-containing eddies.
In neutral conditions, the main eddies are triggered by the mean wind at the top of the PBL, which is driven by the mesoscale pressure gradients.
In unstable conditions, the large-scale energy-containing eddies are given by thermal plumes energized by the buoyancy, i.e., the overall temperature
gradient between the top of the boundary layer and the ground.

\vspace{.1cm}
\noindent
Katul and co-workers mainly investigated the dependence of complexity exponent on stability conditions, again finding a number of empirical observations that are not really conclusive and difficult to put in a theoretical framework \cite{cava_2009,cava_2012,huang_2021}. Cava et al. (2019) \cite{cava_2019} analyzed the evolution of a nocturnal boundary layer from a weakly stable to a very stable boundary layer, highlighting the role of submeso motions. Three phases were considered: a first weakly stratified state at the beginning of the period, characterized by Richardson number $Ri\sim 0.1$, a transition period characterized by the development of a low-level jet (LLJ) where the PBL is highly stable ($Ri>>0.25$), a third period after the complete development of the LLJ, where the SBL present vertical layer, that is:
(i) a shallow layer close to the ground with fully-developed turbulence ($Ri\sim0.1$);
(ii) a quiescent layer under the LLJ nose with weak or intermittent turbulence ($0.2<Ri<0.5$), dominated by submeso motion; 
(iii) a turbulent layer above. 
The analysis was performed in the inertial subrange (ISR) and in the low-frequency range (LFR).
In general:
\begin{itemize}
\item[-]
Temperature appears more intermittent with respect to streamwise velocity in the stable case, 
thus suggesting that small (inertial) scales could be affected by large-scale structures.
\item[-]
Atmospheric stability has minor effects on clustering above the canopy.
\item[-]
Over flat terrain, $< 1$ m from the ground, there are
only slight differences among velocity and temperature
and for different stability conditions.
\item[-] 
In the nocturnal stable boundary layer, large scales were
found to be more heterogeneous than small scales 
along the height.
\end{itemize}

\vspace{.1cm}
\noindent
Regarding clustering features, different values of $\alpha_c$ were found at small and large scales. 
In particular, we have:
\begin{itemize}
\item[-]
Large scales of temperature are more dependent on stability.
whereas velocity is less heterogeneous with respect to stability condition.
\item[-]
Inside the canopy both velocity components and scalars have higher clustering for strongly stable atmosphere.
Typically, we have: $\alpha_c \sim 0.2 - 0.3$ for all variables. 
\item[-]
Clustering increases ($\alpha_c$ decreases) with increasing stability for all turbulent signals. 
\item[-]
Temperature displays higher (lower) clustering than velocity for stable (unstable) conditions.
\item[-]
At small scale, after the development of the LLJ,
the clustering exponent $\alpha_c$ did not change in the layer closer to the ground and in the upper layer. A decrease was observed in the quiescent layer, where was $\alpha_c<0.2$ especially for velocity components. At low frequency range (large scale) $\alpha_c$ dropped for all the variables at all levels, assuming value $\alpha_c\approx 0.1$. We recall that the role of submeso motions is strictly related to external and global intermittency, which play a major role in triggering 
crucial events when PBL is stable
\cite{anfossi_2005,sun_2015}.
\end{itemize}

\vspace{.1cm}
\noindent
Regarding the Oboukhov intermittency exponents of temperature and of related TA, the following results
were found in Refs. \cite{cava_2009,cava_2012,cava_2019,huang_2021}:
\begin{itemize}
\item[-]
For temperature inside the canopy and in unstable/convective atmosphere:
$\gamma_{TA}/\gamma_{s} < 1$, a result somewhat
opposite to that found in Ref. \cite{bershadskii_2004}.

\item[-]
Always for temperature inside the canopy, but for near-neutral and stable conditions:
$\gamma_{TA}/\gamma_{s} \simeq 1$.
\item[-]
For horizontal velocity 
over flat terrain ($< 1$m), 
$\gamma_{TA}/\gamma_{s} > 1$ across all stability regimes.
\end{itemize}

\vspace{.1cm}
\noindent
Regarding the TC exponent $\mu$, a double-regime
often emerges in the IET-PDF with different shapes 
at small (inertial) and large scales. The shapes, and also the emergence of double-regime, depends on the stability condition, on the height and on the experimental
site. It results that the IET-PDFs are not
always  well fitted by a power-law decay and, further,
there is no general consensus about the temporal
complexity exponent $\mu$ in the PBL.
In particular, we have \cite{cava_2009,cava_2012,cava_2019,chamecki_2013,chowdhuri_pf2020,huang_2021}:
\begin{itemize}
\item[-]
Within the crown region of vegetated canopies, the emerging double-regime was more
evident for velocity and in unstable conditions:
a inverse power-law decay in the short IET range and a
log-normal distribution in the large IET range.
\item[-] 
In the trunk region, no double-regime was seen and the log-normal distribution gave good fittings for both
temperature and velocity and for all stability 
conditions.
\item[-]
There was no clear fitting shape of the IET-PDF
for temperature in strongly stable conditions inside the canopy and near the ground.
\item[-]
Analyzing temperature in a dataset from the surface layer of an unstable boundary layer, Chowduri et al. (2020) \cite{chowdhuri_jfm2020} found
power-law decay at small scales and a strecthed exponential cut-off at large scales.
\item[-]
In the nocturnal boundary layer \cite{cava_2019} the IET-PDF is well described by a power-law distribution mainly in the short-time range.
\end{itemize}

\vspace{.25cm}
{\bf The challenge of dissimilarity}

\vspace{.25cm}
\noindent
The transport properties of velocity and scalars, including
heat transport related to temperature, are found to display substantial different behaviors depending on the stability condition of the PBL. In particular, scalar transport increases,
while momentum transport decreases, in passing from neutral to
unstable conditions.
This was found to be strictly related to the 
topological change of large-scale turbulent (coherent) structures, 
that are substantially modified by buoyancy \cite{li_2011}.
In fact, as said above, the neutral condition is associated with hairpin vortices and hairpin packets dominating the vertical transport fluxes, while the unstable condition display the typical convective thermal plumes that extend vertically over all the PBL \cite{watanabe_2019,khanna_1998}. 
In general, scalar intermittency is usually found to be higher than velocity intermittency \cite{chambers_1984,katul_1997,sreenivasan_1997,warhaft_2000}.
Being transition events related to the birth-death of self-organized turbulent structures, it is clear that these different topologies of turbulent structures among velocity and scalar transport are reflected in a dissimilarity of their event-based complexity features.

\vspace{.1cm}
\noindent
Regarding clustering, temperature is often found to display more clustering at large scales with respect to the velocity counterpart \cite{cava_2009,huang_2021}.
In fact, the asymptotic regime $\alpha=0.5$ (no clustering) is reached at much larger scales for temperature. This behavior becomes more evident going deeper into the canopy.
This is related to the ramp-cliff patterns of temperature, which are known to increase intermittency and, inside canopies, are usually associated with sources and sinks (e.g., heat sources/sinks for temperature), which play a major role in the deepest layer of the canopy
\cite{cava_2004,cava_2008,cava_2008b,cava_2009,cava_2012}.
In fact, these ramp-like temporal patterns are associated with metastable self-organized structures emerging from the direct dynamical interaction, at relatively large scales, between the degrees of freedom of the fluid flow and the surrounding boundaries. 
The boundaries affect the flow through both the geometrical constrains and the source/sink distribution, these last one especially affecting scalar dynamics and, thus, diffusion/transport properties of the turbulent flow, a regime known as near-field approximation \cite{cava_2009}.
For this reason, in the large scale regime, clustering and intermittency of temperature and other scalars such as humidity, are much more enhanced with respect to the velocity field inside the canopy. Due to ramp-like patterns, this enhancement is even more evident in stable conditions. \\
%
Regarding the clustering exponent at small scales,
we have \cite{cava_2009,cava_2012,cava_2019,huang_2021}:
\begin{itemize}
\item[-]
Inside the canopy and for all stability conditions, at small scales clustering was found to be higher (lower $\alpha_c$) for temperature than for velocity. This finding is
more evident for stable condition and is in
agreement with Ref. \cite{sreenivasan_2006}.
A similar, but weaker, effect was also found over flat terrain very near to the ground \cite{huang_2021}.
\item[-] 
Temperature assumes values of $\alpha_c$ much lower than
$0.5$ (no clustering) for a more extended range of
small scales with respect to velocity, till the value $\alpha_c \simeq 0.1$ for stable conditions and in the lowest levels of the canopy. 
\item[-] 
Temperature clustering has the highest dependence 
on the height from the ground.
\item[-]
Similarly to large scales, clustering is higher (lower
$\alpha_c$) for temperature than for velocity for all stability conditions and for all heights.
\end{itemize}
%
%
The limit value $\alpha_c = 0.1$ is in agreement with the theoretical prediction of Sreenivasan
and Bershadskii (2006) \cite{sreenivasan_2006}
for fully developed turbulence far from boundaries ($Re_T \rightarrow \infty$), even if the above prediction regards the clustering of velocity turbulent fluctuations. On the contrary,  a maximum value $\alpha=0.07$ for temperature clustering was predicted.

\vspace{.1cm}
\noindent
Regarding the dissimilarity of Oboukhov intermittency, the following results were found:
\begin{itemize}
\item[-]
Inside the canopy, $\gamma_{TA}$ for both streamwise
and spanwise velocities was found to be higher than for scalars, e.g., temperature, for all stability conditions.
\item[-]
Inside the canopy, the values of $\gamma_{TA}$ for temperature were found to decrease with height, while the dependence of velocity on height is weak.
\item[-]
For both velocity components inside canopy,
$\gamma_{TA}/\gamma_s > 1$. 
\item[-]
For temperature $\gamma_{TA}/\gamma_s \sim 1$ at small scale at ground ($<1$ m) and inside the canopy in strongly stable conditions. 

\item[-]
For streawise velocity at small scales and near the ground ($< 1$m), 
it results\footnote{
Similarly to previous papers, also these authors claim
that, for velocity fluctuations, ``amplitude variations
lower intermittency'', i.e., MI mitigate Oboukhov
intermittency of the original signal. Conversely,
for temperature, MI does not have a role in the
Oboukhov intermittency, which is then claimed to
be associated with clustering.
} 
$\gamma_{TA}/\gamma_s \sim 1.5$.
\item[-]
In the stable boundary layer, before the development of a LLJ: $\gamma_{TA}/\gamma_s>1$ for $u,w$ and
$\gamma_{TA}/\gamma_s \approx 1$ for $T$ at small scales;
$\gamma_{TA}/\gamma_s>1$ for $u$ and $\gamma_{TA}/\gamma_s \approx 1$ for $w$ and $T$ at large scales.
\item[-]
In the nocturnal stable boundary layer, after the development of the LLJ, $\gamma_{TA}/\gamma_s$ increases for all signals at large scales and at all vertical levels.
\item[-]
Temperature at large scales has a stronger dependence on atmospheric stability with respect to velocity.
\end{itemize}

\vspace{.25cm}
{\bf To be SOC or not to be SOC ?}

\vspace{.25cm}
\noindent
The link with SOC \cite{bak_1987,bak_1988,jensen_1998,munoz_2018}
and turbulence, which 
is discussed since Ref. \cite{bershadskii_2004}, 
is an interesting perspective in the studies
about intermittency in atmospheric turbulence \cite{cava_2009,cava_2012,cava_2019,huang_2021}.
The interesting aspect to be studied is given by
the range of validity of Eq. (\ref{beta_ta_vs_mu_oboukhov}) and
by the existence of possible variations to this formula.

Bershadskii et al. (2004) \cite{bershadskii_2004}
verified this relationship in both laboratory experiments and field measurements, but it was more
recently found that it is not always valid
in turbulence inside a canopy sublayer or very near to the ground \cite{cava_2009,cava_2012,cava_2019,huang_2021}.
This seems to occur, in particular, under stable conditions,
where the temperature fluctuations $T^\prime$ does not
obey Eq. (\ref{beta_ta_vs_mu_oboukhov}).
Interestingly, Huang et al. (2021) \cite{huang_2021} found
a statistically significant statistical regression with
coefficients that are not the same as in Eq. (\ref{beta_ta_vs_mu_oboukhov}), also finding different
behaviors at large and small (inertial) scale regimes.
However, these same authors claim that even the
improvement of the relation 
(\ref{beta_ta_vs_mu_oboukhov})
is not appropriate to describe the temperature in the
stable case\footnote{
We recall that the relationship (\ref{beta_ta_vs_mu_oboukhov}) is already a 
modification of the classical SOC relation given by Eq. (\ref{beta_ta_vs_mu_soc}) to include Oboukhov intermittency.
}.\\
Cava et al. (2019) \cite{cava_2019} estimated
the linear regression
between measured $\beta_{TA}$ and calculated $\beta_{TA}$, using both Eqs. \ref{beta_ta_vs_mu_soc} and \ref{beta_ta_vs_mu_oboukhov}.
They found higher correlation coefficients after the development of the LLJ, but with similar coefficients and statistical significance for 
both comparisons.
The authors deduce that clustering and external intermittency have a role greater than the internal intermittency, and that the SOC relation (Eq. \ref{beta_ta_vs_mu_soc}) is a reasonable model.\\
Interestingly, stable PBL turbulence is 
the most intricate case. There is a lack in understanding what kind of self-organized structures emerge,
if they are waving motion, turbulence, laminar or at the transition between laminar and turbulent conditions.
Further, there could be some ambiguities arising in the 
dynamical origin of self-organized structures revealed by
event detection methods, in particular
the telegraph approximation (TA). In fact,
the crucial events could be associated to internal, external or global intermittency.
This is in agreement with a claim by Huang et al. (2021)
\cite{huang_2021}: ``the modified SOC relation ...
 displays signatures of excess intermittency in [temperature] beyond the [Oboukhov] intermittency exponent'' ($\gamma_{TA}$ in our notation).
%

\vspace{.25cm}
{\bf Renewal theory and diffusion scaling in atmospheric turbulence}

\vspace{.25cm}
\noindent
Paradisi and co-workers investigated the existence of crucial
events in turbulent flows near the ground over flat terrain by
applying the velocity increment method
\cite{paradisi_epjst09,paradisi_npg12,paradisi_romp12,paradisi_jpcs2015} and the EDDiS approach.
The main results of this analysis are the following:
\begin{itemize}
\item[(i)]
The Renewal Aging analysis applied to both vertical velocity
and PM2.5 aerosol concentration signals gave a slighlty 
aging reduction in the Experimental Aging with respect to the Renewal Aging for aging times of $2000$ sec., thus proving the presence of a small memory in the time series, which could be
interpreted as a residual effect of slow modulation 
\cite{allegrini_2006_model,allegrini_csf07,bianco_cpl07,paradisi_ijbc08,paradisi_cejp09}
from large scale meteorological persistent motions, being most of the effect encoded in the
local mean and variance used to normalize the signal in Eq. (\ref{norm_fluct}).
The TC exponent seems slightly smaller for PM2.5 than for vertical velocity. This could
be related to the presence of local source and/or to internal
degrees of freedom, e.g., nucleation and coagulation processes.
However, statistically significant results about the TC exponents 
were not found and this low robustness could be related to the use of the IET distributions, which were more recently found to be
highly affected by noisy events, also including false positives
related to the threshold choice. 
\item[(ii)]
The EDDiS method was applied to the three velocity components, limiting the analysis to the AJ and SJ rules  and the DFA exponent $H$ \cite{paradisi_npg12}.
The detected values of $H$ were found to be very robust with
respect to changes in the threshold $D_0$.
For streamwise and spanwise velocity fluctuations $u/v$: $H \simeq 0.93-0.96$ and for vertical velocity fluctuations $w$: $H \simeq 0.95-0.98$.
\item[(iii)]
The renewal assumption was further verified on the vertical component by applying Eq. (\ref{eddis_check_renewal}) \cite{paradisi_romp12}, i.e., by looking at the combination of $H$ and $\delta_*$ estimated from the AJ rule\footnote{
The renewal assumption is also in agreement with so-called
{\it surface renewal} models proposed by Higbie and, later,by Perlmutter \cite{higbie_1935_surf_ren,perlmutter_1961_surf_ren}
(see also \cite{luo_2021_surf_ren,ruckenstein_1963_surf_ren}) and
exploited in the atmospheric turbulence framework by Katul et al. (2006) \cite{katul_2006_surf_ren} as a possible model for 
external intermittency.
}.
Thus, on the basis of the renewal assumption, the estimation of 
the TC/IDC exponents are given by $\mu=2.08-2.14$ for 
horizontal velocity fluctuations $u/v$ and $\mu=2.04-2.1$ for
vertital velocity fluctuations \cite{paradisi_npg12}. This also
follows from the constrain $\mu\ge 2$ given by the finding
$H_{_{AJ}}=\delta_{_{AJ}} = 0.5$. This constrain was also
confirmed in Paradisi et al. (2012) by means of a fractal analysis of the event sequence \cite{paradisi_npg12}.
\item[(iv)]
Even if not explicitly stated, the previous analyses were carried 
out in neutral conditions. 
A comparison between neutral and stable conditions was carried
out on the three velocity components and showed a
decrease in the temporal complexity, i.e., an increase in the 
TC/IDC exponent going from the neutral ($\mu = 2.02-2.06$)  to the stable condition ($\mu=2.24-2.28$) \cite{paradisi_jpcs2015}.
%
%
\end{itemize}
The comparison between this EDDiS approach and the 
velocity increment events and other studied based on 
other complexity measures and on the TA signal is not clear
and should need further investigation.

\vspace{.25cm}
{\bf Wind gusts as crucial events}

\vspace{.25cm}
\noindent
Cheng et al. (2014) \cite{cheng_2014} applied the approach
proposed in Paradisi et al. (2012) \cite{paradisi_npg12} to
wind gusts.
According to these authors, wind gusts can be interpreted as
self-organized flow structures given by strong winds
after a cold front with downward vertical velocity associated
with a peak in the horizontal velocity and {\it vice versa}
\cite{cheng_2011,cheng_2012}.
%
Their findings about the exponent $H$ seem not definitive and
have a large variability depending on height: $H\sim 0.54-0.8$
for horizontal velocity and $H\sim 0.67-0.87$.
Interestingly, the short-term noise emerged at time scales up to
$5-10$ sec., whereas in Paradisi et al. (2012) \cite{paradisi_npg12} noise appeared up to about $1$ sec.

\vspace{.2cm}
\noindent
Babi\'c et al. (2016)\cite{babic_2016} 
use the VITA method to detect coherent structures during strong wind events (bora wind occurring in a topographically complex site near the eastern coast of the Adriatic Sea) and relate them to the features of measured heat and momentum flux. In particular, they found that for wind speed greater than $12 m/s$ observed during nearly neutral conditions,
the vertical momentum flux at a height of $22m$ was greater than that at the adjacent levels. The authors conclude that the sweep part of the sweep-ejection cycle could potentially cause the observed momentum flux divergence.

\section{Concluding remarks}
\label{concl_sect}

The goal of this review was to put a first bridge between turbulence studies exploiting event-based complexity approaches, such as the clustering and Oboukhov intermittency exponents,
and studies about intermittent complex systems, where concepts and ideas developed in the fields of non-equilibrium statistical physics, probability theory and dynamical system theory are jointly exploited.
Fluid turbulence is not only one of the most important unsolved problems of classical mechanics and a challenge for mathematical physicists, but, outside topics involving biological systems and living matter, it is also a basic example of a non-living self-organized complex system displaying non-trivial intermittency at several time and space scales.

\vspace{.1cm}
\noindent
These studies can have also some impact on applications, in 
particular on the modeling of turbulent transport.
For example, spatial and temporal clustering is directly related to an increase in the interaction among different “fluid elements or particles”. This enhancement is crucial in the transport properties of the turbulent flow and, thus, in the (advective) diffusion and transport of several quantities: humidity and water droplets, temperature, inert gases; chemicals and pollen plumes; aerosol or inertial particles, e.g., sand storms, being “inertial” here referred to particles that, at variance with passive scalars, have a finite mass\footnote{
 We recall that a passive scalar is a particle that is ideally “docked” to a fluid particle, thus following exactly its same motion dynamics in the turbulent flow. Inertial, or aerosol, particles have a finite mass that determine, at each time
 instant, a small, but crucial, separation from the fluid particle trajectory due to inertia. Thus, inertial 
 particles move across different fluid particles.
}.
As an example, it is known that there is a strict interplay between clustering features of turbulence at small scales and the inertia 
of particles that affect the tendency to  agglomerate of inertial/aerosol particles themselves.
Further, intense events associated with ejections, sweeps or other self-organized turbulent structures, determine very rapid movements of self-organized portion of fluid with high vorticity from one region to another one. This clearly affects the transport properties of the turbulent flow.
%

\vspace{.15cm}
\noindent
As far as we understand, the application of the event-based complexity paradigm to turbulence is still in its early stages, and
future investigations should probably seek to contextualize the event-based approaches to turbulence in the framework
of intermittent complex systems and look at them from the 
``metastable self-organization'' viewpoint (see Section \ref{intermit_complex_sect}).

\vspace{.1cm}
\noindent
An interesting research topic is the complexity of turbulent
flows over complex terrain or very near to the ground, that was briefly discussed in Section \ref{intermit_pbl_sect}.
The temporal complexity features were
investigated in different studies and a high variability in the
complexity exponent was found, in particular about the shape and the double-regime of the IET-PDF.
In Ref. \cite{cava_2009} the authors derived well fitted
shapes that, depending on height and stability condition, could display double-regime (power-law at
small and log-normal at large scales) or single-regime
(log-normal), while in Ref. \cite{cava_2012} they
found a power-law decay with exponential cut-off for
unstable conditions. The authors of Ref. \cite{chamecki_2013} found 
a result similar to that of Ref. \cite{cava_2012}, but on cornfield canopies instead of tree canopies. More
recently, in Ref. \cite{chowdhuri_pf2020} a net power-law decay with a stretched exponential cut-off
was found in the unstable surface layer. 
At variance with these studies, where the TA signal was used to
investigate the event-based complexity features,
Paradisi et al. \cite{paradisi_npg12,paradisi_romp12,paradisi_jpcs2015} proposed an approach based on renewal theory \cite{cox_1970_renewal}.
In this approach, by one hand, the detected RTEs could be compatible with surface renewal models \cite{higbie_1935_surf_ren,perlmutter_1961_surf_ren,katul_2006_surf_ren} 
and, on  the other hand, on EDDiS method, i.e.,
a method that can treat noisy events in a proper way and give a robust estimation of both diffusion exponents and TC/IDC
exponents, often referred to as {\it persistence} in atmospheric turbulence studies.
This allowed to find some interesting results but the link to
TA-based complexity analyses need to be further investigated.

\vspace{.1cm}
\noindent
In summary, there is the lack of a general theoretical framework giving a correct interpretation of the experimental analyses, which does not give universal behaviors in the complexity  exponents. 
In particular, parameterizations, i.e., dependence on site geometry, synoptic and mesoscale meteorological variables and atmospheric stability, are still not understood.\\
Among the open questions, we recall the ones that, in our opinion, are most interesting:
\begin{itemize}
\item[-]
The interplay between crucial transition events and the concepts of internal, external and global intermittency in atmospheric 
turbulence is still not clear. This is a crucial open question
in the case of strongly stable PBL.
Thus, have the RTEs the same geometrical features at the small inertial scales of internal intermittency and at the large scale of global intermittency? In other words, is there a unified 
picture for self-organized turbulent structures?
Until now, the answer seems to be negative, but this needs a deeper understanding.
\item[-]
It is still not clear if the lack of a net scaling/multiscaling could be also related to 
the superposition/alternance of different mechanisms working at the same scales (internal, external, global).
\end{itemize}
%
%
Thus, by one hand, it should be necessary to understand the reliability of event detection algorithms by a comparison between different methods. The correct relationship with the
underlining self-organized metastable structures should also deserve some more attention.
On the other hand, it would be suitable to investigate the
inter-relationship between the different event-based complexity exponents by means of methodological/theoretical studies involving stochastic models mimicking the dynamics
of turbulent structures or through numerical simulations of fluid dynamics basic equations.

\subsection{A personal memento by Paolo Paradisi}

I went to know Paolo Grigolini in Pisa during the summer of 2002, during
my first year as a researcher in the Lecce branch of the Institute of Atmospheric Sciences and Climate, National Research Council (Consiglio Nazionale delle Ricerche, CNR), where I used to work on stochastic models for turbulent diffusion of pollutants in the atmosphere.\\
I met him to discuss about a possible scientific collaboration and I remember a walk from the Department of Physics in Largo Bruno Pontecorvo to the Pisa Research Area in Giuseppe Moruzzi street, where the CNR Research Area of Pisa is still located today. I was going to give a
seminar on anomalous diffusion, Continuous Time Random Walks and fractional diffusion equations.\\
During the walk we were immersed in this beautiful sunny atmosphere alongside the panorama given to us by the aqueduct, known as ``Acquedotto Mediceo dei Condotti'', which runs along the path, and Paolo Grigolini introduced me, or tried to introduce me, to an extensive philosophical picture that involved what, according to him, could be the true deep meaning of the crucial events in natural phenomena.
At the time, I didn't fully understand his speech, but today I know that we were talking about the role of intermittency in complexity, that is to say, the role of transition events between self-organized metastable states in the cooperative dynamics of a complex system.\\
Today my understanding about the role of crucial, or critical, events is much improved, even if the speech that Paolo Grigolini gave me in that walk is largely forgotten and what little I have left in my memory I must probably understand still today.
In fact, when I gave my seminar, which aroused interest and a good discussion with Paolo Grigolini and other colleagues, Paolo Grigolini himself clarified the role of events in the problem in the consequent solution that I had presented, underlining that the role of events was central in my solution, but that this aspect was not still clear to me. 
In fact, as a good mathematical physicist, I used to interpret events as a theoretical simplification that is, an approximated assumption of a real behavior. This is actually not a wrong interpretation, but it represents a somewhat partial and reductive vision.\\
Regarding the topic of this chapter, the event-based approach to turbulence, me and Paolo Grigolini had also an interesting discussion on how to concretize our collaboration and, having specified my interest in turbulent diffusion, in an instant Paolo Grigolini found the connection with the well-known Manneville-Pomeau map \cite{manneville_1980,pomeau-manneville_1980}, a nonlinear model which, despite its simplification, captures the essence of turbulent bursting.

Our collaboration, which mainly concerned physical problems other than turbulence, then began in 2004 and continued until 2009. 
In those years I used to make frequent research visits
at both Center of Nonlinear Sciences in Denton (Texas)
and Department of Physics in Pisa (Italy).
This is a period of my life that I remember with pleasure from the point of view of both scientific research and personal acquaintances, not only with Paolo Grigolini but also with many of his and other students, partly experiencing the typical lifestyle of an American university.
In 2009, during the famous Grigolini's Friday afternoon meetings in Pisa, we started some discussions with some
researchers working in the fascinating world of neurosciences. As a consequence, the work on brain events begun, a research line we were immediately enthusiastic about, and that is nowadays still inspiring me.

\vspace{.1cm}
\noindent
I am therefore delighted to have this opportunity to thank
Paolo Grigolini for the contribution he made to my personal research path on anomalous diffusion and complex systems, which, thanks to him, I believe to be also the route of many others. This joined, in some sense, the other important contribution to my personal research view that comes from the collaboration with Prof. Francesco Mainardi of the Physics Department in Bologna
(see \cite{paradisi_caim15_ctrw_fract}), collaboration that began with the degree thesis and continued for a few years later. A path whose deep contents I was still partly unaware of and which today I am able to delineate more clearly, also thanks to the collaboration with Paolo Grigolini.



%
%
%


\nocite{*}
\bibliography{bib}

\end{document}